\documentclass[10pt]{iopart}

\usepackage{iopams}  
\usepackage{graphicx}
\usepackage{amssymb}
\usepackage{epstopdf}
\usepackage{float}
\usepackage{color}
\usepackage{subfigure}
\usepackage{multirow}
\usepackage{relsize}
\usepackage{booktabs}
\expandafter\let\csname equation*\endcsname\relax
\expandafter\let\csname endequation*\endcsname\relax

\usepackage{amsmath}
\usepackage[ngerman, english]{babel}
\usepackage{mathtools}

\newcommand{\cev}[1]{\reflectbox{\ensuremath{\vec{\reflectbox{\ensuremath{#1}}}}}}

\newcommand{\be}{\begin{equation}}
\newcommand{\ee}{\end{equation}}

\newcommand{\pd}[1]{\frac{\partial}{\partial{#1}}}

\newcommand{\bra}[1]{\langle{#1}|}
\newcommand{\ket}[1]{|{#1}\rangle}
\newcommand{\bkt}[2]{\langle{#1}|{#2}\rangle}

\newcommand{\dd}[1]{\mathrm{d}{#1}}

\newcommand{\ddn}[2]{\mathrm{d^{#1}}{#2}}

\newcommand{\+}[1]{\ensuremath{\mathbf{#1}}}

\def\editm{}
\newcommand{\specialcell}[2][c]{\begin{tabular}[#1]{@{}c@{}}#2\end{tabular}}
\usepackage{etoolbox}
\makeatletter
\newcommand{\mainmatter}{%
  \setcounter{footnote}{0}%
  \patchcmd{\@makefntext}{\fnsymbol}{\arabic}{}{}%
  \patchcmd{\@thefnmark}{\fnsymbol}{\arabic}{}{}%
  \def\@makefnmark{\textsuperscript{\arabic{footnote}}}%
}
\makeatother
\makeatletter
\newcommand\appendix@section[1]{%
  \refstepcounter{section}%
  \orig@section*{Appendix \@Alph\c@section: #1}%
  \addcontentsline{toc}{section}{Appendix \@Alph\c@section: #1}%
}
\let\orig@section\section
\g@addto@macro\appendix{\let\section\appendix@section}
\makeatother
\begin{document}


%
%
%
\ioptwocol
%



\title{Trilobites, butterflies, and other exotic specimens of long-range Rydberg molecules}

\author{Matthew T Eiles$^{1,2}$}

\address{$^1$Department of Physics and Astronomy, Purdue University, West Lafayette, Indiana, 47907, USA.}
\address{$^2$Max-Planck-Institut f\"{u}r Physik komplexer Systeme, N\"{o}thnitzer Str. 38, 01187 Dresden, Germany}
\ead{\mailto{meiles@pks.mpg.de}}

\begin{abstract}     
This Ph.D. tutorial discusses ultra-long-range Rydberg molecules, the exotic bound states of a Rydberg atom and one or more ground state atoms immersed in the Rydberg electron's wave function. This novel chemical bond is distinct from an ionic or covalent bond, and is accomplished by a very different mechanism: the Rydberg electron, elastically scattering off of the ground state atoms, exerts a weak attractive force sufficient to form the molecule in long-range oscillatory potential wells.  In the last decade this topic has burgeoned into a vibrant and mature subfield of atomic and molecular physics following the rapidly developing capability of experiment to observe and manipulate these molecules. This tutorial focuses on three areas where this experimental progress has demanded more sophisticated theoretical descriptions: the structure of polyatomic molecules, the influence of electronic and nuclear spin, and the behavior of these molecules in external fields. The main results are a collection of potential energy curves and electronic wave functions which together describe the physics of Rydberg molecules. Additionally, to facilitate future progress in this field, this tutorial provides a general overview of the current state of experiment and theory.

\end{abstract}
%

\mainmatter
\section{Introduction}
\label{sec:intro}
Long-range Rydberg molecules (LRRMs), first predicted nearly two decades ago \cite{Greene2000} and subsequently observed almost a decade later \cite{Bendkowsky}, are a stunning highlight of the nearly century-long study of the interactions between Rydberg atoms and neutral systems. In 1934 Edoardo Amaldi and Emilio Segr\'{e}\footnote{Who, along with Oscar D'Agostino, Ettore Majorana, Bruno Pontecorvo, and Franco Rasetti were known as the ``\editm{Via Panisperna} Boys'', a research group led by Enrico Fermi and most known for their foundational work in nuclear physics.} observed that Rydberg atoms could be excited even when immersed in a dense gas of other atoms (``perturbers''), although at an energy shifted from that of an isolated Rydberg atom \cite{Segre}.  Curiously, depending on the species of perturber, these line shifts could be either blue or red detuned from the atomic line. This behavior qualitatively contradicted the classical model, which treats the surrounding gas of polarizable atoms as a dielectric material and predicts only a red shift. Fermi resolved this mystery by developing a quantum scattering theory which introduced foundational concepts like the scattering length and zero-range pseudopotential \cite{Fermi}. His model recognized that a ground state atom's polarization potential extends over only a fraction of the Rydberg electron's enormous de Broglie wavelength, and thus it only adds a scattering phase shift to the Rydberg wave function\footnote{This standard ``origin story'' of Rydberg molecules fails to acknowledge simultaneous independent measurements performed in Rostock by F\"{u}chtbauer and coworkers \cite{Fuchtbauer1,Fuchtbauer2,Fuchtbauer3}, as was graciously pointed out to me by T. Stielow and S. Scheel from that same university}. The resulting energy shift is equivalent to that provided by a delta function potential located at the perturber and proportional to the electron-atom scattering length. This simple concept became the basis for many subsequent studies of the interactions between Rydberg atoms and neutral systems, especially after Omont generalized it to incorporate energy-dependent scattering lengths and arbitrarily high partial waves \cite{Omont}. With this pseudopotential researchers studied such diverse phenomena as collisional broadening, Rydberg quenching, $l$-changing collisions, and charge transfer  \cite{LebedevFabrikant,LebFabrikant2,Fabrikant1986,Masnou-Seeuws}. Many properties of these phenomena exhibited oscillatory behavior, reflecting oscillations in the Rydberg wave function through the delta function potential.

From a different direction -- the study of excited molecular states -- hints of this same oscillatory behavior were also found, such as in the Born-Oppenheimer potential energy curves (PECs) of some excited heteronuclear dimers \cite{DuGreene87,DePrunele2,DuGreene89}. Although the energy regime of these molecular states is unlike the regime where the Fermi pseudopotential was first designed, the pseudopotential still semi-quantitatively reproduced PECs calculated through more sophisticated approaches \cite{DuGreene87,DePrunele}. Thus, by the 1980s, theorists were aware that the Fermi pseudopotential accurately described Rydberg-neutral interactions, and that excited (but not yet ``Rydberg'') molecules existed and possessed some oscillatory features. The key concepts underlying long-range Rydberg molecules (LRRMs) were therefore at hand, but it was not until the realization of Bose-Einstein condensates (BEC) in the mid-1990s that it became possible to fully forge the link between these concepts: \textit{Rydberg molecules} consisting of a Rydberg atom loosely bound to a distant perturber\footnote{ These molecules are distinct from \textit{Rydberg states of molecules} (such as H$_2$ \cite{Herzberg, Jungen}), which are ``typical'' molecules bound by traditional chemical bonds with small internuclear distances, but which have a highly excited electron. These are the molecular analogue of a Rydberg atom.}.   The condensate's high density -- so that two atoms could be found at the right proximity for photoassociation -- and extremely cold temperature -- so that thermal motion would not immediately destroy the fragile molecular bond -- could provide the right conditions. This led Greene and coworkers in 2000 \cite{Greene2000} to make a key leap: they took the Fermi pseudopotential seriously as the foundation for Born-Oppenheimer PECs for a Rydberg and a ground state atom, and showed that this led to a new type of chemical bond. They calculated molecular spectra and found that these molecules were stable at internuclear distances of hundreds of nanometers. Furthermore, in some configurations the molecules would exhibit surprisingly non-trivial properties due to the high Coulomb degeneracy. Most notably, they possess large \textit{permanent electric dipole moments} and electronic wave functions resembling the trilobite fossil of antiquity\footnote{Trilobites were marine anthropods having distinctively ridged exoskeletons with three lobes down the length of the body. They declined into extinction about 250 million years ago, leaving fossils of thousands of different species worldwide.}. Soon after, a second molecular species with a wave function shaped like a butterfly were predicted \cite{HamiltonGreeneSadeghpour,KhuskivadzeJPB}. These zoomorphic colloquialisms have persisted. 

The number of atomic species successfully laser cooled to ultracold temperatures grew rapidly in the last two decades, as did the ability of experimentalists to excite Rydberg atoms in such an environment \cite{BECRb, BECNa, BECCa, BECSr,HydrogenBECPRLs,HydrogenBECPRLs2,Cr,dysprosium,erbium}. It was thus only a matter of time before these molecules, if the predictions were correct, would be observed. Indeed, within a decade of Greene \textit{et al}'s prediction, the group of Tilman Pfau observed vibrational spectra of LRRMs in ultracold rubidium \cite{Bendkowsky}. This confirmed the basic veracity of the theory and sparked considerable interest in the larger community. Within a few years evidence for all the predicted molecular states in a variety of atomic species had been gathered and was accompanied by renewed theoretical interest in these exotic molecules \cite{BoothTrilobite,Butterfly}.
\subsection{Outline and related work}
This tutorial describes the current state of and future prospects for this theory.  It focuses on a pedagogical description that can serve as a foundation for future exploration, and is intertwined with a summary of the experimental impetus for these theoretical developments. Although this tutorial is primarily a review of previously published material, in several places original calculations are presented.   Section  \ref{sec:inter} reviews the theory of electron collisions and spectroscopy in the context of Rydberg atoms and electron-atom scattering phase shifts. This section is tailored for researchers unfamiliar with Rydberg atoms, and can be skipped by the expert reader with the caveat that much of the notation used in later sections is only defined here. Section \ref{sec:primer} describes the features common to nearly all LRRMs and provides the theoretical ``skeleton'' for the following sections, which focus on three particularly interesting aspects of the molecular structure. First, section \ref{sec:polyintro} elucidates the structure and experimental signatures of polyatomic LRRMs. Next, Section \ref{sec:spinintro} incorporates all relevant spin degrees of freedom. These modify the molecular states due to the Rydberg fine structure, the hyperfine structure of the perturber, and the relativistic spin-orbit splitting of the electron-atom scattering. These are necessary for a theoretical description of similar accuracy to what is now experimentally attainable. Finally, section \ref{sec:fieldstudies} reviews how these molecules interact with and can be controlled by electric or magnetic fields. These can be applied externally in the laboratory or generated by the dipole moments of other LRRMs. Section \ref{sec:conclusions} concludes with speculations for the future. 

Four other publications have similar aims as this tutorial. Ref. \cite{marcassareview} overlaps section \ref{sec:primer} and also describes a second class of highly excited molecules called Rydberg macrodimers \cite{Boisseau,Farooqi,DeiglmayrRydRyd,RaithelRydRyd}. These are formed by two Rydberg atoms bound weakly together by long-range multipole interactions and have bond lengths exceeding one micron\footnote{ Note that, unless otherwise specified, we always mean the ``trilobite''-type of Rydberg molecule bound by the Fermi pseudopotential rather than these two other similarly named molecules.}. Ref. \cite{CsReview} reviews experiments on both types of molecules in Cs, and Ref. \cite{RydbergRev} reviews experiments in the high density regime, complementing the perspective given in Sec. \ref{sec:polyintro}. Ref. \cite{Hosseinreview} also reviews both types of molecules, and it discusses some of the spin effects reviewed here. Several other relevant reviews are either highly specific (Ref. \cite{Gaj2016} describes experimental aspects of field control) or quite generic (Ref. \cite{DunningRev} covers Rydberg states of alkaline-earth atoms and Ref. \cite{GallagherPillet} discusses Rydberg interactions). This tutorial is designed to complement these reviews by tailoring the discussion to a more pedagogical focus.  We use atomic units throughout except when specified.

 \section{Physics background: high energy Rydberg atoms and low energy collisions}
\label{sec:inter}
At first glance, the principal components of LRRMs appear rather paradoxical. An atom in an ultracold gas, cooled to just a few hundred nanokelvin, absorbs several electron volts of energy from a laser beam. The highly excited electronic wave function swells, spanning several thousands of angstroms in diameter. The electron, storing nearly all of its energy in the Coulomb potential between it and the distant positively charged ionic core of the atom, has very low velocity and hence a large de Broglie wavelength. As such its interaction with a perturber -- whose spatial extent is dwarfed by the Rydberg wavelength -- can be described by the Fermi pseudopotential. In this way the highly energetic and spatially extended Rydberg atom and the ultra-low-energy collision between an electron and a point-like perturber conspire to form LRRMs. Of particular advantage to the theorist is that this three-body system contains only two-body interactions. The simplest, between the ionic core and the perturber, is given by the potential
\be
\label{eq:polarizationotential}
V_\text{ion-atom}=-\frac{\alpha}{2R^4},
\ee
where $\alpha$ is the perturber's polarizability and $R=|\vec R|$ is the internuclear distance.  The first half of this section reviews the second, and  strongest, interaction: the Coulomb attraction between the electron and the ionic core. The second half discusses the electron-perturber interaction, in particular its determination via Fermi's model by scattering phase shifts.  This section relies on scattering and quantum defect theory covered in greater detail in e.g. Refs. \cite{OrangeRMP, FanoRau,FriedrichBook}.

\subsection{Rydberg atoms}
\label{sec:interryd}
We study first Rydberg atoms, notorious for their exaggerated properties such as large size, long lifetime, and powerful long-range interactions \cite{GallagherBook,Stebbings}. None of these properties can be characterized accurately without knowing the Rydberg spectrum of that particular atom. This spectrum can either be regular and predictable, as in the alkali atoms, or highly complex due to mixing between intertwined Rydberg series, as in the alkaline earth atoms. The powerful theoretical tools of multichannel quantum defect theory (MQDT) and eigenchannel $R$-matrix theory can be used to disentangle and interpret the spectra of the outer valence electron(s) \cite{ OrangeRMP,FanoRau, FanoJOSA, AymarReview1984, AymarTelmini1991,CookeCromer,Seaton}. In this section we will use quantum defect theory to understand both classes of spectra. A solid grasp of the spectrum of one-electron Rydberg states of alkali atoms is essential to understanding any aspect of Rydberg physics. We also provide a glimpse into the rich physics of two-electron Rydberg spectra, hinting at the diverse range of behavior exhibited by atoms beyond the first column of the periodic table.
\subsubsection{Spectra of alkali atoms.}
\label{subsec:alkaliryd}

Alkali atoms are ubiquitous in ultracold laboratories due to the conceptual and experimental simplicity of their sole valence electron.  This electron, when excited to a Rydberg state, only interacts with the other electrons over a very small region of its total volume. It is shielded from the full attraction of the atomic nucleus by the other, more tightly bound, electrons. It is thus an excellent approximation to include the shielding, polarization, and exchange effects of the deeply bound electrons within a model potential $V(r)$, which approaches the Coulomb potential as $r$ increases and the effects of the shell electrons fade away. 

The Rydberg wave function $\Psi_{nlm}(\vec r)$ is an eigenfunction of the time-independent Schr\"{o}dinger equation,
\begin{align}
\label{eq:ryd1}
E\Psi_{nlm}(\vec r)&=\left(-\frac{1}{2}\nabla_r^2 + V(r)\right)\Psi_{nlm}(\vec r);\\
\Psi_{nlm}(\vec r) &= \frac{u_{nl}(r)}{r}Y_{lm}(\theta,\phi)\label{eq:ryd2}.
\end{align}
We use spherical coordinates since $V(r)$ is spherically symmetric.  This results in a separable wave function where $Y_{lm}(\theta,\phi)$ is a spherical harmonic. The quantum numbers $l$ and $m$ give the orbital angular momentum and its projection on the $z$ axis, respectively. The principal quantum number $n$ is related to the energy $E$, and fixes the number of radial nodes of a bound state of Eq. \ref{eq:ryd1} to be $n_r = n-l-1$. Together with $l$ and $m$ it defines a complete set of quantum numbers and uniquely defines the eigenenergies $E_{nl}$. All that remains is to solve the radial Schr\"{o}dinger equation,
\be
\label{rydbergham}
0=\left(-\frac{1}{2}\frac{d^2}{dr^2}+\frac{l(l+1)}{2r^2}+V_l(r)-E_{nl}\right)u_{nl}(r).
\ee 
We have specialized here to an $l$-dependent model potential $V_l(r)$\footnote{The dependence of this potential operator on $l$ makes it non-local, but this creates no problems in our treatment.}. Many parameterizations of this potential exist. We have used that of Ref. \cite{Marinescu},
\begin{align}
\label{eq:modelpotential}
V_l(r) &= -\frac{Z_l(r)}{r} - \frac{\alpha_c}{2r^4}\left[1 - e^{-(r/r_c)^6}\right],\\
Z_l(r) &= 1 + (Z - 1)e^{-a_1r}-r(a_3 + a_4r)e^{-a_2r},
\end{align}
where $Z$ is the nuclear charge, $a_1$, $a_2$, $a_3$, and $a_4$ are fit parameters, $\alpha_c$ is the static dipole polarizability of the positive ion, and $r_c$ cuts off the unphysical behavior of the $r^{-4}$ potential at the origin. Ref. \cite{Marinescu} determined these parameters by fitting calculated $E_{nl}$ from Eq. \ref{rydbergham} to experimentally obtained atomic energy levels. Once $V_l(r)$ is determined one can solve Eq. \ref{rydbergham} numerically to find the energies of higher lying Rydberg states. However, this quickly grows tedious as the energy levels become densely spaced while the wave functions grow spatially diffuse. It also provides little to no information about the general properties of the spectrum. An analytic theory is thus necessary.

Quantum defect theory exploits one central idea:  the highly excited Rydberg electron traverses a large domain of space, defined by $r\ge r_0$, where $V_l(r)$ is synonymous with the Coulomb potential. In this region of space the two linearly independent solutions $f_l(r)$ and $g_l(r)$ of Eq. \ref{rydbergham} are given analytically in terms of confluent hypergeometric functions\footnote{The properties of these functions are rather complicated, but they are unnecessary for the present discussion.} \cite{OrangeRMP,Seaton,StrinatiGreene}. For positive energy $E = \frac{k^2}{2}$ their asymptotic behavior at large $r$ is 
\begin{align}
\label{eq:fposenco}
f_l(r)&\to (2/\pi k)^{1/2}\sin\left(kr+\frac{1}{k}\ln r + \eta_l\right),\\
\label{eq:gposenco}
g_l(r)&\to -(2/\pi k)^{1/2}\cos\left(kr+\frac{1}{k}\ln r + \eta_l\right).
\end{align}
These functions are energy normalized and include the Coulomb phase $\eta_l$ \cite{OrangeRMP}. Since they are linearly independent, any valid wave function for $r>r_0$ must be a linear combination of these two functions since the non-Coulombic potential is restricted to $r<r_0$. This superposition is written
\be
\label{eq:positiveenergysol}
u_{nl}(r) = \mathcal{N}\left[f_l(r) - \tan\delta_l g_l(r)\right],r\ge r_0,
\ee
where $\mathcal{N}$ is a normalization constant and $\delta_l$, as in standard scattering theory, is the phase shift for the $l$th partial wave. This phase reflects the mixing of solutions caused by the non-Coulomb part of $V_l(r)$, since $f_l(r)$ is the physical solution (obeying the proper boundary conditions) for the pure Coulomb potential.  The ``quantum defect'' is related to this phase through $\mu_l=\delta_l/\pi$. It is readily determined by  computing\footnote{Using, for example, Numerov's algorithm.} the solution $F(r)$, $0\le r\le r_0$, of Eq. \ref{rydbergham} subject to the boundary condition $F(0)=0$. $F(r)$ is obtained using a small-scale numerical calculation since this region of space is much smaller than the span of the actual Rydberg wave function, and additionally it can be computed at a small, but arbitrary, positive energy. Since we are interested in highly excited Rydberg states at energies satisfying $1\gg|E|$, we expect over this small $r$ range that $V(r)$, with its massive Coulomb forces, dominates the total energy: $|V(r<r_0)|\gg  |E|$. As a result $F(r)$ is insensitive to the exact value, and even the sign, of the energy for small $r$. Matching $F(r)$ along with its derivative $F'(r)$\footnote{Throughout this tutorial, a primed function represents its derivative with respect to its full argument, $f'(x) = \frac{df(x)}{dx}$.} to the asymptotic solutions at $r_0$ determines $\mu_l$:
\be
\label{eq:defnmu}
\tan\pi\mu_l = \left.\frac{F'(r)f_l(r) - F(r)f_l'(r)}{F'(r)g_l(r)-F(r)g_l'(r)}\right|_{r=r_0}.
\ee

Eqs. \ref{eq:fposenco}-\ref{eq:defnmu} have set the form of the wave function and, by requiring continuity at $r_0$, applied one of its two boundary conditions. Furthermore, since $F(r)$ is nearly independent of $E$ and the energy-normalized solutions used to construct the wave function are ``nearly analytic'' in energy \cite{Seaton}, $\mu_l$  is a very smooth function of energy. We have thus combined a small numerical calculation over the region of complicated non-Coulomb potential with the analytic Coulomb functions valid outside of this range to obtain a parameter, the quantum defect $\mu_l$, which is essentially constant for all Rydberg states of interest (those having $n\gtrapprox10$). The next step is to determine the quantized Rydberg eigenspectrum in terms of $\mu_l$ by analytically imposing the second boundary condition: the bound-state wave function must be normalizable.

We first need the asymptotic wave function (Eq. \ref{eq:positiveenergysol}) at negative energy, $E = -\frac{k^2}{2} = \frac{(i\kappa)^2}{2}$. The regular and irregular functions  can be obtained by careful analytic continuation $k\to i\kappa$ of the exact solutions\footnote{There are considerable technical details involved due to a small non-analyticity in the solutions at $E = 0$ and the desire for real solutions both below and above threshold. Refs. \cite{OrangeRMP,Seaton,StrinatiGreene} present several different approaches using the properties of confluent hypergeometric functions or WKB theory arguments. It is for this reason that one cannot simply analytically continue the asymptotic solutions in Eqs. \ref{eq:fposenco} and \ref{eq:gposenco}, although one can get the essence by considering how the $kr$ and $\frac{1}{k}\ln r$ terms lead to terms of the form $e^{\pm \kappa r}$ and $r^{\pm \nu}$ in Eqs. \ref{eq:fnegenco} and \ref{eq:gnegenco}.}, giving
\begin{align}
\label{eq:fnegenco}
f_l&\to \left(\pi\kappa\right)^{-1/2}\left(\sin\beta D^{-1}r^{-\nu}e^{\kappa r} - \cos\beta Dr^\nu e^{-\kappa r}\right),\\
\label{eq:gnegenco}
g_l&\to -\left(\pi\kappa\right)^{-1/2}\left(\cos\beta D^{-1}r^{-\nu}e^{\kappa r} +\sin\beta Dr^\nu e^{-\kappa r}\right),
\end{align}
where  $\nu = \kappa^{-1}$ and $\beta = \pi(\nu -l)$. $D$ is a parameter\footnote{The explicit form of $D$,  which depends on $\nu$ and $l$, is not needed in any of what follows}. The asymptotic function $u_{nl}(r)$ follows by inserting these expressions into  Eq. \ref{eq:positiveenergysol}:
\begin{align}
u_{nl}(r)
\to\frac{\mathcal{N'}}{\cos\pi\mu_l}&\Bigg[\sin(\beta + \pi\mu_l)D^{-1}r^{-\nu} e^{\kappa r}\\&-\cos(\beta+\pi\mu_l)Dr^\nu e^{-\kappa r}\Bigg].\nonumber
\end{align}
To ensure that $u_{nl}(r)$ is normalizable the diverging $e^{\kappa r}$ term of this solution must be totally eliminated. This is possible if its coefficient, $\sin(\beta + \pi\mu_l)$, vanishes. Thus,  
\be
\pi(\nu - l+\mu_l) = N\pi,\,\,N\in\mathcal{Z}.
\ee
We define $n=N+l$, and hence,
\be
\label{eq:quantumdefect1}
E_{nl} = -\frac{\kappa^2}{2} = -\frac{1}{2\nu^2} = -\frac{1}{2(n-\mu_l)^2}.
\ee
This quantization condition is the famed Rydberg formula defining the infinite number of bound state energies of Eq. \ref{rydbergham}. To improve this formula one can add small correction terms to the quantum defect to compensate for the weak energy dependence ignored in this derivation (see Eq. \ref{eq:quantumdefects}).  We obtain a simple analytic expression for the corresponding eigenfunctions by noticing that Eq. \ref{eq:positiveenergysol} can now be written
\be
\label{eq:newsuperposition}
u_{nl}(r) = -f_l(r)\cos\beta - g_l(r)\sin\beta.
\ee
For non-zero quantum defects the coefficient of $g_l(r)$ does not vanish, and hence this linear combination diverges as $r\to 0$. However, for our application -- \textit{long-range} Rydberg molecules -- we never need the exact Rydberg wave function at such small distances\footnote{And if this is ever required, then a numerical solution of Eq. \ref{rydbergham} is straightforward using Numerov's algorithm since the eigenenergy is already determined.}. The superposition in Eq. \ref{eq:newsuperposition} is related to the Whittaker function\footnote{WhittakerW$[\dots]$ in Mathematica.} \cite{OrangeRMP,WhitWats} through
\be
\label{fdefradial}
u_{nl}(r) = \frac{W_{\nu,l+1/2}\left(\frac{2r}{\nu}\right)}{\sqrt{\nu^2(\Gamma(\nu+l+1)\Gamma(\nu-l)}}.
\ee
This wave function is normalized so that $1 = \int_{r_0}^\infty |u_{nl}(r)|^2\dd{r}$, i.e. we ignore the tiny contribution from $0\le r \le r_0$ that is present in the exact wave function.\footnote{The expression in Ref. \cite{OrangeRMP} is energy-normalized. To go between normalization conventions one can multiply the energy-normalized functions by $(\nu^3+\frac{d\mu_l}{d\varepsilon})^{-1/2}$. We have used the approximate conversion factor $\nu^{-3/2}$ since the quantum defects are essentially constant in energy. } 

The key point of this discussion is that, since the interaction of a Rydberg electron with the ionic core extends over such a small range, the key differences between its spectrum and that of hydrogen are encapsulated by a few essentially energy-independent quantum defects. These can be determined numerically, using Eq. \ref{eq:defnmu}, or fit to measured energies. As a result, Rydberg wave functions and energy levels are excellently described entirely analytically using Eqs. \ref{fdefradial} and \ref{eq:quantumdefect1}, respectively.

   \begin{table*}[ht]
\resizebox{\textwidth}{!}{%
\begin{tabular}{|| c|| c c| c|| c c| c|| c c||}
\hline
 {\bf Li} & $\mu(0)$ & $\mu'(0)$ &{\bf Na} & $\mu(0)$ & $\mu'(0)$ &{\bf K} & $\mu(0)$ & $\mu'(0)$  \\ 
  \hline \hline
  $s_{1/2}$& 0.3995101 & 0.0290 & $s_{1/2}$& 1.347964 & 0.060673 & $s_{1/2}$& 2.1801985 & 0.13558   \\
 $p_{1/2}$ & 0.0471780 & -0.024 & $p_{1/2}$& 0.855380 & 0.11363  & $p_{1/2}$& 1.713892 & 0.233294 \\
 $p_{3/2}$ & 0.0471665 & -0.024 & $p_{3/2}$& 0.854565 & 0.114195 &$p_{3/2}$& 1.710848 & 0.235437 \\
 $d_{3/2}$ & 0.002129 & -0.01491 & $d_{3/2}$& 0.015543 & -0.08535 & $d_{3/2}$& 0.2769700 & -1.024911  \\
 $d_{5/2}$ & 0.002129 & -0.01491& $d_{5/2}$& 0.015543 & -0.08535 & $d_{5/2}$& 0.2771580 & -1.025635 \\
 $f_{5/2}$ & -0.000077 &0.021856 & $f_{5/2}$& 0.0001453 & 0.017312 & $f_{5/2}$& 0.010098 & -0.100224   \\
 $f_{7/2}$ & -0.000077 &0.021856 & $f_{7/2}$& 0.0001453 & 0.017312 & $f_{7/2}$& 0.010098 & -0.100224     \\
  \hline
 {\bf Rb} & $\mu(0)$ & $\mu'(0)$ &{\bf Cs} & $\mu(0)$ & $\mu'(0)$ & {\bf Sr} & $\mu(0)$ & $\mu'(0)$\\ 
  \hline \hline
  $s_{1/2}$& 3.1311804 & 0.1784 & $s_{1/2}$& 4.049325 & 0.2462 & $5sns^3$S$_1$ &3.371 & 0.5\\
 $p_{1/2}$ & 2.6548849 & 0.2900 & $p_{1/2}$& 3.591556 & 0.3714  & $5snp^3$P$_{2(1)}$ & 2.8719 (2.8824) & 0.446(0.407) \\
 $p_{3/2}$ & 2.6416737 & 0.2950 & $p_{3/2}$& 3.559058 & 0.374  & $5snp^3P_0$ & 2.8866 & 0.44\\
 $d_{3/2}$ & 1.34809171 & -0.60286& $d_{3/2}$& 2.475365 & 0.5554  & $5snd^3D_{3(2)}$ &2.612(2.662) & $-41.4(-15.4)$\\
 $d_{5/2}$ & 1.34646572 & -0.59600& $d_{5/2}$& 2.466210 & 0.067  & $5snd^3D_1$ & 2.673 &  -5.4\\
 $f_{5/2}$ & 0.0165192 & -0.085 & $f_{5/2}$& 0.033392 & -0.191    & $5snf^3F_{4(3)}$ & 0.120(0.120) & -2.4(-2.2)\\
 $f_{7/2}$ & 0.0165437 & -0.086 & $f_{7/2}$& 0.033537 & -0.191 & $5snf^3F_2$ & 0.120 & -2.2  \\
  \hline \hline
  & $\alpha$ (a.u.) & $\alpha_c$ (a.u.)&   & $\alpha$ (a.u.)  & $\alpha_c$ (a.u.) & & $\alpha$ (a.u.)  & $\alpha_c$ (a.u.)\\
  \hline
  {\bf Li} & 164.9$^{a},164.2^b$ & 0.1923 \protect\cite{Marinescu} & {\bf Na}  & 165.9$^a$,162.7$^b$ & 0.9448  \protect\cite{Marinescu} & {\bf K} & 307.5$^a$,290.6$^b$ & 5.3310\protect\cite{Marinescu}\\
    {\bf Rb} & 319.2$^{b}$ & 9.12  \protect\cite{Gallagher2016a}& {\bf Cs} &401$^b$ & 15.544 \cite{GallagherCS} & {\bf Sr}&186$^c$ & 86  \protect\cite{SrPlusPol} \\
\hline
\end{tabular}}
   \caption{Quantum defect parameters (used in Eq. \ref{eq:quantumdefects}) $\mu$, $\mu'$  and polarizabilities $\alpha$, $\alpha_c$ for the alkalis and strontium. 
 The quantum defects are from Ref. \protect\cite{GoyLi} and Ref. \protect\cite{NiemaxLi} (Li), Ref. \protect\cite{NiemaxLi} (Na and K), Ref. \protect\cite{LiRb,JamilRb} (Rb), Ref. \protect\cite{GoyCs,MoreCs} (Cs), and Refs. \protect\cite{SrQD1,SrQD2,SrQD3,SrQD4,DingSpec} (Sr). The different polarizabilities are calculated$^a$ using the model potential or measured in $^b$Ref. \protect\cite{updatedPols,Miffre2006,CsPolHigh,Polarizabilities} and$^c$ Ref. \protect\cite{SrPol}. Due to the multichannel physics discussed in the text the quantum defects for Sr are not guaranteed to be energy-independent over the entire Rydberg series, and should be confirmed for a given Rydberg state. }
 \label{tab:datatable2}
\end{table*}

A more realistic treatment of the Rydberg atom must include relativistic fine structure effects. One of these, the $p^4$ correction to the kinetic energy, is automatically included in the model potential since it is fit to empirical energies which intrinsically include this shift. More complicated is the spin-orbit coupling between the electron's orbital and spin angular momenta, which can be included with an additional model potential of the form \cite{GreeneAymar},
\be
V_{so}^{slj}(r) = \frac{(g-1)\alpha_\text{FS}^2}{2}\frac{1}{r}\frac{dV_l(r)}{dr}\left[\tilde{V_l}(r)\right]^{-2}\vec s_1\cdot\vec l,
\ee
where $\tilde{V_l}(r) = 1 - \frac{(g-1)\alpha_{FS}^2}{2}V_l(r)$, $\alpha_\text{FS}$ is the fine-structure constant, $\vec s_1$ is the electron spin operator, and $g$ is the electron $g$-factor. This potential couples the orbital and spin angular momenta, but it is diagonal in the total angular momentum $\vec j = \vec s_1+ \vec l$ and its projection $m_j$. We can generalize the Rydberg formula, Eq. \ref{eq:quantumdefect1}, to include the spin-orbit splitting by making it $j$-dependent:
   \be
\label{eq:Rydbergformula}
E_{n(s_1l)jm} = -\frac{1}{2(n-\mu_{(s_1l)j}(n))^2}.
\ee
To better match observed energy levels we use quantum defects with a linear energy dependence,
\be
\label{eq:quantumdefects}
\mu_{(s_1l_1)j}(n) = \mu_{(s_1l_1)j}(0) + \frac{\mu'_{(s_1l_1)j}(0)}{\left[n - \mu_{(s_1l_1)j}(0)\right]^2}.
\ee
Table \ref{tab:datatable2} displays these quantum defect parameters for low-$l$ Rydberg levels. For higher angular momenta, the quantum defects are determined almost entirely by the core polarization $\alpha_c$. Their values,
 \begin{align}
 \mu_l(n) &= \left(\frac{\alpha_c[3n^2 - l(l+1)]/4}{n^2(l-1/2)l(l+1/2)(l+1)(l+3/2)}\right)\label{eq:coredefects},
  \end{align}
  are given by calculating perturbatively the first-order energy shift of the polarization potential. 
  The fine structure splitting for these nonpenetrating high-$l$ ($l>3$) states is given by the formula obtained for hydrogen using the Dirac equation,
  \be
  \label{eq:finestrucsplitting}
  \Delta E_{n(s_1l)jm} = -\frac{\alpha^2}{2n^3}\left(\frac{1}{j+1/2} - \frac{3}{4n}\right).
  \ee
  Eqs. \ref{eq:quantumdefects}- \ref{eq:finestrucsplitting} thus fully specify the Rydberg spectrum of an alkali atom. 
 To complete this discussion we give the spin-dependent electronic wave function,
  \be
\label{eq:jdepefuncs}
\Psi _{n(ls_1)jm_j}(\vec r) = \sum_{m,m_1}C_{lm,s_1m_1}^{jm_j}\frac{u_{nlj}(r)}{r}Y_{lm}(\hat r)\chi_{m_1}^{s_1},
\ee 
where $\chi_{m_1}^{s_1}$ is the Rydberg electron's spin state. $u_{nlj}(r)$ is equivalent to $u_{nl}(r)$ but with $j$-dependent $\nu$.

\subsubsection{The multichannel Rydberg spectra of two-electron atoms.}
\label{MQDTrydbergs}
The previous section showed that a set of nearly energy-independent quantum defects defines the Rydberg spectrum  of alkali atoms. We now introduce the basic concepts of multichannel Rydberg systems by considering the spectra of atoms with two valence electrons.  Interest in such atomic species dates back decades \cite{OrangeRMP,AymarReview1984,GreeneAymar}, and has recently resurged in theoretical studies of Rydberg interactions \cite{Vaillant,LRinteractionsFrancisBooth,Eiles2015} and in ultracold Rydberg spectroscopy of atoms such as Sr, Ho, and Yb \cite{DingSpec,SrMeas2,holmium,YtRyd}. The discussion here is intended to spark increased interest in the possibilities of these atoms in the context of LRRMs and to present the reader with a more complete picture of the richness and intricacy of Rydberg atoms.
 
The non-relativistic Hamiltonian for the two valence electrons is
\be
\label{twoelec1}
H_{2e} = -\frac{1}{2}\nabla_{r_c}^2 - \frac{1}{2}\nabla_r^2 + V_{l_c}(r_c) + V_{l}(r) + \frac{1}{|\vec r_c - \vec r|}.
\ee
We have labeled the position operators of the two electrons $\vec r_c$ and $\vec r$ to clarify the following essential point: we focus on energy regimes far below the double ionization threshold, and hence the two-electron wave function $\Psi(\vec r_c,\vec r)$ vanishes when $r_c>r_0$. The size of $r_0$ is set by the spatial extent of the excited core states we wish to include, but typically is a few tens of atomic units. The model potential $V_{l_i}(r_i)$ is similar to Eq. \ref{eq:modelpotential}, except that it is modified to represent a doubly, rather than singly, charged positive ion\footnote{Ref.  \cite{OrangeRMP} gives some explicit expressions and fit parameters for the alkaline-earth atoms.}. 

To efficiently describe the six-dimensional $\Psi(\vec r_c,\vec r)$ we define a set of \textit{channel functions}, $\Phi_i(\omega)$, where $\omega$ refers to all coordinates except for $r$, and $i$ is a set of quantum numbers defining each channel. By definition, $\Phi_i(\omega)$ is an eigenfunction of a smaller Hamiltonian, $H'$, involving only the coordinates $\omega$.  Since $H'$ is spherically symmetric, it shares eigenstates with $\vec l_c^2$, $\vec l^2$, $\vec L^2$, and $L_z$, where $L$ is the total orbital angular momentum. The eigenvalue equation for the channel functions $\Phi_{i=l_clLM,\epsilon_{l_c}}(\omega)$ is
\begin{align}
\nonumber\left(\left[-\frac{\nabla_c^2}{2} + V_{l_c}(r_c)\right] + \frac{\vec l^2}{2r^2}\right)&\Phi_{l_clLM,\epsilon_{l_c}}(\omega) \\= \left(\epsilon_{l_{c}} +  \frac{l(l+1)}{2r^2}\right)&\Phi_{l_clLM,\epsilon_{l_c}}(\omega),
\end{align}
where $\epsilon_{l_c}$ is the eigenenergy of the inner electron whose Hamiltonian is contained in the square brackets. These channel functions define a complete set of basis functions to expand the full wave function into:
\be
\label{channelexp}
\Psi_{i'}(\vec r_c,\vec r) = \mathcal{A}\sum_ir^{-1}\Phi_i(\omega)F_{ii'}(r).
\ee 
 $\mathcal{A}$ denotes the antisymmetrization operator and $F_{ii'}(r)$ is the $i'$th linearly independent radial wave function in the $i$th channel.  A matrix of radial solutions $F_{ii'}$ is necessary because, after imposing boundary conditions at the origin but before imposing boundary conditions at infinity, an $N$-channel Schr\"{o}dinger equation has $N$ independent solutions in each channel. These unknown radial functions are found by projecting $H_{2e}\Psi(\vec r_c,\vec r)$ onto the channel functions $\Phi_i(\omega)$ to obtain  the \textit{coupled-channel equations},
\begin{align}
\label{eq:outersolns}
 \nonumber&\sum_{j=1}^N\Bigg[\left(-\frac{1}{2}\frac{d^2}{dr^2} +\frac{l_{i}(l_{i}+1)}{2r^2}+V_{l_i}(r)+\frac{1}{2\nu_i^2}\right)\delta_{ij}\\& \,\,\,+\int\frac{\Phi_i^*(\omega)\Phi_j(\omega)\dd{\omega}}{|\vec r_c-\vec r|}\Bigg] F_{ji'}(r) = 0,\,\,r\le r_0
 \end{align}
 where we have truncated to $N$ channels. Exchange effects recrease rapidly for $r >r_0$ and are neglected in these equations.
 The channel-dependent quantum number $\nu_i$ defines the energy of the outer electron via  $(E - \epsilon_{i})=-\frac{1}{2\nu_i^2}$, where $E$ is the total energy. 
 
Now that the channel structure of the wave function is laid out, we can generalize the single channel equations. For $r>r_0$, $V_{l_i}(r) = -\frac{2}{r}$, and using the multipole expansion of $1/|\vec r_c - \vec r|$ we find that the coupling term in the second line of Eq. \ref{eq:outersolns} is, to first order, a diagonal Coulomb potential $\frac{1}{r}\delta_{ij}$\footnote{Higher multipolar coupling terms can be ignored for now provided $r_0$ is not too small, and they can treated perturbatively later if necessary.}. Thus, the coupled channel equations decouple into radial Coulomb-Schr\"{o}dinger equations asymptotically. Following Eq. \ref{eq:positiveenergysol}, we express $F_{ii'}(r)$ as a linear combination of $f_i(r)$ and $g_i(r)$, the two linearly independent solutions in channel $i$. The multichannel generalization of Eq. \ref{eq:positiveenergysol} is \cite{OrangeRMP}
\begin{align}
\label{mqdtlongrange}
\Psi _{i^{\prime }}&=\mathcal{A}\sum_{i}\Phi _{i}(\omega )\left[f_i(r)\delta_{ii'}-g_i(r)K_{ii'}(r)\right],r>r_0,
\end{align}
where $\underline{K}$, the \textit{reaction matrix}, is related to the phase shift matrix through the equation $\underline{K} = \tan\underline{\delta}$ \footnote{We could also use the scattering matrix $\underline{S} = e^{2i\underline{\delta}}$ to set up this derivation. By converting this exponential function into trigonometric form we can identify the relationship between the scattering and reaction matrices,
\be
\underline{K} = \frac{i(\underline{I} - \underline{S})}{\underline{I} + \underline{S}}. \nonumber
\ee
We prefer the $K$-matrix formalism because all arithmetic is explicitly real.}.
The form of this equation provides physical intuition for the mathematical statement above about the number of linearly independent solutions. As the Rydberg electron, in channel $i$, careens into the ion and interacts with the inner electron, it swaps angular momentum and energy with this electron and exits the interaction region in channel $i'$ through the matrix element $K_{ii'}$. 

As in the single-channel case, we impose long-range boundary conditions by eliminating the exponential growth of $f_i$ and $g_i$ as $r\to\infty$. Using Eqs. \ref{eq:fnegenco} and \ref{eq:gnegenco} reveals the relevant exponential terms, $f_i\to D\sin\beta_ie^{\kappa_ir}r^{-\nu_i}$ and $g_i\to-D\cos\beta_ie^{\kappa_ir}r^{-\nu_i}$. As before, $\underline\beta = \pi(\underline\nu - \underline l)$.  We utilize the flexibility to choose any superposition of linearly independent solutions $\Psi_{i'}$ to form a wave function $\Psi = \sum_{i'}\Psi_{i'}B_i'$. At large $r$,
\be
\label{sdada}
\Psi\to  De^{\underline{\kappa}r}r^{-\underline{\nu}}(\sin\underline\beta + \cos\underline\beta \underline K)\vec B,
\ee
where all matrices except for $\underline{K}$ are diagonal. 
This expression must vanish, and so $(\sin\underline\beta + \cos\underline\beta \underline K)\vec B=0$. This system of equations has a non-trivial solution if the determinant vanishes,
\be
\label{deteqn}
\det(\tan\underline\beta + \underline K) = 0.
\ee  This defines a relationship between the $\nu_i$ which, in combination with energy conservation fully determines the energies. After Eq.  \ref{deteqn} is solved we can set $\underline K\vec B= -\tan\underline\beta\cdot\vec B$, and hence
\begin{align}
\label{mqdtlongrange}
\Psi &=\mathcal{A}\sum_{i}\Phi _{i}(\omega )\left[f_i(r)\cos\pi\beta_{i}+\sin\pi\beta_{i}g_i(r)\right]\frac{B_{i}}{\cos\beta_{i}}. 
\end{align}
This linear combination of $f_i$ and $g_i$ is just the channel $u_{nl}(r)$ function, and so:
\begin{align}
\label{MQDTwf}
\Psi(\vec r_c,\vec r) &= \sum_i\frac{1}{r}\Phi_i(\omega)u_{n_il_i}(r)\frac{-B_i}{\cos\beta_i},\,\,r>r_0
\end{align}
This wave function, due to channel coupling from the electron-electron interaction, is a mixture of channel functions weighted by the coefficients $-B_i/\cos\beta_i$.

\subsubsection{Determination of the $K$ matrix and Lu-Fano plots.} 
Eq. \ref{deteqn} and \ref{MQDTwf} show that we can obtain the energies and wave functions of multichannel Rydberg states from the $K$-matrix through a similar, but algebraically more involved, process as in the single channel case. We now turn to the practical matter of how to obtain $\underline{K}$, focussing on a semi-empirical method which also illustrates some important concepts of these multichannel Rydberg states.  Ref.  \cite{OrangeRMP} explains the nearly \textit{ab initio} determination of $\underline{K}$ using the $R$-matrix method, which is also briefly summarized in the context of electron-atom scattering in the following section. 

Of critical importance is the representation that diagonalizes or approximately diagonalizes the Hamiltonian also  diagonalizes $\underline K$. In the previous discussion we constructed channels using the $LS$-coupling scheme: the orbital ($l_c$ and $l$) and spin ($s_c$ and $s$) angular momenta of the two electrons are coupled separately to form $L$ and $S$. These are subsequently coupled to form the total angular momentum $J$, and the channel functions are $\ket{(l_cl)L(s_cs)S]JM_J}$.  $H$ is approximately diagonal in this coupling scheme since non-relativistic effects are so far ignored, and therefore so is $\underline{K}$, $K_{ii^{\prime
}}^{(LS)}=\delta _{ii^{\prime }}\tan \pi \mu _{i}$.

We illustrate this with an example: the $J=0$ Rydberg states of silicon, which has the ground state configuration Ne $3s^2 3p^2$. For each parity there are two relevant $LS$-coupled channels: $^3P^e(d)$, $^3P^e(s)$, $^1S^o(p)$, and $^3P^o(p)$, where $(l)$ labels the Rydberg electron's angular momentum and $^{2S+1}L^\pi$ is the standard term symbol. An approximately energy-independent $K$-matrix is extracted from measured energy levels for these four configurations \cite{BrownGinterGinter,BGG2,BGG3}. These quantum defects are nearly constant in energy over several low-lying excited states, confirming the basic principle of quantum defect theory. 

This approach is disrupted by the spin-orbit splitting of $\Delta E=35.7$meV between the $j_c = 1/2$ and $j_c = 3/2$ states of the Si$^+$ ion, where $j_c=l_c+s_c$. When the Rydberg electron is near the core this splitting is dominated by the strong electrostatic and exchange interactions, and since the long-range potential far from the core is still a purely diagonal Coulomb potential one might naively think that this splitting has essentially no effect on the Rydberg spectrum. However, energy conservation requires that the total energy $E$ of the system be partitioned between the two electrons, and hence the channel quantum numbers are defined relative to these two different thresholds $\epsilon_{3/2}$ or $\epsilon_{1/2}$:
\be
E = \epsilon_{3/2} - \frac{1}{2\nu_{3/2}^2} = \epsilon_{1/2} - \frac{1}{2\nu_{1/2}^2}.
\ee
Clearly, the kinetic energy available to the Rydberg electron depends very sensitively on the state of the inner electron, which in turn causes the electron to accumulate phase at very different rates depending on the $j_c$ state of the core. $LS$ coupling is fundamentally unable to include this non-perturbative effect as $j_c$ is not a good quantum number in this coupling scheme, and so we must use a different set of quantum numbers for the long-range behavior of the Rydberg electron. The $jj$-coupling scheme
represented by the ket $|[(l_{c}\frac{1}{2})j_{c}(l\frac{1}{2}%
)J_e]jM_{J}\rangle $ can accomplish this since it explicitly labels the $j_c$ and $j=l+s$.

This is an example of a very generic problem in atomic and molecular physics: we have a physical system obeying different symmetries, and therefore described by different sets of quantum numbers, in different regions of space. It can be tackled using the powerful technique of a frame transformation \cite{ChrisFra,LeeLu,OrangeRMP}. In the present case, it is only once $r\sim1000a_0$ that the channel radial wave functions begin to dephase. We can still use $LS$-coupling at small $r$ to obtain a $K$-matrix, and then in the region where both coupling schemes are roughly equivalent we can simply project the wave function in the $LS$-coupling scheme onto the $jj$-coupling scheme. In the present case, this projection is effected by a  \textquotedblleft geometric\textquotedblright\ orthogonal frame
transformation matrix $U_{ij}$. This recoupling matrix rotates the $K$-matrix from the $LS$ coupling scheme to the $jj$-coupled representation \cite{LeeLu} and is given by standard angular momentum algebra \cite{Varsh}:
\begin{align*}
U_{ij} &= \sqrt{\lfloor j_c \rfloor\lfloor j_e \rfloor\lfloor L\rfloor\lfloor S\rfloor}\left\{\begin{array}{ccc} l_e &s_e &j_e\\l_c & s_c & j_c\\ L & S & J\end{array}\right\},
\end{align*}
where $\lfloor x\rfloor = (2x+1)$ and $\{\dots\}$ is a Wigner 9J Symbol. The $jj$-coupled $K$ matrix is obtained via $%
K_{ii^{\prime }}^{(jj)}=\sum_{jj'}U_{ij}K_{jj^{\prime }}^{(LS)}U_{j^{\prime }i^{\prime
}}^{\dagger }$.

\begin{figure}[t]
\begin{centering}
{\normalsize \includegraphics[width = \columnwidth]{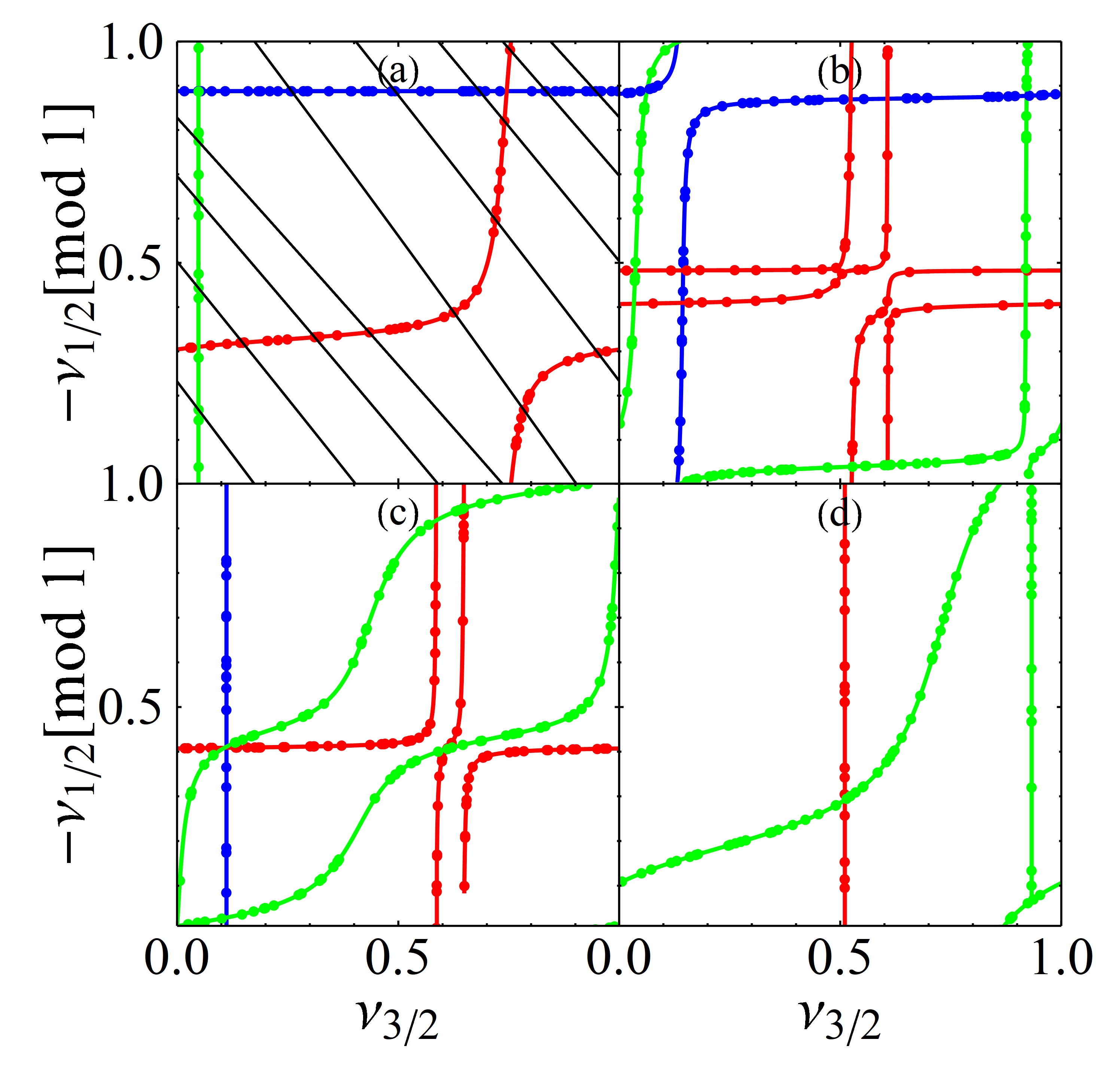} 
\vspace{-20pt}
}
\caption{Lu-Fano plots for a) $J=0$, b) $J=1$, c) $J=2$, and d) $J=3$
symmetries. Blue points are $l\approx 0$ odd parity; red are $l \approx 1$ even parity, and green are $l \approx 2$ odd parity.
Intersections of the solid curves (Eq. \ref{deteqn}) with the diagonal lines (Eq. \ref{qnumrel}; only a few representative ones are shown) give the positions of bound states (points). This figure is modified from Ref. \cite{Eiles2015}. }
\label{lufanoJs}
\end{centering}
\end{figure}

We thus transform the $LS$-coupled $K$-matrix obtained from experimental energy levels into $jj$-coupling, and then solve Eq. \ref{deteqn} to obtain the Rydberg series leading to each ionization threshold. These series are labeled by the principal quantum numbers in each channel, which are related by energy conservation,
\be
\label{qnumrel}
\nu_{1/2}(\nu_{3/2}) =\left(\nu_{3/2}^{-2}-2\Delta E\right)^{-1/2}.
\ee
In a system with two thresholds a Lu-Fano plot, shown in Fig. \ref{lufanoJs}, graphically illustrates the behavior of the quantum defects. The solutions of Eq. \ref{deteqn} are colored curves, while Eq. \ref{qnumrel} determines the black lines. We show only a few representative ones in Fig. \ref{lufanoJs}a. At the intersections of these curves lie bound states \cite{GreeneKim,FrancisGreene}. Since the only relevant information contained in the quantum defect is its non-integer part, we collapse all energy levels onto a single curve by plotting $\nu_{3/2}$ and $\nu_{1/2}$ modulo one. 

The Lu-Fano plots contain a great deal of information about the channel couplings and behavior of this Rydberg system. In the $J=0$ case exemplified here, we see that the even parity Rydberg series (where the Rydberg electron is either in an $s$ or a $d$ state) are essentially two uncoupled Rydberg series, since the bound states lie on straight lines. This means that the quantum defect in one channel is independent of the other, and these series are effectively single-channel. The odd parity curves, on the other hand, are not flat; a pronounced avoided crossing reveals strong channel interactions. For bound states around this avoided crossing the mixing coefficients in Eq. \ref{MQDTwf} will be significant, leading to wave functions which mix angular momentum as well as levels of radial excitation, since states with very different principal quantum numbers mix. These channels mix strongly because an energy level (if they could be treated independently) in one channel is nearly degenerate with one in a channel corresponding to the other threshold. Away from this avoided crossing, the curves are approximately flat: these are energetically isolated Rydberg states that are predominantly single-channel.  With these tools for multichannel systems in hand: the graphical analysis provided by the Lu-Fano plot, the powerful set of approximations contained in the frame transformation, and the  multichannel spectrum and wave functions determined by Eqs. \ref{deteqn} and \ref{MQDTwf}, one can determine the rich spectrum of multichannel Rydberg systems.

\subsection{Electron-atom scattering phase shifts}
\label{sec:interphases}
We now study the scattering of a very low energy electron from a neutral atom.  Through the partial wave decomposition this process is described by a collection of phase shifts.  At low energy only a few partial waves are relevant, and the Fermi pseudopotential described in the following section utilizes this simplicity to parametrize the interaction of a Rydberg electron with an atom in terms of just $s$ and $p$-wave phase shifts. Since these phases directly determine the properties of LRRMs, it is paramount that they be computed accurately. This section outlines this calculation and discusses the properties of these phase shifts in the alkali atoms relevant to LRRMs. 
\subsubsection{Details of the calculation.}
We keep the basic philosophy undergirding the previous section: the multidimensional coordinate space can be partitioned into two regions. In a small volume around the atomic core the system's dynamics are complicated due to the strong interactions between the scattered electron and the atomic electrons. We use the $R$-matrix method to compute the logarithmic derivative of the wave function on the surface of this volume \cite{OrangeRMP}.  The phase shift is extracted upon matching this to the correct long-range solutions. We give only an abridged discussion of this calculation, and the reader may consult Refs. \cite{OrangeRMP, TaranaCurikLi,EilesHetero,tennyson,PanStaraceGreene1} for more details. Here we describe only the relatively simple (but most relevant to LRRM) scenario of alkali atom-electron scattering. 

The first step is to compute single-electron wave functions satisfying Eq. \ref{rydbergham} and which vanish at $r=0$ and $r=r_0$.  For moderately large $r_0$, $r_0\approx40a_0$, the first few eigenstates are the physical atomic states. Because of the hard-wall boundary condition at $r_0$, the rest of the spectrum consists of positive energy solutions that, while not corresponding to any physical states, give a complete set of states to represent continuum scattering states. With these ``closed'' functions we can  accurately describe the total wave function within the $R$-matrix volume. We also calculate two ``open'' functions which are non-zero at $r_0$; these describe the part of the wave function corresponding to the scattering electron, which is non-zero at the surface of the $R$-matrix volume.

A two-electron basis $y_k$, satisfying the proper symmetry of the state under consideration, is constructed from these one-electron functions. The Hamiltonian is identical to Eq. \ref{twoelec1}, except $V_{l_i}(r_i)$ are the model potentials for a singly-charged ion defined in Eq. \ref{eq:modelpotential}.  For the light alkali atoms we ignore fine structure, and hence for $s$ and $p$-wave scattering we must compute four scattering phase shifts for the symmetries $^1S$, $^1P$, $^3S$, and $^3P$. Relativistic effects become  important in the heavier atoms Rb and Cs, splitting $J=L+S$ levels into a fine structure. These phase shifts have $J$ labels: $^1S_0$, $^1P_1$, $^3S_1$, and $^3P_{0,1,2}$.  The $^1S$ symmetry is particularly important as each alkali atom has a bound anion of this symmetry, and in order for the computed ground state of $H_{2e}$ to reproduce the correct electron affinity a dielectronic polarization potential,
\be
\label{eq:dielpolarizability}
V_{pol}=-\frac{\alpha}{r_1^2r_2^2}(1 - e^{-(r_1/r_c)^3})(1 - e^{-(r_2/r_c)^3})P_1(\hat r_1,\hat r_2),
\ee
must be added to $H_{2e}$ \cite{Laughlin}. $V_{pol}$ describes how one electron influences the other by polarizating the positively charged core. $P_l$ is a Legendre polynomial, and $r_c$ is a fitting parameter. Without Eq. \ref{eq:dielpolarizability} the model Hamiltonian overpredicts the electron affinity. This can have a strong influence on the phase shifts, particularly the resonant $P$-wave shifts \cite{ChrisJJ}, and must be included. 

The logarithmic derivative $b$ for a given scattering energy $E$ can be obtained via a variational calculation using the trial wave function $\Psi = \sum_ky_kC_k$ \cite{OrangeRMP,ChrisRmat1}. This requires solving a generalized eigenvalue equation, $\underline\Gamma \vec C = \underline{\Lambda}\vec Cb$, where 
\be
\Gamma_{kl} = 2(EO_{kl} - (H_{2e})_{kl}-L_{kl}),
\ee
and $\Lambda_{kl} = \int y_ky_l\delta(r - r_0)\dd{V}$. $O_{kl}$ and $\Lambda_{kl}$ are volume and surface overlap matrix elements, respectively, and $(H_{2e})_{kl}$ is a matrix element of the Hamiltonian. $L_{kl}$ is a matrix element of the Bloch operator, $\frac{1}{2r}\delta(r - r_0)\frac{\partial}{\partial r}r$. Even though many basis states are involved in constructing these matrices, the overlap matrix $\underline{\Lambda}$ is singular because most basis functions have no surface amplitude. Only as many eigenvalues as there are open channels are non-zero; in particular for elastic low-energy scattering we have only one open channel, the atomic ground state.  Once $-b = F'(r)/F(r)$ is obtained, the phase shifts are extracted immediately from Eq. \ref{eq:defnmu} with the proper long range solutions $f_l$ and $g_l$. 

Typically one expects that the long-range solutions for an electron in the field of a neutral object correspond to a free electron,
\be
\label{asymptoticsolns}
f_l(r) = \sqrt{\frac{2}{\pi k}}krj_{l}(kr),\,\,g_l(r) = -\sqrt{\frac{2}{\pi k}}kry_{l}(kr),
\ee 
where $j_n(x)$ and $y_n(x)$ are the spherical Bessel and Neumann functions, respectively. However, if the off-diagonal coupling elements in Eq. \ref{eq:outersolns} are still significant at $r_0$ then it is not yet valid to match to these diagonal solutions. The coupled channel equations can be adiabatically diagonalized, decoupling the channels at long-range but introducing a polarization potential \cite{WatanabeGreene}. The quantum defect theory has been generalized to this $r^{-4}$  potential \cite{WatanabeGreene} and can be used to analytically match the functions. We opt instead to numerically propagate wave functions in the polarization potential from $r_0'$ inward to $r_0$. At $r_0'$ the polarization potential is vanishingly small and the functions in Eq. \ref{asymptoticsolns} provide the initial conditions.  Once obtained,  the phase shifts define the energy-dependent $s$-wave scattering length and $p$-wave scattering volume,
\begin{align}
\label{asdef}
a_s[k(R)]&=\left(-\frac{\tan\delta_s[k(R)]}{k(R)}\right)\\
\label{apdef}
a_p^3[k(R)] &= \left(-\frac{\tan\delta_p[k(R)]}{[k(R)]^3}\right),
\end{align}
respectively.

     \subsubsection{Phase shifts and scattering lengths.}

 \begin{figure}[b]
\includegraphics[width= 1\columnwidth]{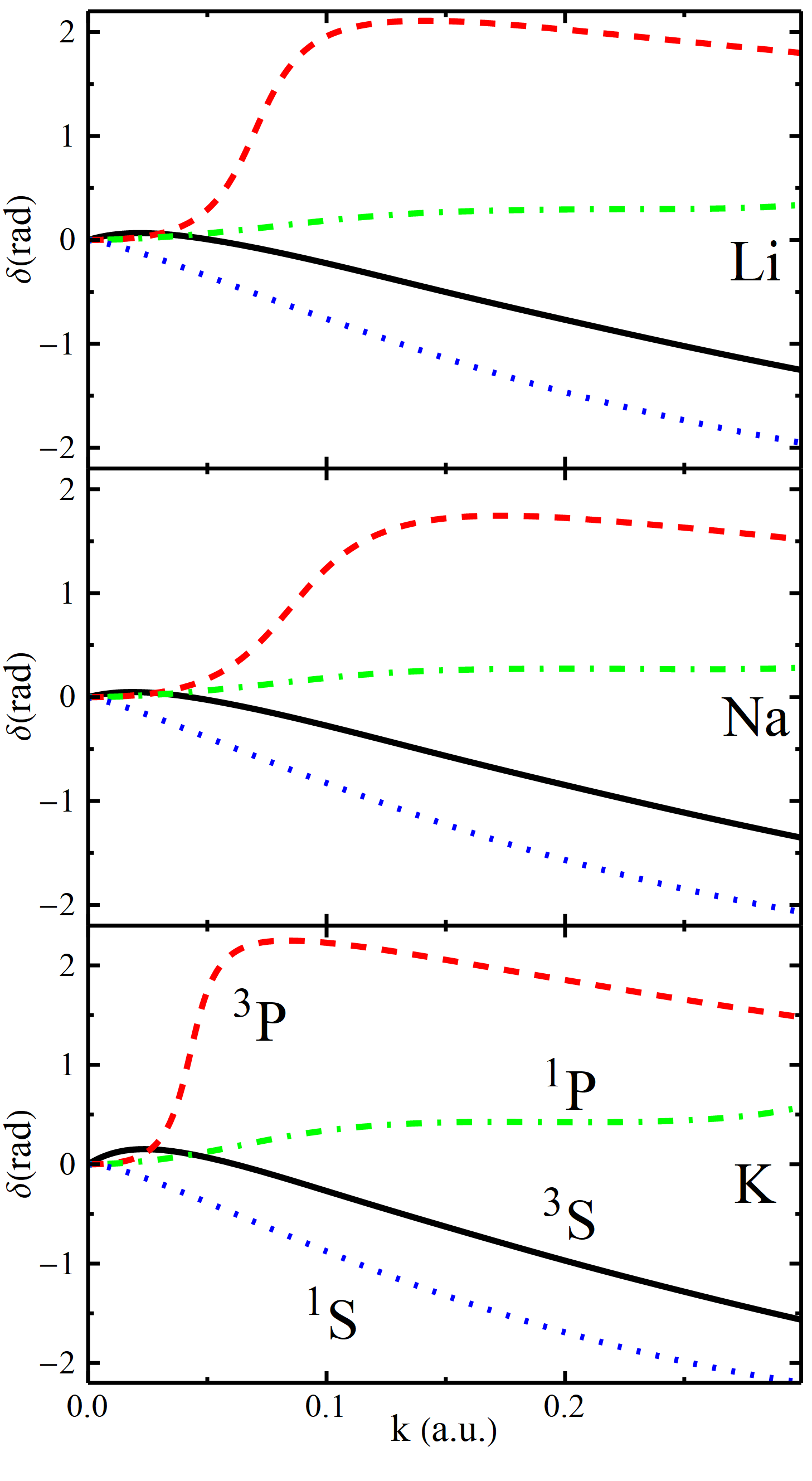}
\caption{\label{fig:phaseshifts} Alkali atom phase shifts for $^1P$ (green, dot-dashed), $^3P_J$ (red, dashed), $^3S$ (black, solid), and $^1S$ (blue, dotted), ignoring the spin-orbit splitting of the $^3P$ states. This figure is modified from Ref. \cite{EilesHetero}.}
\end{figure}
Fig. \ref{fig:phaseshifts} shows electron-atom phase shifts for the lighter alkali atoms calculated with this approach \cite{EilesHetero}. They share many similar features between species. 
The $^3S$ phase shift is positive near zero energy, signalling a negative zero-energy scattering length necessary for Rydberg molecule formation. The point where it changes sign identifies the location of a Ramsauer-Townsend zero. The $^3P$ phase shift exhibits a shape resonance, as the scattering electron is temporarily trapped behind the centrifugal barrier. This shape resonance leads to a divergence in the scattering volume; as a result the $p$-wave interaction includes a far larger contribution than expected based purely on the Wigner threshold law\cite{BahrimThumm,BahrimThumm2,Wigner1948}. 
The $^1S$ and $^1P$ phase shifts are comparitively featureless. Since the $^1S$ symmetries support an anionic bound state, the positive zero-energy scattering length of this symmetry is unsurprising.

No calculation is capable of converging results at zero energy, so an effective range expansion is employed to extrapolate to zero energy. The energy dependence of the $s$-wave scattering length for a long-range polarization potential is well described by the effective range formula \cite{OmalleyRosenbergSpruch}
\begin{align}
a(k) &\approx a(0) + \frac{\alpha \pi}{3}k+\frac{4}{3}a(0)k^2\ln(1.23\sqrt{\alpha}k)\\&\nonumber+\left(\frac{R_e}{2} + \frac{\sqrt{\alpha}\pi}{3} - \frac{\sqrt{\alpha^3}\pi}{3a(0)^2}\right)a(0)^2k^2 \\&- \frac{\pi}{3}\alpha k^3\left(a(0)^2 + \frac{7\alpha}{117}\right)+\cdots\nonumber,
\end{align}
which has  two adjustable parameters, the zero-energy scattering length $a(0)$ and an effective range parameter $R_e$. The first two terms in this expression, linear in $k$, have been used occasionally in the Rydberg molecule community to approximate the full phase shifts. We recommend against this rather crude procedure as this linear approximation rapidly and strongly deviates from the actual values. The scattering volume diverges as $k\to 0$. This is irrelevant in Rydberg molecules as $k\to 0$ implies an infinitely extended wave function, rather than the physical Rydberg wave function. For numerical stability we simply extrapolate the scattering volume to some finite value as $k\to 0$; the PECs calculated below are independent of the specific extrapolation.

\begin{center}
\begin{table}[t]
\begin{tabular}{||c|| c| c  || }
\hline
& | & | \\
  &$a_s^T(a_0)$&$a_s^S(a_0)$ \\ 
  \hline \hline

 Li& \specialcell{$-7.12^a,-7.43^b$,\\$-5.66^d,-6.7^c$} & \specialcell{$3.04^a,2.99^b$,\\$3.65^d,3.2^c$} \\ 
\hline
 Na &  $-6.19^a,-5.9^d,-5.7^c$ & $4.03^a,4.2^d,4.2^c$ \\ 
\hline
 K &   $-15^d,-15.4^g,-14.6^c$ & $0.55^d,0.57^g,0.63^c$ \\
  \hline
   Rb  &   \specialcell{$-16.1^i,-16.9^{g,c},-13^j$,\\$-19.48^l,-14\pm 0.5^m$} & $0.627^{i,c},2.03^g$ \\
  \hline
   Cs &   \specialcell{$-21.7^i,-22.7^g$,\\$-17^j,-21.8\pm 0.2^k$} & \specialcell{$-1.33^i,-2.40^g$,\\$-3.5\pm 0.4^k$} \\
  \hline
    Sr &   $-18^{n,\dagger},-13.2\pm 0.1^o$ & \\
  \hline
\end{tabular}
\caption{A summary of theoretical and experimental values (extracted from molecular spectroscopy) of the zero-energy scattering lengths for the triplet (T) and singlet (S) symmetries of the alkali atoms as well as for the ground state of Sr. These values are from: a) \cite{LiNaNorcross}, b) \cite{TaranaCurikLi}, c) \cite{EilesHetero}, d)\cite{Karule}, e)  \cite{JohnstonBurrow}, f) \cite{BFKNa}, g) \cite{Fabrikant1986}, h)  \cite{Moores1976}, i) \cite{RbCsFr}, j) \cite{BahrimThumm}, k) \cite{Sass},l) \cite{quantumreflection}, m) \cite{AndersonPRL}, n)  \cite{BartschatSadeghpour}, o) \cite{DeSalvo2015}.  
}
 \label{tab:phases}
\end{table}
 \end{center}
    
   \begin{figure}[tbp]
{\normalsize 
\begin{center}
\includegraphics[width=0.9\columnwidth]{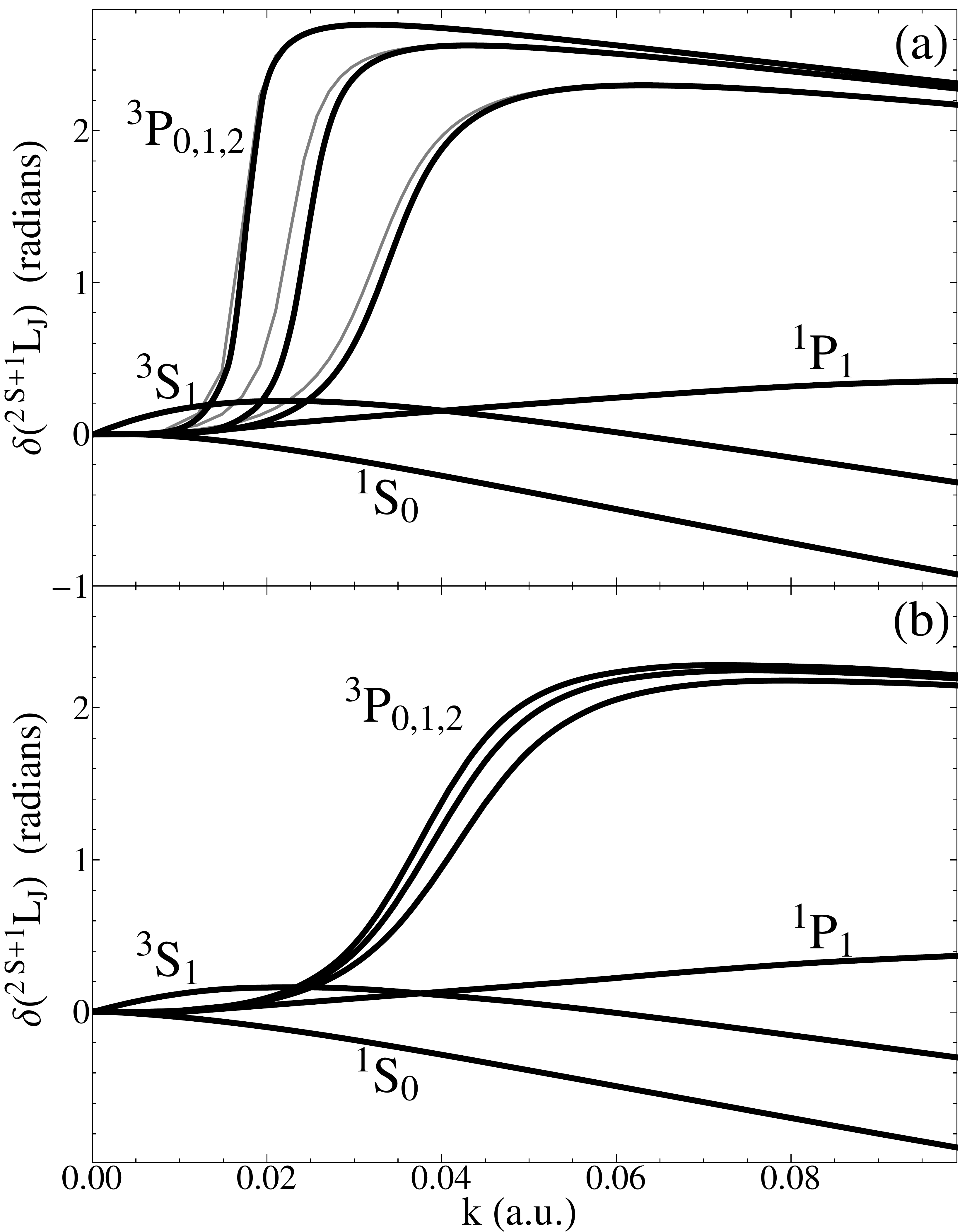}
\end{center}
}
\caption{Scattering phase shifts for Cs (a) and Rb (b), extracted from Ref. \cite{KhuskivadzePRA}. In panel a the unshifted phases are shown as faint curves; the thick curves were shifted slightly to better reflect experimentally observed resonance positions.  This figure is taken from Ref. \cite{EilesSpin}.  }
\label{fig:phases}
\end{figure}

The zero-energy scattering lengths presented in Table \ref{tab:phases} are crucially important for LRRMs, as they set the overall strength of the molecular bond.  The discrepancies in these values, which differ by 10-20\% between reference, are presumably caused by differences in the model Hamiltonian or the level of accuracy in determining its spectrum, approximate long-range potentials, or the zero-energy extrapolation. Electron-atom scattering lengths are extremely difficult to measure, and so one promising application of the vibrational spectroscopy of LRRMs is to extract their values from the spectrum \cite{CsReview,MacLennan}. 

We did not calculate \textit{relativistic} phase shifts for the heavier alkali atoms, Rb and Cs.  Refs \cite{KhuskivadzePRA,BahrimThumm} are the standard references for their phase shifts, reproduced in Fig. \ref{fig:phases}. In most of this tutorial we neglect the spin-orbit splitting of the $p$-wave interaction \footnote{In Rb this is qualitatively acceptable, with a few notable acceptions; on the other hand the PECs of Cs molecules are not even qualitatively correct without including this splitting.}.  The non-relativistic phase shifts of Ref. \cite{BahrimThumm} are used for these calculations, and we have verified that the theoretical approach described agrees with these values. When spin-orbit effects are included, as in Sec. \ref{sec:spinintro}, we use phase shifts from Ref. \cite{KhuskivadzePRA} with a slight modification: the Cs $^3P_J$ phases are shifted by $\sim1$ meV to align the resonance positions with experimental values \cite{pwaveresonance1,pwaveresonance2}. Since no direct experimental measurements of the Rb resonance positions yet exist, we did not modify these phase shifts. 

 Low-energy phase shifts for other atomic species are unfortunately very uncommon in the literature. To the best of our knowledge the Ca, Sr, and Mg phase shifts published in Ref. \cite{BartschatSadeghpour} are the only sufficiently high-resolution calculations of low-energy phase shifts available. Refs. \cite{Saha,scattNoble,Schwartz} provide zero-energy scattering lengths for He, the noble gas atoms, and H, respectively, but without energy dependent phase shifts or higher partial waves these have limited quantitative utility in the context of LRRMs. In addition to higher resolution calculations for these species, scattering length calculations for more complex atoms -- such as the lanthanide species recently in vogue in ultracold experiments -- would be a highly desirable goal for theory.

 \section{A Rydberg molecule primer}
 \label{sec:primer}
 The previous section described the spectrum of a Rydberg atom and the  phase shifts describing low energy electron-atom collisions. We now unite these two concepts, using the Fermi pseudopotential, to describe the Born-Oppenheimer PECs of Rydberg molecules.  By presenting only the simplest description of these molecules, focusing on alkali atoms and neglecting electronic and nuclear spin, we draft a blueprint which will then allow us to systematically introduce further complexity in later sections.

\subsection{Fermi pseudopotential} 
\label{subsec:fermi}
In principle, the properties of LRRMs could be calculated using standard techniques from quantum chemistry. One could solve the Schr\"{o}dinger equation for an electron in the modified Coulomb field of the atomic nucleus along with the polarization potential of the perturber to obtain the Born-Oppenheimer PECs. In practice this approach is excessively difficult. As the previous section revealed, the electron-atom interaction is sensitive not just to the long-range polarization potential but also to the detailed electron-electron interactions, exchange, and correlation that occur by the perturber. Treating these at an \textit{ab initio} level is extremely challenging. Additionally, the inherent two-center nature of the problem makes a full calculation extremely imposing due to the lack of symmetry. Finally, the overwhelming number of Rydberg states accessible at these high energies and the vast spatial dimensions these wave functions occupy quickly discourage attempts to converge numerical calculations. 

In contrast, the Fermi pseudopotential almost immediately provides accurate results. Its predictions have been verified in a multitude of experimental contexts and in comparison with alternative theoretical methods of increasing complexity \cite{DuGreene87,DuGreene89,DePrunele,KhuskivadzeJPB,KhuskivadzePRA,Lebedev,TaranaCurik, GadeaDickinson,JCPabinitioRydMol}. Rather than re-deriving the Fermi pseudopotential here, we instead survey the literature surrounding this topic. A derivation of the $s$-wave pseudopotential most closely tied to the Rydberg context can be found in Fermi's original paper \cite{Fermi}\footnote{A more recent presentation is found in \cite{RydbergRev}.}. Fermi's approach relied on the nature of the zero-energy wave function and its relationship to the scattering length; Omont generalized this by expanding the Rydberg wave function into plane waves near the perturber \cite{Omont}. Independently, Huang and Yang formulated an equivalent pseudopotential for hard sphere scattering in the context of many-body physics\cite{HuangYang}. Their pseudopotential was valid for all partial waves, but unfortunately contained an algebraic mistake for $l>0$ that created substantial confusion in the community once researchers began studying $p$-wave scattering in detail. Before this discrepancy was fully resolved several groups found alternative derivations, and the results and methodologies of these papers may be useful for the Rydberg molecule community \cite{Deutsch,Derevianko,Idziaszek}. Several of these approaches contain explicitly a regularization operator in the pseudopotential. This is necessary for exact calculations using the three-dimensional delta function operator, due to its highly singular nature, but since we never encounter irregular wave functions in the perturbative calculations employed for LRRMs we can ignore this operator. 

Omont formulates the pseudopotential as
 \begin{align}
\label{omontgeneralization}
V_\text{fermi}(\vec r, \vec R) &= 2\pi\sum_{l=0}^\infty(2l+1)\delta^3(\vec r - \vec R)\\&\times\left(-\frac{\tan\delta_l[k(R)]}{k(R)}\right)P_l\left(\frac{\cev{\nabla}\cdot\vec\nabla}{[k(R)]^2}\right).\nonumber
\end{align}
With the origin at the Rydberg core, $\vec r$ and $\vec R$ are the position operators of the electron and perturber, respectively. The backwards vector symbol on the gradient operator implies that it acts on the bra in a matrix element, thus making the operator Hermitian. In this plane wave approximation $k(R)$ is the semiclassical momentum of the Rydberg electron,  $k(R) = \sqrt{2\left(\frac{1}{R} +E\right)}$, where $E$ is the electron's energy. In principle $k(R)$ should be determined self-consistently by iteratively recalculating the electronic eigenenergies until they become stable, but so far the small errors implied by this semiclassical momentum have not demanded this more complicated approach. This definition of $k(R)$ implies two ambiguities: which electronic energy should be chosen when calculating matrix elements $\bra{i}V\ket{j}$ when the electronic state energies $E_i$ and $E_j$ differ, and what happens in the classically forbidden region where $k(R)<0$? In all of our calculations we set $E = -\frac{1}{2n_H^2}$, where $n_H$ is the principal quantum number of the nearest hydrogenic manifold, to eliminate these ambiguities at the expense of neglecting the small effect of quantum defects on $k(R)$. This also fixes the classical turning point $R_\text{out} = 2n_H^2$ for all electronic states. For $R>R_\text{out}$ we either set $k(R>R_\text{out}) = k(R_\text{out})$ or $k(R>R_\text{out}) = -\sqrt{2\left|\frac{1}{R} +E\right|}$ and smoothly interpolate the phase shift from positive to negative $k(R)$. Although both of these approaches are unphysical, they result in smooth PECs\footnote{ Care should be taken on this point when quantitatively comparing PECs from different references. In particular, the behavior near zero energy leads to irrelevant kinks near the classically turning point in some references, and the choice of total electron energy frequently varies.}-- a primarily aesthetic choice since the classical turning point lies outside of the potential wells which support bound states, and thus has little effect on the spectrum.

It is convenient to recast Eq. \ref{omontgeneralization} to include only $s$ and $p$ partial waves; this will pave the way for a more concise notation for later sections as well. The index $\xi$ differentiates the four terms: $\partial_\xi = 1$, $\partial_r$, $\frac{1}{r}\partial_\theta$, and $\frac{1}{r\sin\theta}\partial_\phi$ for $\xi=1$, $2$, $3$, and $4$, respectively. It additionally denotes
\begin{align}
a_\xi[k(R)] &= a_s[k(R)],\xi=1\\
&=3a_p^3[k(R)],\xi>1,\nonumber
\end{align}
where the scattering length/volume were defined in Eqs. \ref{asdef} and \ref{apdef}. We can rewrite Eq. \ref{omontgeneralization} for $s$ and $p$ partial waves as
\be
\label{fermicompact}
V_\text{fermi}(\vec R,\vec r) = 2\pi\sum_{\xi = 1}^4a_\xi[k(R)]\cev{\partial_\xi} \delta^3(\vec r-\vec R)\vec{\partial_\xi},
\ee
where the individual potential terms define  $V_\xi(\vec R,\vec r) = a_\xi[k(R)]\cev{\partial_\xi} \delta^3(\vec r-\vec R)\vec{\partial_\xi}$. 
\subsection{Rydberg molecule potential energy curves}
Within the standard Born-Oppenheimer framework, the potential energy curves (PECs) for Rydberg molecules (unless otherwise stated, these are assumed to be dimers) are the eigenenergies of the electronic Hamiltonian depending parametrically on the internuclear coordinate $\vec R$,
\be
 H\Psi(\vec r;\vec R) = E(\vec R)\Psi(\vec r;\vec R). 
\ee
The most rudimentary electronic Hamiltonian consists of the Rydberg electron's Hamiltonian $H_0$ and the Fermi pseudopotential,
\be
 H =  H_0 + V_\text{fermi}(\vec R,\vec r).
\ee
The Fermi pseudopotential is valid if the neutral atom has a strictly perturbative effect on the Rydberg wave function. It is therefore sufficient to use perturbation theory to compute the PECs as well. The zeroth order states are the Rydberg wave functions satisfying $H_0\ket{nlm} = -\frac{1}{2(n-\mu_l)^2}\ket{nlm}$. These electronic states are shifted by the Fermi pseudopotential, giving rise to the PECs. The nuclear Hamiltonian, $H_\text{nuc}(\vec R)=-\frac{1}{2M}\nabla_R^2 + E(\vec R)$, is solved  afterwards to find the vibrational spectrum of the molecule with reduced mass $M$. 

In general, a diatomic molecule only possesses cylindrical symmetry, and as such the only conserved quantum number is the projection $\Omega$ of the total angular momentum onto the internuclear axis, which we set parallel to $\hat z$.  Molecular states of cylindrical symmetry are classified by their $\Omega$ value ($\Omega=0$ is a $\Sigma$ state, $|\Omega|=1$ is a $\Pi$ state, etc.).  The $V_1$ and $V_2$ operators are only non-zero if  $\bkt{\vec r=r\hat z}{nlm}\ne 0$, which is only true for $m=0$ states\footnote{$|Y_{lm}(0,\phi)|^2=\frac{2l+1}{4\pi}\delta_{m,0}$.}. The resulting molecules are therefore classified as $\Sigma$ states. The $V_3$ and $V_4$ operators change, through their angular derivatives, $Y_{l|m|=1}(\theta,\phi)$ functions into $Y_{l0}(\theta,\phi)$ functions, and therefore correspond to  $\Pi$ symmetry.  Since we have only included $s$ and $p$ wave pseudopotentials we only have $\Sigma$ and $\Pi$ symmetries.

\subsubsection{Low-l Rydberg molecules.}
\begin{figure}[t]
 \begin{centering}
 \begin{center}
\includegraphics[width= 0.9\columnwidth]{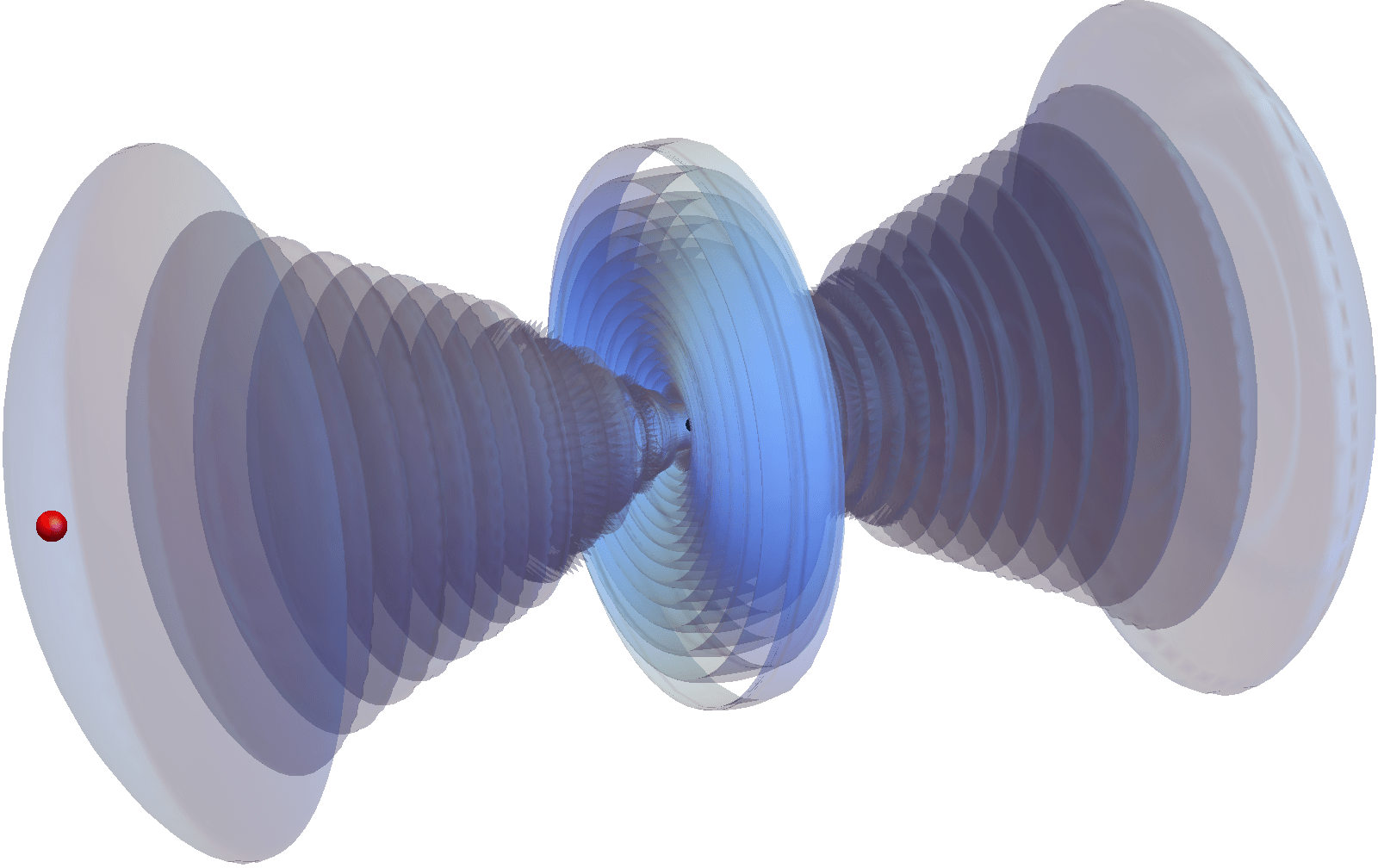}
\end{center}
\end{centering}
\caption{ A probability isosurface plot of a Rydberg $nD$ state. Potential wells are located in the lobes of the wave function, which can trap the perturber (red sphere).  }
\label{fig:ndfunc}
\end{figure}

We first calculate the PECs associated with electronic Rydberg states having low angular momentum $l\le l_\text{min}$, where typically $l_\text{min}=2$. Since these states have finite quantum defects they are energetically isolated, and so the molecular PECs associated with these non-degenerate levels are computed by evaluating $\langle nlm| V_\text{fermi}(\vec r,\vec R)|nlm\rangle$ at each value of $R$\footnote{This breaks down for atoms with a $p$-wave shape resonance which can couple many electronic states together, and later we will develop a more sophisticated method.}. The resulting $\Sigma$ and $\Pi$ PECs,
  \begin{align}
  \label{lowlNone}
  E^\Sigma_{l_\text{min}}(R)& =2\pi\left(\frac{2l+1}{4\pi}\right) \Bigg(a_s[k(R)][u_{nl}(R)/R]^2\\&+3a_p^3[k(R)]\left|\frac{du_{nl}(R)/R}{dR}\right|^2\Bigg), \nonumber\\
  E^\Pi_{l_\text{min}}(R) &= 6\pi a_p^3[k(R)]\left(\frac{(2l+1)(l+1)l}{8\pi}\right)\left[\frac{u_{nl}(R)}{R^2}\right]^2,
\end{align}
depend only on $R=|\vec R|$. 
There are two degenerate $\Pi$ PECs, and they are far weaker than the $\Sigma$ PECs because of an additional $R^{-2}$ factor. Crucially, these PECs depend on the product of two terms. The first, the scattering length/volume, determines their overall strength and repulsive or attractive nature.  The second is the radial probability density ($s$-wave) and its gradient ($p$-wave), which cause the PECs to oscillate as a function of $R$. For negative scattering lengths the perturber is therefore trapped in the lobes of the Rydberg wave function, sketched in Fig. \ref{fig:ndfunc} for a Rydberg $nD$ state. This linking of the oscillations in the atomic wave function with oscillations in the PECs is one of the distinguishing features of LRRMs in marked contrast to covalent or ionic bonds. Eq. \ref{lowlNone} reveals that the depth of these potentials increases with $l$ since the electronic density can focus along the internuclear axis at higher $l$.  As such, we anticipate that high-$l$ Rydberg states can form very deeply bound molecules due to this probability enhancement.
 
\subsubsection{High-$l$ Rydberg molecules: ``trilobites'' and ``butterflies''.}
As $l$ increases the quantum defects rapidly shrink (see Table \ref{tab:datatable2}). The strength of the Fermi pseudopotential dwarfs the energy splitting between states, and they can be treated as exactly degenerate, just as in the hydrogen atom. Eq. \ref{lowlNone}, derived using non-degenerate perturbation theory, is therefore clearly inadequate for $l>l_\text{min}$. We must use \textit{degenerate} perturbation theory to compute PECs in this high-$l$ subspace. The perturbed eigenstates will be superpositions of the many degenerate unperturbed states, which can therefore combine very effectively into a perturbed wave function that extremizes the potential and no longer resembles the unperturbed states \footnote{For this reason one should always keep in mind the maxim that ``degenerate perturbation theory is non-perturbative.''}.
 
The degenerate subspace includes all Rydberg wave functions $\ket{nlm}$ having $l>l_\text{min}$ and identical $n$. We illustrate here the diagonalization of the potential within this subspace using a single $V_\xi$. For brevity, we define a shorthand\footnote{When only $\phi_{nlm}(\vec r)$ is used without an index, it is assumed that $\xi=1$ and this is the standard hydrogenic wave function.} for the wave function and spherical gradient components:
\be
\label{estatedef}
\phi^\xi_{nlm}(\vec r) =\partial_\xi\phi_{nlm}(\vec r).
 \ee 
The matrix elements of $V_\xi(\vec R,\vec r)$ are thus proportional to
\begin{align}
\label{matelforshow}
\bra{nlm}\cev{\partial_\xi} \delta(\vec r-\vec R)\vec{\partial_\xi}\ket{nl'm'}&= \left[\phi_{nlm}^\xi(\vec R)\right]^*\phi_{nl'm'}^\xi(\vec R)
\end{align}
This defines a rank 1 {\it separable matrix} and so, despite its large ($\sim n^2$) dimension in this representation it has only one non-zero eigenvalue,
\be
\label{evalpert}
E(\vec R) = a_\xi[k(R)]\sum_{l,m}\left[\phi^\xi_{nlm}(\vec R)\right]^*\phi^\xi_{nlm}(\vec R).
\ee
The summation is over $|m|\le l$ and $l_\text{min}<l<n-1$.  The corresponding (un-normalized) perturbed wave function is
\begin{align}
\label{eqeigenstatesgeneral}
\Psi_{\xi}(\vec R,\vec r) &= \sum_{l,m}\left[\phi_{nlm}^\xi(\vec R)\right]^*\phi_{nlm}(\vec r).
\end{align}
These formulas betray a recurring pattern: repeatedly we have to sum a product of Rydberg wave functions or their derivatives. It is therefore very useful to study the following object, which we call the \textit{trilobite overlap}, in depth:
\be
\label{eqtomdef}
\+{\Upsilon_{pq,n}^{\pmb\alpha\pmb\beta}}= \sum_{l,m}\left[\phi^\alpha_{nlm}(\vec R_p)\right]^*\phi^\beta_{nlm}(\vec R_q).
\ee
The full generality of this formula will be useful throughout this tutorial. With this notation we express Eq. \ref{evalpert} as $E(\vec R) = a_\xi[k(R)]\+{\Upsilon_{RR,n}^{\pmb\xi\pmb\xi}}$ and Eq. \ref{eqeigenstatesgeneral} as $\Psi_\xi(\vec R,\vec r) =\+{\Upsilon_{Rr,n}^{\pmb\xi\pmb1}}$, with the prescription that a subscript $R$ or $r$ implies $\vec R_p=\vec R$ or $\vec R_p=\vec r$, respectively. The generic subscripts ``$p$'' and ``$q$'' imply that $\+{\Upsilon_{pq,n}^{\pmb\alpha\pmb\beta}}$ is evaluated at $\vec R_p$ and $\vec R_q$, two specific points in space. The trilobite overlap's structure is thus reminiscent of a Green's function or a two-point correlation function.

Surprisingly,  the trilobite overlap can be analytically summed provided $l_\text{min}=0$, i.e. neglecting quantum defects\footnote{This introduces only small errors which can be subtracted later if necessary}. This was accomplished by Chibisov and coworkers \cite{ChibisovPRL2,Chibisov2000} and used occasionally in their study of LRRMs \cite{KhuskivadzePRA}. We feel that the utility of this summation has not been appreciated in the following literature, and therefore provide here a sketch of the derivation and key results\footnote{The author recently became aware of even more under-appreciated articles in the mathematical chemistry literature which seem to have also been unknown to Chibisov and coworkers: Refs. \cite{Blinder1993,Bartell1996} present these same formulas several years before Ref. \cite{ChibisovPRL2}.}.

All of these summations can be derived from the unnormalized trilobite state,
\begin{align}
\+{\Upsilon_{Rr,n}^{11}}=\sum_{l = 0}^{n-1}\sum_{m = -l}^{m=l}[\phi_{nlm}(\vec R)]^*\phi_{nlm}(\vec r).
 \label{sumformgoal}
\end{align}
  This expression is akin to the Coulomb Green's function, defined as a summation (which extends into an integral over continuum states) over the complete set of orthogonal Coulomb functions:
\be
G(\vec r, \vec R,E) = \sum_{nlm}\frac{\phi^*_{nlm}(\vec r)\phi_{nlm}(\vec R)}{E-E_n}.
\ee
We can neglect all continuum states and even all discrete states except for one degenerate manifold by evaluating the Green's function at a bound state energy:
\be
\label{gff1}
G(\vec r, \vec R, E \to E_n) \approx \frac{1}{E-E_n}\sum_{lm}\phi^*_{nlm}(\vec r)\phi_{nlm}(\vec R).
\ee 
This formula shows that we can evaluate this sum provided that we can obtain the Green's function by another means and properly cancel out the divergent $\frac{1}{E-E_n}$ term. Hostler and Pratt \cite{HostlerPratt} derived a closed form expression for the Coulomb Green's function, 
\begin{align}
\label{eq:hostlerprattgreenfunction}
G(\vec r,\vec R,E) &= \frac{\Gamma(1 - \nu)}{2\pi|\vec r - \vec R|}\left(\pd{(x/\nu)}-\pd{(y/\nu)}\right) \\&\times W_{\nu,1/2}(x/\nu)M_{\nu,1/2}(y/\nu),\nonumber
\end{align}
in terms of the variables $(x,y) = r + R \pm |\vec r - \vec R|$ and Whittaker functions $M_{\nu,1/2}(\tau)$ and $W_{\nu,1/2}(\tau)$.  These are related by
\begin{align}
\label{eq:irregularwhittaker}
&\Gamma(1 - \nu)M_{\nu,1/2}(\tau)\\ &= (-1)^{1 + \nu}\frac{\Gamma(1 - \nu)}{\Gamma(1 + \nu)}W_{\nu,1/2}(\tau) + (-1)^\nu W_{-\nu,1/2}(-\tau).\nonumber
\end{align}
As $\nu$ approaches an integer $n$, as occurs at a bound state, $\Gamma(1 - \nu)$ diverges as
\be
\Gamma(1 - \nu)|_{\nu\to n }=\frac{(-1)^n}{n^3\Gamma(n)}\frac{1}{E - E_n},
\ee
where $E = -(2\nu^2)^{-1}$ and $E_n = -(2n^2)^{-1}$. Now, by matching Eqs. \ref{gff1} and \ref{eq:hostlerprattgreenfunction} as $E \to E_n$, we have
\begin{align}
\label{eq:matchingGF}&\frac{1}{E-E_n}\sum_{lm}\phi^*_{nlm}(\vec r)\phi_{nlm}(\vec R)\\ &=\nonumber\frac{1}{E - E_n}\Bigg[\frac{(-1)^n}{n^3\Gamma(n)}\frac{1}{2\pi|\vec r - \vec R|}\left(\pd{(x/\nu)}-\pd{(y/\nu)}\right)\\&\left.\,\,\,\,\nonumber\times W_{\nu,1/2}(x/\nu)\frac{ (-1)^{1 + \nu}}{\Gamma(1 + \nu)}W_{\nu,1/2}(y/\nu)\right|_{\nu=n}\Bigg].
\end{align}
Since both sides of this equation diverge identically as $1/(E-E_n)$, the summation on the left is equivalent to the bracketed term on the right.
We insert the standard hydrogen radial wave function $u_{nl}(r)$ using Eq. \ref{fdefradial}, and evaluating the derivatives Eq. \ref{eq:matchingGF} simplifies to
\begin{align}
  \label{chibisovtrilobite}
\+{\Upsilon_{Rr,n}^{11}}&= \frac{u_{n0}'(t_-)u_{n0}(t_+) - u_{n0}(t_-)u_{n0}'(t_+)}{4\pi\Delta t},
\end{align}
where $t_{\pm} =\frac{1}{2}\left(R+r\pm\sqrt{R^2 + r^2 -2Rr\cos\gamma}\right)$,  $\Delta t = t_+ - t_-$, and $\gamma$ is the angle between $\vec R$ and $\vec r$. Differentiation of Eq. \ref{chibisovtrilobite} with respect to $R$, $\theta$, or $\phi$ generates the three types of butterfly orbitals $\+{\Upsilon_{Rr,n}^{\pmb\xi 1}}$:
\begin{align}
\label{chibisovRbutterfly}
\+{\Upsilon_{Rr,n}^{21}} &=\frac{(r\cos\gamma-R)\mathcal{F}(t_+,t_-)}{8 \pi\Delta t^3}\\&\nonumber+\frac{u_{n0}(t_+)u_{n0}''(t_-)-u_{n0}(t_-)u_{n0}''(t_+)}{8 \pi\Delta t}\\
\label{chibisovthetabutterfly}
\+{\Upsilon_{Rr,n}^{31}}&= \cos\theta_R\cos\varphi_R\Upsilon_x\\&\nonumber + \cos\theta_R\sin\varphi_R\Upsilon_y-\sin\theta_R\Upsilon_z \\
\label{chibisovphibutterfly}
\+{\Upsilon_{Rr,n}^{41}}&= -\sin\varphi_R\Upsilon_x + \cos\varphi_R\Upsilon_y,\end{align}
where 
\begin{align}
\label{eq:symproperties}
\begin{pmatrix}\Upsilon_x\\\Upsilon_y\\\Upsilon_z\end{pmatrix} = \frac{r\mathcal{F}(t_+,t_-)}{8\pi(\Delta t)^3}\begin{pmatrix}\sin\theta_r\cos\varphi_r\\\sin\theta_r\sin\varphi_r\\\cos\theta_r\end{pmatrix}
\end{align}
and
\begin{align}\mathcal{F}(t_+,t_-)&=- 2(\Delta t)  u_{n0}'(t_+) u_{n0}'(t_-)\\&-u_{n0}(t_-)[2 u_{n0}'(t_+) - (\Delta t) u_{n0}''(t_+)]\nonumber \\ &+  u_{n0}(t_+)[ 2u_{n0}'(t_-) +(\Delta t)  u_{n0}''(t_-)].\nonumber
\end{align}

  \begin{figure*}[ht]
 \begin{center}
\includegraphics[width=\textwidth]{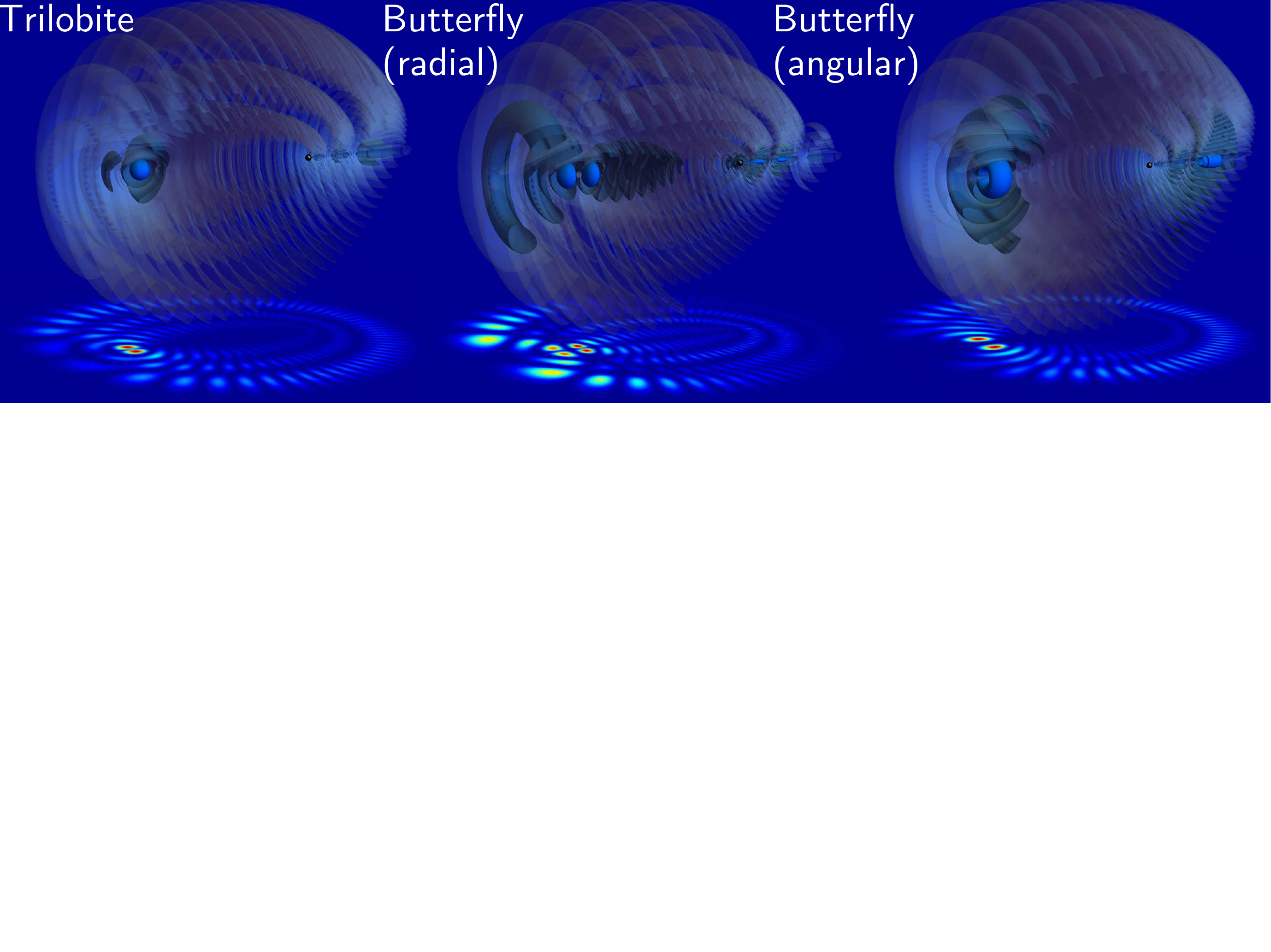}
\vspace{-200pt}
\caption{Trilobite and butterfly states for $n=30$ and $R=1232$, plotted as isosurfaces and as a density plot. For the isosurfaces, the deep blue color is a cut of $|\Psi|^2 = X$ and a full rotation about $\hat z$ is plotted. The grayer surfaces are for $X/10$ and $X/100$, respectively, and are only plotted for half a rotation about $\hat z$ to reveal the inner structure. The density plot for $\rho|\Psi(\rho,z)|^2$ is shown in cylindrical coordinates aligned parallel to the isosurfaces. The scale factor $\rho$ emphasizes the nodal structure, but introduces an additional zero along $\hat z$.  The angular butterfly lacks the $\sin\phi$ or $\cos\phi$ modulation seen in Eq. \ref{eq:symproperties}. }
 \label{fig:trilobitebasis}
\end{center}
 \end{figure*}
\noindent
The diagonal elements $\+{\Upsilon_{pp,n}^{\pmb\xi\pmb\xi}}$ are obtained by carefully evaluating equations (\ref{chibisovtrilobite} - \ref{chibisovphibutterfly}) as $\vec R_p$ approaches $\vec R_q$ using L'Hopital's rule\footnote{We eliminate second and third derivatives using the radial Schr\"{o}dinger equation}:
\begin{align}
\label{diagonalTrilo}
\+{\Upsilon_{RR,n}^{11}}&= \frac{(2n^2 - R)\left(u_{n0}(R)/n\right)^2+ Ru_{n0}'(R)^2}{4\pi R}\\
\label{diagonalRbutterfly}
\+{\Upsilon_{RR,n}^{22}}&= \+{\Upsilon_{RR,n}^{33}}-\frac{u_{n0}(R)}{12\pi R^3}\left[3Ru'_{n0}(R)+2u_{n0}(R)\right]\\
\label{diagonalAngbutterfly}
\+{\Upsilon_{RR,n}^{33}}&=\+{\Upsilon_{RR,n}^{44}}\\&\nonumber
= \frac{4\pi R(2n^2 - R)\+{\Upsilon_{RR,n}^{11}} - n^2u_{n0}'(R)u_{n0}(R) }{12\pi n^2R^2}.
\end{align}
This analysis confirms that the two $\Pi$ butterflies are degenerate. Surprisingly, although both terms in the numerator of Eq. \ref{diagonalAngbutterfly} oscillate with $R$, $\+{\Upsilon_{RR,n}^{33}}$ does not (see Fig. \ref{fig:Rbmanifold}).  This is particularly intriguing in light of the summation representation of $\+{\Upsilon_{RR,n}^{33}}$,
\be
\+{\Upsilon_{RR,n}^{33}}=\sum_{l=0}^{n-1}\left[\frac{u_{nl}(R)}{R^2}\right]^2\frac{(2l+1)(l+1)(l)}{8\pi},
\ee
which shows that the coefficients $c_l = (2l+1)(l+1)l$ guarantee perfect cancellation of all oscillations in the radial wave functions.

These eigenstates are shown in Fig. \ref{fig:trilobitebasis}. Their appearance in cylindrical coordinates (plotted as ``shadows'' in Fig. \ref{fig:trilobitebasis} and more explicitly as surface plots in the insets of Fig. \ref{fig:Rbmanifold}) calls to mind a trilobite fossil or a butterfly with spread wings, respectively\footnote{Whether this is an epiphany or an apophany is up to the reader.}. Their distinctive nodal patterns have attracted interest due to their deep connections to periodic orbit theory and to a near separability of the diatomic Hamiltonian in elliptic coordinates \cite{Granger,Lambert}. One underappreciated point is the extent to which these wave functions localize the wave function near the perturber. The trilobite representats the delta function in this truncated degenerate subspace, and as seen in Fig. \ref{fig:trilobitebasis} its density is focused into a small region around the perturber. The radial butterfly is also localized around the perturber, but as it  maximizes the gradient in the $z$ direction it has a node directly on the perturber. As $\Sigma$ states, the trilobite and $R$-butterfly orbitals are invariant under rotation around the $z$ axis, and Eqs. \ref{chibisovtrilobite} and \ref{chibisovRbutterfly} are correspondingly independent of $\phi_r$. The angular butterfly molecules have a node along the internuclear axis in accordance with their $\Pi$ symmetry, as reflected by the $\sin\varphi_r$ and $\cos\varphi_r$ modulation factor in Eq. \ref{eq:symproperties}. This factor is dropped in Fig. \ref{fig:trilobitebasis} to facillitate the visualization.  Further details of the symmetry properties of these butterfly states, as pertaining to the symmetries of polyatomic molecules, are discussed in Ref. \cite{JPBdens}.

\subsubsection{Beyond perturbation theory.}
Thus far we have computed the molecular states separately for low-$l$ and high-$l$ Rydberg atoms. The accuracy of the resulting PECs (Eqs. \ref{lowlNone} and \ref{evalpert}), computed in first order perturbation theory, is compromised for several reasons:
\begin{itemize}
\item The matrix elements of $V_\xi$ were calculated and diagonalized separately,  ignoring any coupling between trilobite and butterfly states.
\item The coupling between the trilobite/butterfly states and low-$l$ Rydberg states, which can become large depending on the strength of the perturbation compared to the energy gap due to the quantum defects, was ignored.
\item The $p$-wave shape resonance creates an unphysical divergence in the PECs that catastrophically reduces their accuracy. This must be remedied by including couplings between different $n$ manifolds adjacent to the manifold of interest. The resulting level repulsion constrains this divergence and gives sensible results \cite{HamiltonGreeneSadeghpour}.    
\end{itemize} 
To address these problems we expand the exact wave function into the complete and orthonormal basis of Rydberg wave functions, truncating only in the number of $n$-manifolds we include:
\begin{align}
\tilde\Psi(\vec r; \vec R) = \sum_{n=n_1}^{n=n_2}\sum_{l=0}^{l=n-1}&\sum_{m=-l}^{m=l}c_{nlm}(\vec R)\phi_{nlm}^1(\vec r).\label{exp1}
\end{align}
The expansion coefficients $c_{nlm}$ are determined variationally by diagonalizing the Hamiltonian, $\bra{n'l'm'}H_0 + V_\text{fermi}\ket{nlm}$.  This is the standard approach, and we will refer to it as the ``Rydberg basis'' method. Typically an expansion such as this converges provided the number of basis states is not truncated too low. However, its accuracy in this context is a matter of some controversy since the delta function potential is not formally convergent \cite{Fey}.  Care must be taken in choosing the number of $n$ manifolds due to this non-convergent behavior, and will be discussed more in Section \ref{sec:spinintro}. 

This approach requires the diagonalization of a $\mathcal{M}n^2$-dimensional matrix, where $\mathcal{M}$ is the number of $n$ manifolds included.  As Eq. \ref{matelforshow} reveals, there is a huge redundancy as the matrix of a given $V_\xi$, of dimension $n^2$, has only a single eigenstate given in terms of the $l=0$ radial wave function only. The evaluation of so many high-$l$ basis states in order to diagonalize $H$ in the full Rydberg basis seems particularly wasteful. 

In this tutorial we develop an alternative method inspired by Ref. \cite{Rost2006}, which considered the trilobite states, rather than the Rydberg basis, as the fundamental unperturbed basis. In the context of polyatomic molecules Ref. \cite{JPBdens} included coupling terms between the butterfly and trilobite states. Here we fully generalize this concept to include all couplings between different $l$ states and $n$ manifolds. This is, to our knowledge, the first time this approach has been presented. We refer to this approach as the ``trilobite basis'' method as its core idea is that we can replace Eq. \ref{exp1} with a new trial wave function which collapses all of the redundant degenerate high$-l$ states into just four states per $n$ manifold: one trilobite and three butterfly ($\xi = 2,3,4$) states. This trial wave function contains these, along with the few non-degenerate low-$l$ states:
\begin{align}
\label{trialwavetrilo}
\Psi(\vec r; \vec R) = \sum_{n=n_1}^{n=n_2}\Bigg[\sum_{l=0}^{l=l_\text{min}}&\sum_{m=-l}^{m=l}c_{nlm}(\vec R)\phi_{nlm}^1(\vec r)\\&+\sum_{\xi=1}^4\mathcal{C}_{n\xi}(\vec R)\+{\Upsilon_{Rr,n}^{\pmb\xi1}}\Bigg].\nonumber
\end{align}
We solve for the coefficients $c_{nlm}(\vec R)$ and $\mathcal{C}_{n\xi}(\vec R)$ by projecting $H\Psi(\vec r;\vec R)$ onto each basis function. Since the trilobite states are not orthogonal,
\be
\int\left[\+{\Upsilon_{pr,n}^{\pmb\alpha1}}\right]^*\+{\Upsilon_{qr,n'}^{\pmb\beta1}}\ddn{3}{r} = \+{\Upsilon_{pq,n}^{\pmb\alpha\pmb\beta}}\delta_{nn'},
\ee
this results in a generalized eigenvalue equation $\underline{H}\vec c = E(\vec R)\underline{\Lambda}\vec c$. The Hamiltonian matrix has a block structure. The first block,
\begin{align}
\label{matelementdimer1}
&\bra{\+{\Upsilon_{Rr,n}^{\pmb\alpha 1}}}H_0+V_\text{fermi}\ket{\+{\Upsilon_{Rr,n'}^{\pmb\beta 1}} }\\&=-\frac{1}{2n^2}\+{\Upsilon_{RR,n}^{\pmb\alpha\pmb\beta}}\delta_{nn'}+ 2\pi\sum_{\xi=1}^4a_\xi\+{\Upsilon_{RR,n}^{\pmb\alpha\pmb\xi}}\+{\Upsilon_{RR,n'}^{\pmb\xi\pmb\beta}}, \nonumber
\end{align}
couples trilobite and butterfly states together. The next block couples the non-degenerate low-$l$ states
\begin{align}
&\bra{\phi_{nlm}^1}H_0 + V_\text{fermi}\ket{\phi_{n'l'm'}^1} \\&= -\frac{\delta_{nn'}\delta_{ll'}\delta_{mm'}}{2(n-\mu_l)^2} + 2\pi \sum_{\xi=1}^4a_\xi\phi_{nlm}^\xi(\vec R)^*\phi_{n'l'm'}^\xi(\vec R).\nonumber
\end{align}
Finally, we have an off-diagonal block coupling trilobite and butterfly states to the low-$l$ functions, 
\begin{align}
&\bra{\+{\Upsilon_{Rr,n}^{\pmb\alpha 1}}}H_0+V_\text{fermi}\ket{\phi_{n'l'm'}^{1} }\nonumber\\&
= 2\pi \sum_{\xi=1}^4a_\xi\+{\Upsilon_{RR,n}^{\pmb\alpha \pmb\xi}}\phi_{n'l'm'}^\xi(\vec R).
\label{offdiagcouptrilo}
\end{align}
The overlap matrix $\underline{\Lambda}$ has a trilobite block given by $\+{\Upsilon_{RR,n}^{\pmb\alpha\pmb\beta}}\delta_{nn'}$, a purely diagonal low-$l$ block,  $\delta_{nn'}\delta_{ll'}\delta_{mm'}$, and no off-diagonal blocks. In the calculations presented below which use this approach, we set $n_1=n-1$, $n_2=n+1$ and $l_{<} = 3$ to account for the non-negligible $\mu_f$.

\subsection{Alternative approaches}
Before we examine the PECs, we briefly mention several alternative approaches. The earliest such approach, predating LRRMs and  developed in the context of collisional broadening, is the Borodin and Kazansky (BK) model  \cite{BKmodel}. This yields PECs that agree in shape and magnitude with the Fermi model, but lack the oscillatory nature from the electronic density. They are determined purely by the phase shifts:
\be
\label{BKmodel}
E^{BK}_l(R) = -\frac{1}{2n^2}+\frac{1}{2}\left(n-\frac{\delta_l[k(R)]}{\pi}\right)^{-2}.
 \ee 
 This approximation provides a useful comparison when attempting to understand the convergence challenges of the delta function potential \cite{Fey,EilesSpin}, since these approximate PECs do not diverge  when $\delta_p$ rises by $\pi$. This confirms that the $p$-wave shape resonance is converged adequately by level repulsion when multiple $n$ manifolds are included.

Immediately following the first trilobite prediction, Green's function techniques were developed by Greene and coworkers\cite{HamiltonThesis,Crowell} and, essentially simultaneously, by Fabrikant and coworkers \cite{KhuskivadzeJPB,KhuskivadzePRA}. These methods differ in implementation but are built around a similar logic. The basic idea is that the electron, outside of the non-Coulombic region near the Rydberg core or the polarization potential region surrounding the perturber, experiences only a Coulomb potential. It is thus described by hydrogenic wave functions. Although the wave function differs near the Rydberg core and the perturber, its exact form there is irrelevant since (taking the perspective of quantum defect theory as discussed in Sec. \ref{sec:inter}) the wave function outside of this region is determined only by the quantum defects and scattering phase shifts. Thus, the non-Coulomb regions impart new boundary conditions on the wave function, and these can be included readily once the Green's function is known \cite{HostlerPratt}.  The method of Ref. \cite{KhuskivadzePRA} is particularly sophisticated and includes the Rydberg fine structure, singlet and triplet scattering, and even the relativistic $^3P_J$ splitting of the electron-atom phase shifts \cite{KhuskivadzePRA}. For many years, this was the only approach which properly included the fine structure of the phase shifts (Sec. \ref{sec:spinintro} describes a different approach). Tarana and Curik \cite{TaranaCurik} developed an $R$-matrix program which solves directly the two-electron interaction near the perturber before matching to the long-range Coulomb functions. This technique improves slightly upon Ref. \cite{KhuskivadzePRA} since it is available for higher partial waves and does not require input of the scattering phases from an external calculation. It appears to be quite accurate, particularly at small internuclear distances where the Fermi model is unsuitable. Unfortunately so far only fairly low-lying ($n\le 20$) molecular PECs for H$_2$ have been computed with this model, and it would be very useful to extend this line of research to verify the performance of these methods in the alkali atoms.

 \subsection{Potential energy curves}

   \begin{figure*}[h]
 \begin{center}
\includegraphics[width=\textwidth]{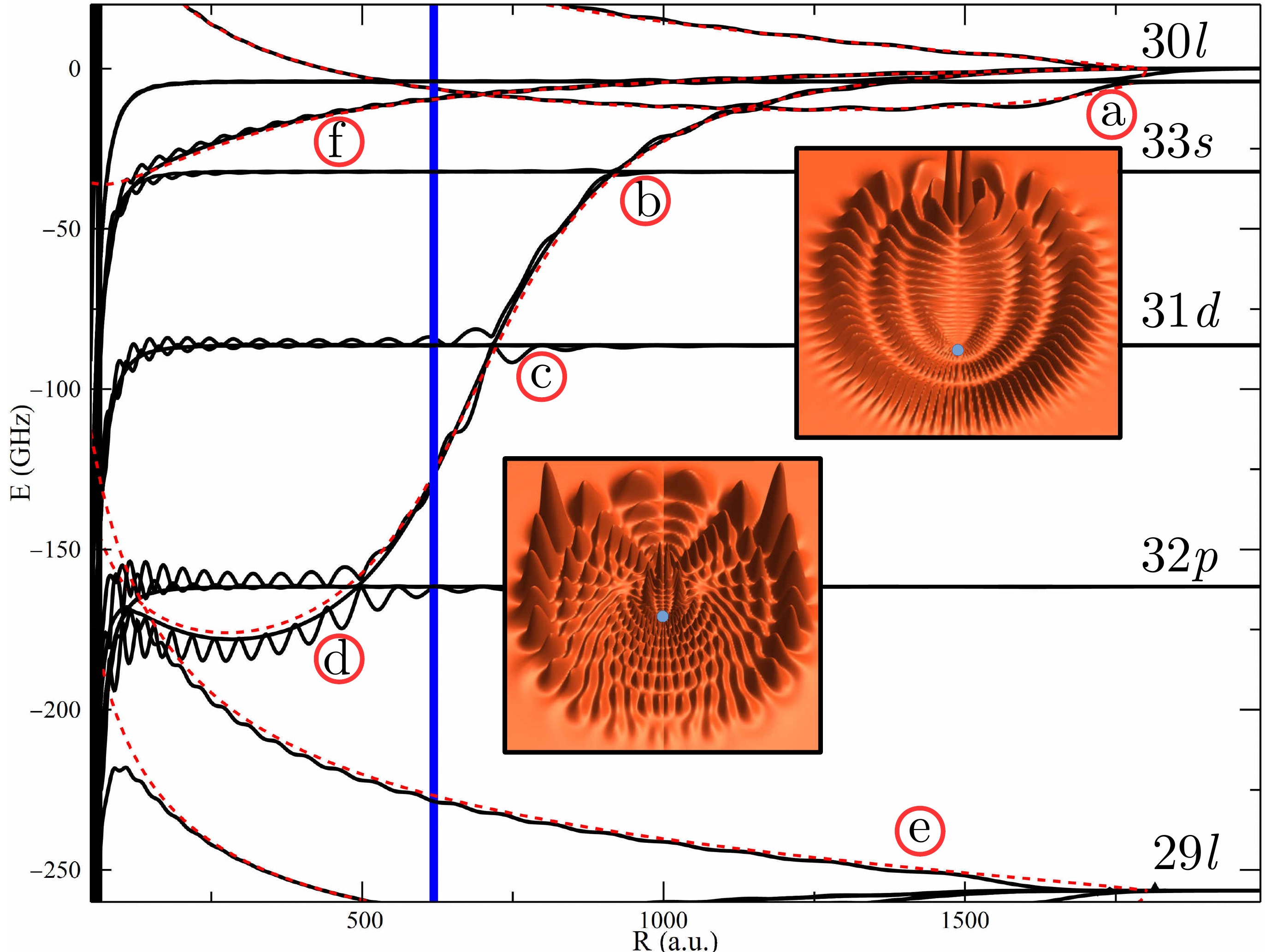}
\caption{The Rb$_2$ PECs obtained by diagonalizing the Fermi pseudopotential (black) or using Eq. \ref{BKmodel} (red,dashed). The location of the $p$-wave divergence is shown as a blue line. Different important regions are highlighted: (a) The trilobite PEC, and the distinctive wave function (plotted in cylindrical coordinates); (b) the $nS$ PEC and its intersection with the butterfly potential; (c) the $nD$ PEC and its intersection with the butterfly potential; (d) the potential wells that support butterfly states and a butterfly wave function in cylindrical coordinates; (e) a the singlet trilobite PEC; (f) the singlet butterfly PEC. }
 \label{fig:Rbmanifold}
\end{center}
 \end{figure*}
  Figure \ref{fig:Rbmanifold} shows the potential energy landscape between the $n=29$ and $n=30$ manifolds for a Rb$_2$ Rydberg molecule. The regularity of the Rydberg spectrum implies that this same picture is repeated largely unchanged between every two Rydberg levels, according to a collection of scaling laws.  The Rydberg level splittings decrease as $n^{-3}$. Away from the $p$-wave shape resonance, the potential wells associated with low-$l$ states get shallower as $n^{-6}$, while the trilobite potential wells, due to the mixing between high-$l$ states, decrease as $n^{-3}$. In contrast, the position of the $p$-wave shape resonance is approximately independent of $n$ while the range of the PECs grows as $n^2$; as a result the relative importance of the $p$-wave shape resonance decreases at higher $n$. Any properties associated with the perturber, such as its zero-energy scattering length or hyperfine splitting, are $n$-independent\footnote{One slight nuance is that the energy-dependence of the scattering length varies with $n$ due to the $n$-dependence of the semiclassical $k(R)$.}.  We now examine Fig. \ref{fig:Rbmanifold} in detail as it shows most of the key features of this unusual class of molecules, and we will expand upon this picture throughout this tutorial. 

The main features are four nearly flat potentials, corresponding to the three non-degenerate low-$l$ states and the manifold of unperturbed states at the hydrogen energy $-1/n^2$, two degenerate smooth $\Pi$-butterfly potentials, and four oscillatory potentials: the triplet and singlet trilobite potentials (labeled (a) and (e), respectively), and the triplet and singlet $\Sigma$ radial butterfly potentials ((d) and (f), respectively). The BK model (dashed red) confirms the accuracy of the Fermi pseudopotential. The insets hold density plots in cylindrical coordinates of the trilobite and butterfly states. The Rydberg core is marked with a blue dot, and the perturber lies under the ``twin peaks'' of probability density. It is clear from the asymmetric bunching of electron density that these have non-zero dipole moments. 

\begin{figure}[b]
 \begin{center}
\includegraphics[width=\columnwidth]{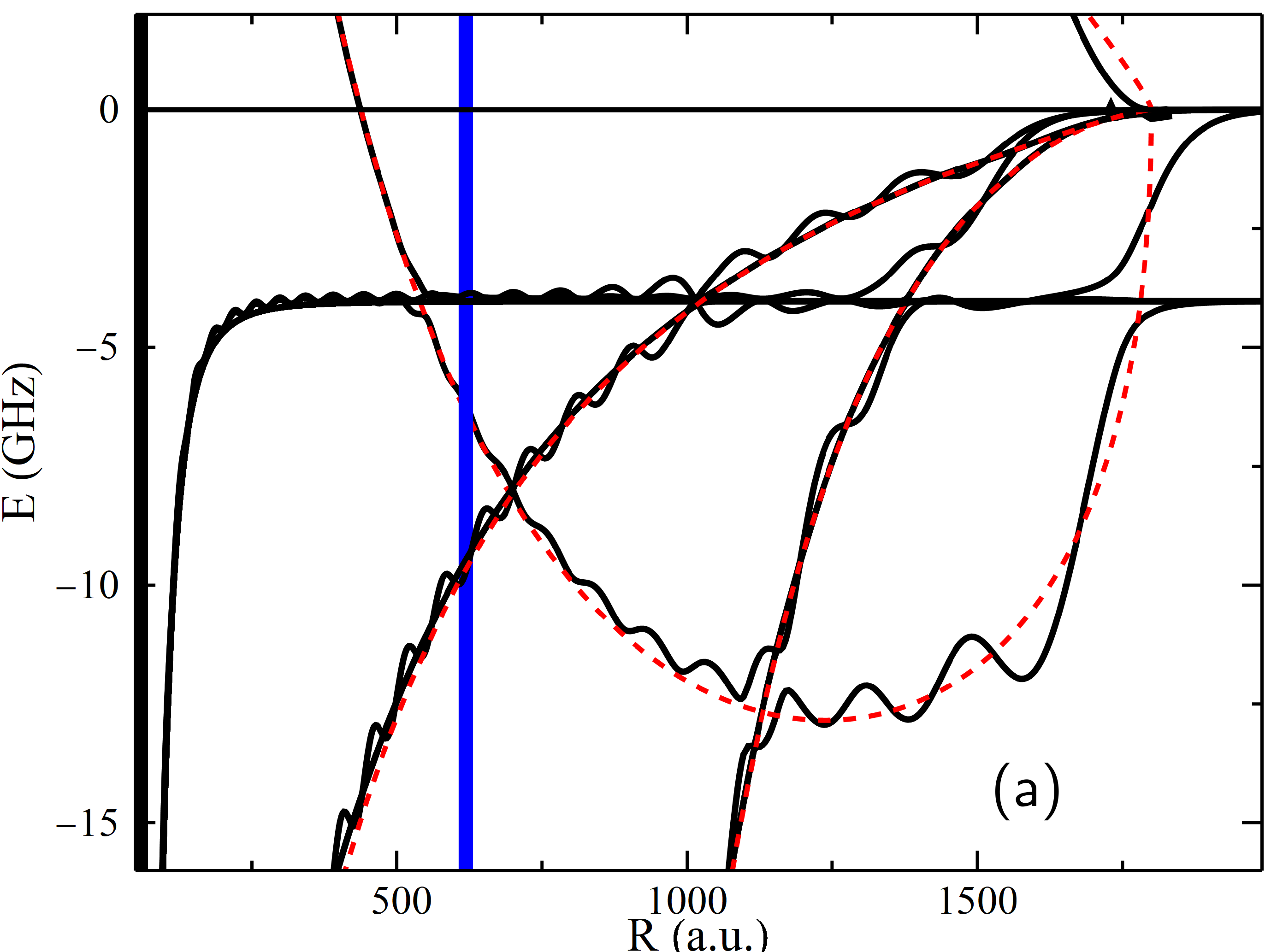}\\
\includegraphics[width=\columnwidth]{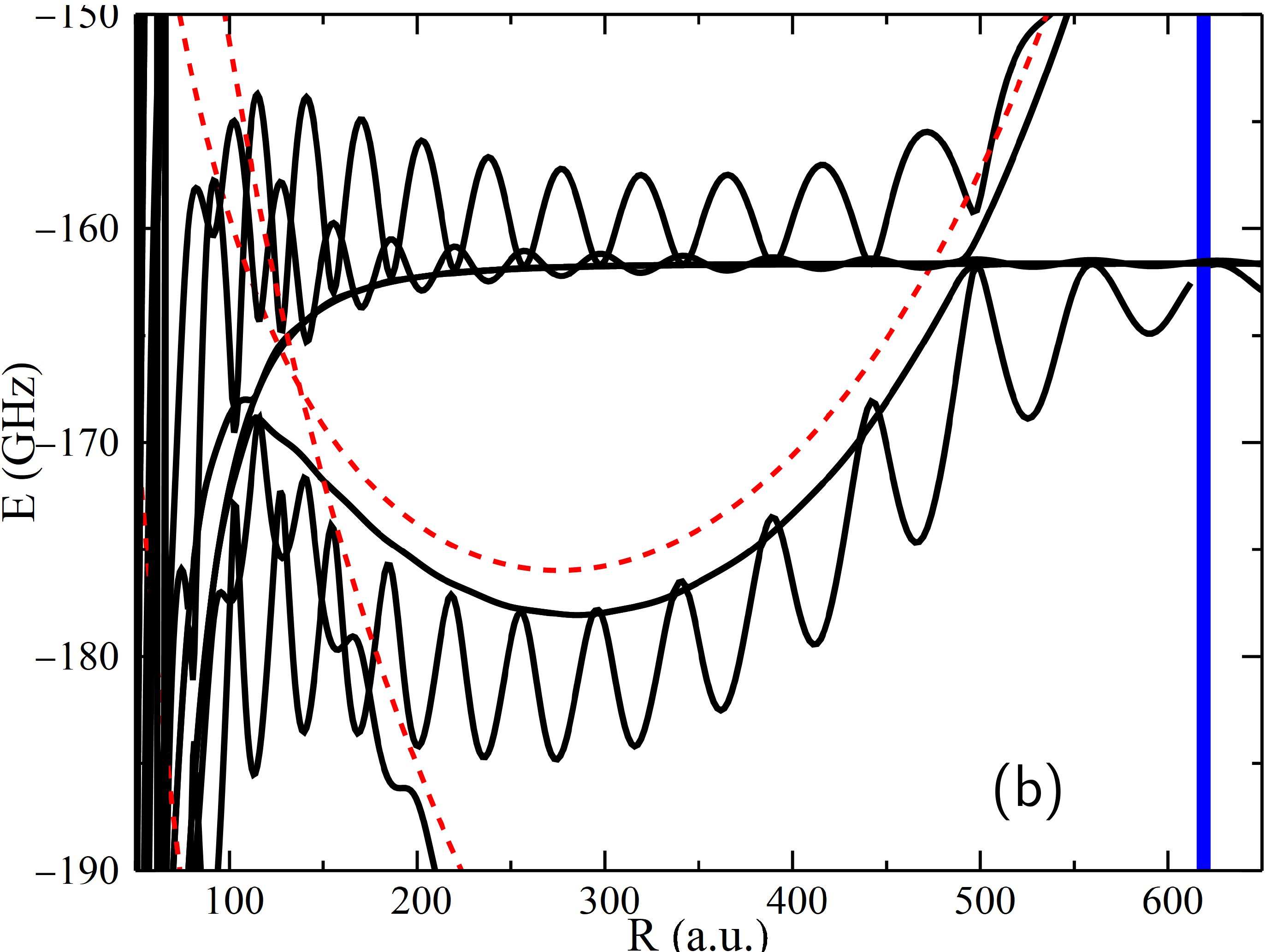}\\
\includegraphics[width=\columnwidth]{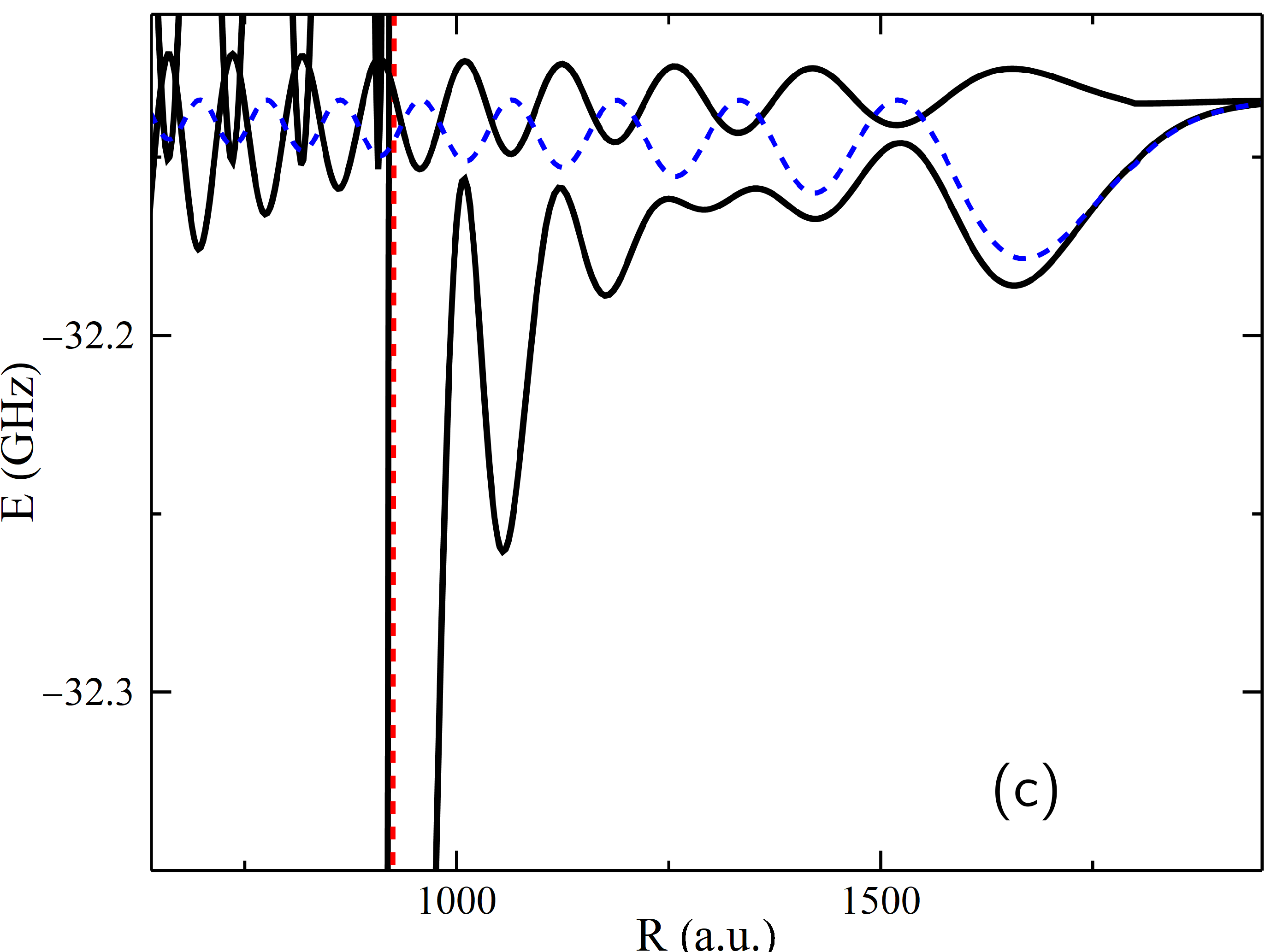}
\caption{Details of the adiabatic potential curves. (a) The triplet trilobite potential curve; (b) the triplet butterfly potential wells; (c) singlet and triplet $nS$ potential curves. The blue vertical line is located at the $p$-wave shape resonance. The dashed red curves show the BK model, while the dashed blue curve in (c) shows the $nS$ potential curve without the influence of the $p$-wave potential. }
 \label{fig:Rbzooms}
\end{center}
 \end{figure}

The triplet trilobite potential curve, marked (a), is several GHz deep and possesses a global minimum due to the change in sign of the phase shift (see Fig. \ref{fig:phases}). The singlet trilobite potential curve, highlighted (e), is basically monotonically increasing because the singlet phase shift exhibits no such Ramsaeur-Townsend minimum. The trilobite curve is shown in more detail in Fig. \ref{fig:Rbzooms}a. The triplet butterfly potential, which also has $\Sigma$ symmetry, dives downward through the trilobite potential and the low-$l$ potentials, creating a series of sharp avoided crossings. Its interaction with the $n=29$ manifold is clearly critical to constrain the divergent scattering volume (the location of this divergence is highlighted as the blue line). A series of butterfly wells are formed at relatively short internuclear distances. The $\Pi$ angular butterfly curves do not couple to any other potentials, and are non-oscillatory as predicted by Eq. \ref{diagonalAngbutterfly}. The butterfly potential wells are highlighted in Fig. \ref{fig:Rbzooms}b. 

Although these states are the most theoretically appealing due to their remarkable wave functions, deep potentials, and large dipole moments, they are experimentally the most challenging to produce. Since they are composed of high-$l$ basis states, dipole selection rules prohibit excitation from the ground state without a three-photon process. It was not until 2015 that a trilobite molecule with a kilo-Debye dipole moment was observed in Cs \cite{BoothTrilobite}. Cesium has a unique advantage over Rb: its $s$-wave quantum defect is very close to an integer ($\mu_s = 4.05$). As such, the trilobite state intersects and couples to the $nS$ potential curve, allowing a two-photon pathway through this admixture. In $n=30$ Rb these states are separated by $>25$GHz, making this coupling extremely weak \cite{PfauSci}. However, following this same logic, formation of butterfly molecules should be possible because the butterfly potential wells are energetically close to the $(n+2)P$ potential curve, and hence have some mixing. Indeed, Rb butterfly states were observed in 2016, photoassociated via single photon excitation \cite{Butterfly}.

In contrast, all of the low-$l$ molecules have been studied extensively since they can be directly coupled to the ground state via single or double photon excitation. Fig. \ref{fig:Rbzooms}c highlights the $nS$ PECs, which are prototypical for the other low-$l$ states in this spin-independent picture. The deep(shallow) curve corresponds to triplet(singlet) scattering. Although the singlet scattering length is positive, small $p$-wave contributions cause the potential wells to fall below the asymptotic energy. The first LRRMs observed were bound in the outermost well of the triplet potential \cite{Bendkowsky}. The dashed blue PEC in Fig. \ref{fig:Rbzooms}c neglects $p$-wave contributions, and shows that the  outermost potential well is determined entirely by $s$-wave scattering.  The butterfly PEC plunges through the $s$-wave potential at about $R=1000$ and, due to this avoided crossing, strongly affects the vibrational states, particularly the excited ones not localized in the outermost potential well. The $p$-wave shape resonance leads to a sharp drop in the potential curve, and the lack of an inner barrier seems to suggest that vibrational states will rapidly decay or even be destroyed. However, it was observed and pointed out in Ref. \cite{quantumreflection} that these states can still exist due to quantum reflection: at exactly the bound state energies the molecular wave function exponentially decays in the plateau region to the left of the shape resonance, reflecting the strange quantum mechanical principle that a potential drop can function similarly to a barrier. 

This narrow $p$-wave crossing also calls into question the accuracy of the adiabatic Born-Oppenheimer approximation since its applicability depends not only on the difference between nuclear and electronic masses but also on the energy separation between PECs. Ref. \cite{UltracoldChem}  studied how these sharp avoided crossings lead to novel chemical pathways and non-adiabatic processes like $l$-changing collisions.  Sr does not have a $p$-wave shape resonance, and hence its PECs more closely resemble the dashed blue curve in Fig. \ref{fig:Rbzooms}c. Strontium LRRMs \cite{DeSalvo2015} are therefore useful to compare with Rb in order to investigate the role of this $p$-wave resonance on the decay channels and lifetimes of these molecules \cite{Camargo2016,WhalenLifetimes,Rblifetimes}. 
  
The other low-$l$ states, asymptotically associated with Rydberg $nP$ and $nD$ states, have very similar potential curves as the $nS$ molecules just discussed. In Rb, $nP$ \cite{Niederprum} and $nD$ \cite{AndersonPRL,MacLennan,PfauKurz}  molecules have been observed, and in Cs $nP$ molecules have been studied\cite{Sass}. Since these Rydberg states have fine structure, which non-trivially couples to the perturber's hyperfine structure, the observed spectra require the full spin-dependent calculation described in Sec. \ref{sec:spinintro} and discuss these states further there. 

Although we do not study it in detail, there is one final step after obtaining these PECs before the properties of LRRMs are known: the nuclear  Schr\"{o}dinger equation must be solved using these adiabatic PECs. This yields the molecular spectrum, and shows that the vibrational states are typically split by several tens of MHz (for $20\le n \le 40$) and have rotational splittings, inversely proportional to their bond length, on the order of kHz which are typically unresolved. The lifetimes of these states are comparable to those of  Rydberg atoms, although somewhat shortened due to additional decay routes provided by the molecular structure. 
 \subsection{Electric dipole moments}
 \begin{figure}[t]
{\normalsize 
\begin{centering}
\begin{center}
\includegraphics[scale =0.08]{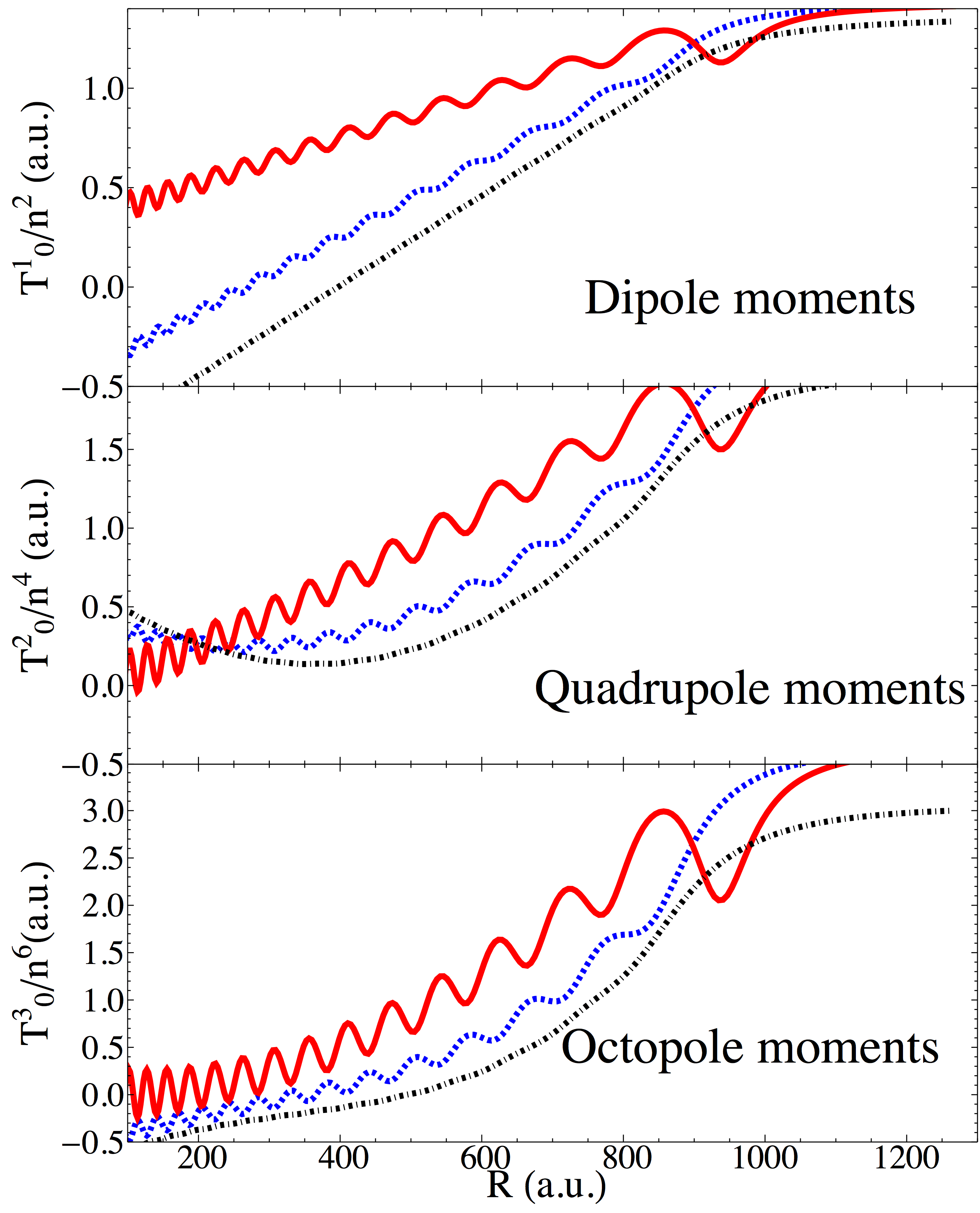}
\end{center}
\end{centering}
}
\caption{ Hydrogenic dipole, quadrupole, and octupole moments for $n =23$, as determined by Eq. \ref{eq:dipolehydrogen}. The trilobite (blue,dashed) and $\Sigma$ butterfly state (red,solid) oscillate as a function of $R$, while the $\Pi$ butterfly state (black,dot-dashed) does not.  This figure is taken from Ref. \cite{EilesSpin}.  }
\label{fig:multipoles}
\end{figure}

The state mixing induced by the perturber creates permanent electric dipole moments exceeding hundreds of Debye. Because of the coupling between the polar trilobite/butterfly states and the low-$l$ states, even these exhibit dipole moments of a few Debye \cite{PfauSci}. These dipole moments have sparked interest in the application of these molecules in dipolar gases and ultracold chemistry. Here we calculate arbitrary multipole moments for the trilobite/butterfly states, $
d_{\alpha}^{k,q} =\bra{\+{\Upsilon_{Rr,n}^{\pmb\xi\pmb1}}}T^k_q\ket{\+{\Upsilon_{Rr,n}^{\pmb\xi\pmb1}}}$. The multipole moments from classical electrostatics \cite{Jackson} are promoted to quantum-mechanical operators: 
\be
T^k_q = -r^kC_{kq}(\hat r),\,\,\,C_{kq}(\hat r) = \sqrt{\frac{4\pi}{2k+1}}Y_{kq}(\hat r).\nonumber
\ee
Here $k$ and $q$ label tensor operator components; $T^1_0$ is the usual dipole operator. 
A straightforward calculation provides
\begin{align}
\label{eq:dipolehydrogen}\langle \xi| T_q^k|\xi\rangle &= \sum_{l,l'}\frac{\left[\phi_{nlm}^\xi(R)\right]^*\phi_{nl'm'}^\xi(R)}{\+{\Upsilon_{RR,n}^{\pmb\alpha\pmb\alpha}}}  R_{nl}^{nl'}(k)\\& \times C_{l'\Omega,kq}^{l\Omega}(-1)^{k-l'}\sqrt{(2l'+1)}\begin{pmatrix}l & l' & k\\ 0 & 0 & 0 \end{pmatrix}.\nonumber
\end{align}
The radial matrix element 
\be
\label{eq:radmatelforfield}
R_{nl}^{nl'}(k) = \int_0^\infty u_{nl}(r)u_{nl'}(r)r^k\dd{r},
\ee 
can be evaluated analytically \cite{radmatel}.  These multipole moments scale as $n^{2k}$, and are displayed  in Fig. \ref{fig:multipoles} up to the octupole moments. At small $R$ the dipole moments of both butterfly symmetries become negative.

This section introduced the foundational concepts  and properties of Rydberg molecules: oscillatory PECs, extremely large bond lengths and, in the trilobite and butterfly cases, highly localized wave functions with exotic nodal structure and large permanent electric dipole moments. All of the resoundingly succesful experimental observations of these molecules, with the exception of satellite peak observations at thermal temperatures\cite{Crowell,NiemaxHet}, have occurred recently -- within the last decade. In the following sections we discuss the theoretical progress made in response to this experimental success.

 \section{Polyatomic Rydberg molecules}
 \label{sec:polyintro}
 
 Rydberg molecules can be formed over the whole range of principle quantum number $n$. For small $n$ they were photoassociated directly from bound Rb-Rb molecules \cite{Carollo1,Carollo2}. As $n$ increases, or equivalently as the atomic density $\rho$ increases, the average number of perturbers $N\sim \rho n^6$ within a Rydberg volume grows rapidly. As $N$ increases, so does the probability that polyatomic molecules -- trimers, tetramers, and so on -- can form. For $N\gg1$, the individual molecular lines smear into a mean field energy shift linear in density and proportional to the zero-energy scattering length, and we return to the scenario originally studied by Amaldi, Segre and Fermi \cite{PfauBEC,RydbergRev}. Between the two extremes of $N = 1$ and $N\gg 1$ resides a range of fascinating phenomena involving polyatomic LRRMs, and this section investigates their structure and properties. The electronic Hamiltonian of Eq. \ref{fermicompact} is expanded\footnote{Atom-atom van der Waals interactions are negligible at the Rydberg-scale internuclear distances we consider here.} to include $N$ perturber atoms located at $\vec R_i=(R_i,\theta_i,\varphi_i)$:
\begin{align}
\label{ham}
H_N({\vec r};\{\vec R_i\}) 
& =  H_0 + 2\pi \sum_{i=1}^N\sum_{\xi = 1}^4V_\xi(\vec R_i,\vec r).
\end{align}
Since the number of spatial degrees of freedom grows rapidly in a polyatomic molecule, calculation of the potential energy surfaces becomes computationally challenging and visualization becomes nearly impossible.  In the following examples illustrating the generic structure of polymers we will study only the breathing mode, i.e. the cut through the potential surfaces where all atoms share a common distance $R$ from the Rydberg core. Although this gives only a small glimpse of the full physical picture, these cuts illustrate most of the interesting physics determining the molecular structure. A more complete analysis can be found in Refs. \cite{FeyKurz,FeyTrimer}, which study the potential surfaces of triatomic LRRMs and reveal the role of stretching modes.

Several experiments have observed polyatomic LRRMs  of the simplest type:  an $nS$ Rydberg excitation. Soon after the first observation of diatomic states, additional spectral lines much deeper than the deepest dimer state were seen \cite{quantumreflection}. These states were identified as trimers for the following reasons. First, the $nS$ state is non-degenerate, and thus we can sum the PEC from Eq. \ref{lowlNone} over all perturbers. Second, since the $nS$ state is spherically isotropic and along the breathing mode all $R_i = R$, there is no dependence on $i$ in Eq. \ref{lowlNone}, and hence the polyatomic PEC is $N E_{l_\text{min}=0}^\Sigma(R)$. Since the outermost potential well is approximately harmonic, this scaling implies that the vibrational states of a ($N+1$)-mer are $N$ times deeper than the dimer state, and are thus clearly identified in an experimental spectrum. In later experiments in both Rb and Sr \cite{MolSpec,Schlag,Whalen2} even more polymer (trimer, tetramer, and pentamer) lines were identified. Excited vibrational states were also assigned in this fashion in the measurements of Ref. \cite{Whalen2}, which span four orders of magnitude in spectral intensity. At higher $n$ or $\rho$ these vibrational lines blend together into an overall shift with a distinctive lineshape. Some aspects of this lineshape in Rb were attributed to the $p$-wave shape resonance \cite{Schlag}, while other results were interpreted via quantum many-body calculations adapted to study Rydberg ``polarons'' \cite{WhalenPoly,Demler,Whalen2}. These observations are fundamentally reliant upon the accuracy of the potential energy surfaces defining the molecular structure, which is our focus here.

 \begin{figure*}[t]
\begin{centering}
\includegraphics[width = \textwidth]{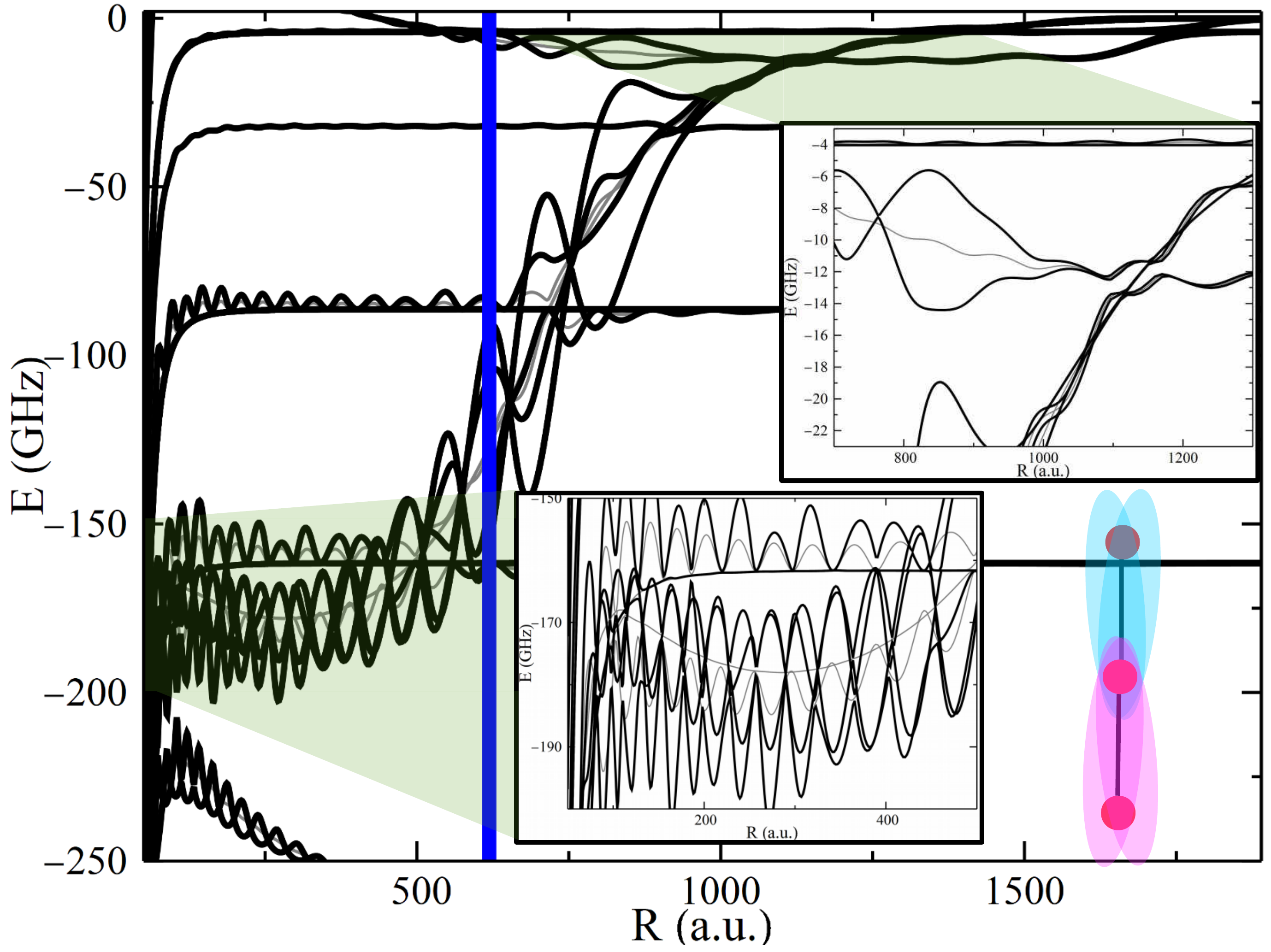}
\caption{{Breathing mode PECs, relative to the $n=30$ manifold, for a Rb$_3$ trimer in a collinear geometry. Only triplet states are shown. The insets highlight the trilobite and butterfly potential wells. The faint gray curves show the dimer PECs. The diagram in the bottom right depicts schematically the dimer orbitals, in pink and blue, stretching to each perturber. The vertical blue line denotes the $p$-wave resonance position.}}
\label{fig:poly1}
\end{centering}
\end{figure*}

\begin{figure*}[t]
\begin{centering}
\includegraphics[width = \textwidth]{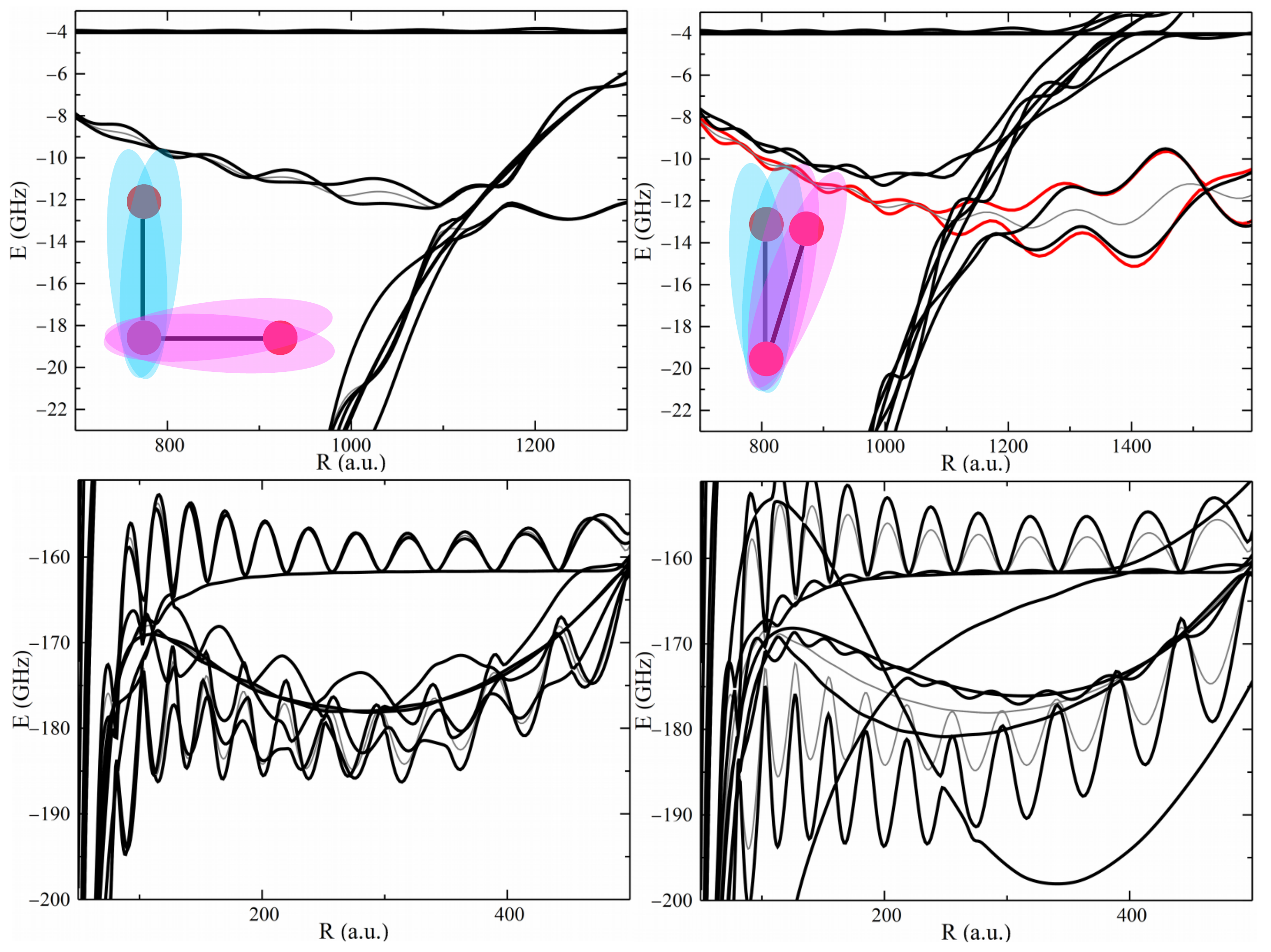}
\caption{{Breathing mode PECs for an $n=30$ Rb$_3$ trimer in two different geometries. Only triplet states are shown. The top panels highlight the trilobite region, while the bottom panels highlight the butterfly potential wells; the left panels are for a right-angle geometry while the right panels are for an angle $\pi/10$. The faint gray curves running through all four figures show the dimer PECs. The red curve neglects butterfly couplings. }}
\label{fig:polymore}
\end{centering}
\end{figure*}

We develop a generic description of polyatomic LRRMs \footnote{Just as Rydberg molecules (of the H$_2$ variety), ultra-long-range Rydberg molecules, and Rydberg-Rydberg macrodimers can be easily confused due to their similar appellations, so it is with polyatomic Rydberg molecules. Several references have investigated a different type of polyatomic molecule formed by replacing the perturber with a polar dimer such as KRb \cite{Rittenhouse2010,Rittenhouse2011, Mayle2012,Rosario2015}, or even with several polar  perturbers \cite{polarperturberpoly}.} using the trilobite orbital method. We enlarge the basis set used in the trial wave function in Eq. \ref{trialwavetrilo} to include trilobite and butterfly states associated with each new perturber:
\begin{align}
\label{trialwavepoly}
\Psi(\vec r; \vec R) = \sum_{n=n_1}^{n=n_2}\Bigg[\sum_{l=0}^{l=l_\text{min}}&\sum_{m=-l}^{m=l}c_{nlm}(\vec R)\phi_{nlm}^1(\vec r)\\&+\sum_{i=1}^N\sum_{\xi=1}^4\mathcal{C}_{ni\xi}\+{\Upsilon_{ir,n}^{\pmb\xi1}}\Bigg].\nonumber
\end{align}
The basis size,  $\mathcal{M} (4N+(l_\text{min}+1)^2)$, grows linearly with $N$. The Hamiltonian matrix elements derived for the dimer readily generalize. The block of high$-l$ trilobite states now contains additional overlap terms between trilobite states associated with each perturber,
\begin{align}
\label{matelementpolymer1}
&\bra{\+{\Upsilon_{pr,n}^{\pmb\alpha 1}}} H_N({\vec r};\{\vec R_i\}) \ket{\+{\Upsilon_{qr,n'}^{\pmb\beta 1}} }\\&=-\frac{1}{2n^2}\+{\Upsilon_{pq,n}^{\pmb\alpha\pmb\beta}}\delta_{nn'}+ 2\pi \sum_{i=1}^N\sum_{\xi = 1}^4a_\xi\+{\Upsilon_{pi,n}^{\pmb\alpha\pmb\xi}}\+{\Upsilon_{iq,n'}^{\pmb\xi\pmb\beta}}.\nonumber
\end{align}
The quantum defect block is unchanged except for an additional summation over perturbers, 
\begin{align}
&\bra{\phi_{nlm}^1}H_N({\vec r};\{\vec R_i\}) \ket{\phi_{n'l'm'}^1} \\&= -\frac{\delta_{nn'}\delta_{ll'}}{2(n-\mu_l)^2} + 2\pi  \sum_{i=1}^N\sum_{\xi = 1}^4a_\xi\phi_{nlm}^\xi(R_i)^*\phi_{n'l'm'}^\xi(R_i),\nonumber
\end{align}
and the off-diagonal coupling blocks are 
\begin{align}
\label{endmat}
&\bra{\+{\Upsilon_{pr,n}^{\pmb\alpha 1}}}H_N({\vec r};\{\vec R_i\}) \ket{\phi_{n'l'm'}^{1} }\\&
= 2\pi  \sum_{i=1}^N\sum_{\xi = 1}^4a_\xi\+{\Upsilon_{p i,n}^{\pmb\alpha \pmb\xi}}\phi_{n'l'm'}^\xi(R_i).\nonumber
\end{align}
The overlap matrix has now a larger trilobite block given by $\+{\Upsilon_{pq,n}^{\pmb\alpha\pmb\beta}}\delta_{nn'}$, but the rest is unchanged.   This matrix is typically one to two orders of magnitude smaller than that required when using the Rydberg  basis, a major increase in computational efficiency. Furthermore, it reveals the essential structure of the trilobite states of these molecules that is difficult to extract from full basis.

This structure is most easily observed by considering just the trilobite states within a single $n$-manifold. Eqs. \ref{matelementpolymer1}-\ref{endmat} then simplify, and the PECs are the $N$ eigenvalues of $\+{\Upsilon_{pq,n}^{11}}$, an $N\times N$ matrix\footnote{This is strictly true only if the scattering length is equal for all perturbers; in fact, the PECs are eigenvalues of a generalized matrix equation $\sum_ia(R_i)\+{\Upsilon_{pi,n}^{11}}\+{\Upsilon_{iq,n}^{11}}\vec c=E\+{\Upsilon_{pq,n}^{11}}\vec c$}. The average value of these eigenvalues for the polyatomic system equals the single eigenvalue of the dimer. The off-diagonal elements of $\+{\Upsilon_{pq,n}^{\pmb1\pmb1}}$,$\alpha\ne\beta$ correspond to the overlap between trilobite states extending from the Rydberg core to different perturbers, and their size determines the splittings between the polyatomic eigenvalues and their average value.  If these off-diagonal elements vanish, the eigenvalues are simply $N$ copies of the dimer eigenvalue $\+{\Upsilon_{RR,n}^{11}}$. The formation of polyatomic states is therefore predicated on the overlap between these trilobite basis states, lending this approach some physical meaning beyond its mathematical effectiveness. The angular dependence of the trilobite wave functions likewise contributes considerably to the energy landscape\cite{FeyTrimer}.  

In Figs. \ref{fig:poly1} and \ref{fig:polymore} we show three sets of breathing mode PECs for a triatomic molecule. This calculation includes three hydrogenic manifolds and, as a result of the computational gains granted by the trilobite overlap method, are the first calculations showing the trimer butterfly potential curves. In the collinear configuration shown in Fig. \ref{fig:poly1} the high level of symmetry leads to large overlap elements, and hence large splittings in the trimer potential curves. These have gerade and ungerade symmetry, and are especially notable in the angular butterfly curves which, due to their initial degeneracy in the diatomic case, couple strongly and are shifted wildly from the dimer limit. At large $R$ the overlap vanishes and the PECs converge to the dimer curve. 

Fig. \ref{fig:polymore} provides more detail about the effect of the molecular geometry. In the 90$^o$ configuration, the trilobite-like states overlap much less than in the collinear configuration, and the potential curves exhibit much smaller splittings. An angular arrangement at a very small angle, $\pi/10$, is shown on the right. The overlaps now increase at larger $R$, unlike the other two configurations, and moreover the trilobite curves to the left of the butterfly crossing no longer oscillate about the dimer potential. The trilobite potentials without any $p$-wave contributions (red) behave as expected, oscillating about the dimer PEC. This shows that the coupling between butterfly and trilobite states is quite sensitive to the geometry; in this configuration the pronounced avoided crossing between states shifts the trilobite potentials on either side of the crossing so that they are higher (left) and lower (on the right) than the dimer value. These strong couplings are also manifest in the butterfly potentials in the lower right panel, which evince exagerrated, yet slow, oscillations. 

\begin{figure*}[ht]
\begin{center}
\subfigure[]{\label{hoodoo2}\includegraphics[scale = 0.4035]{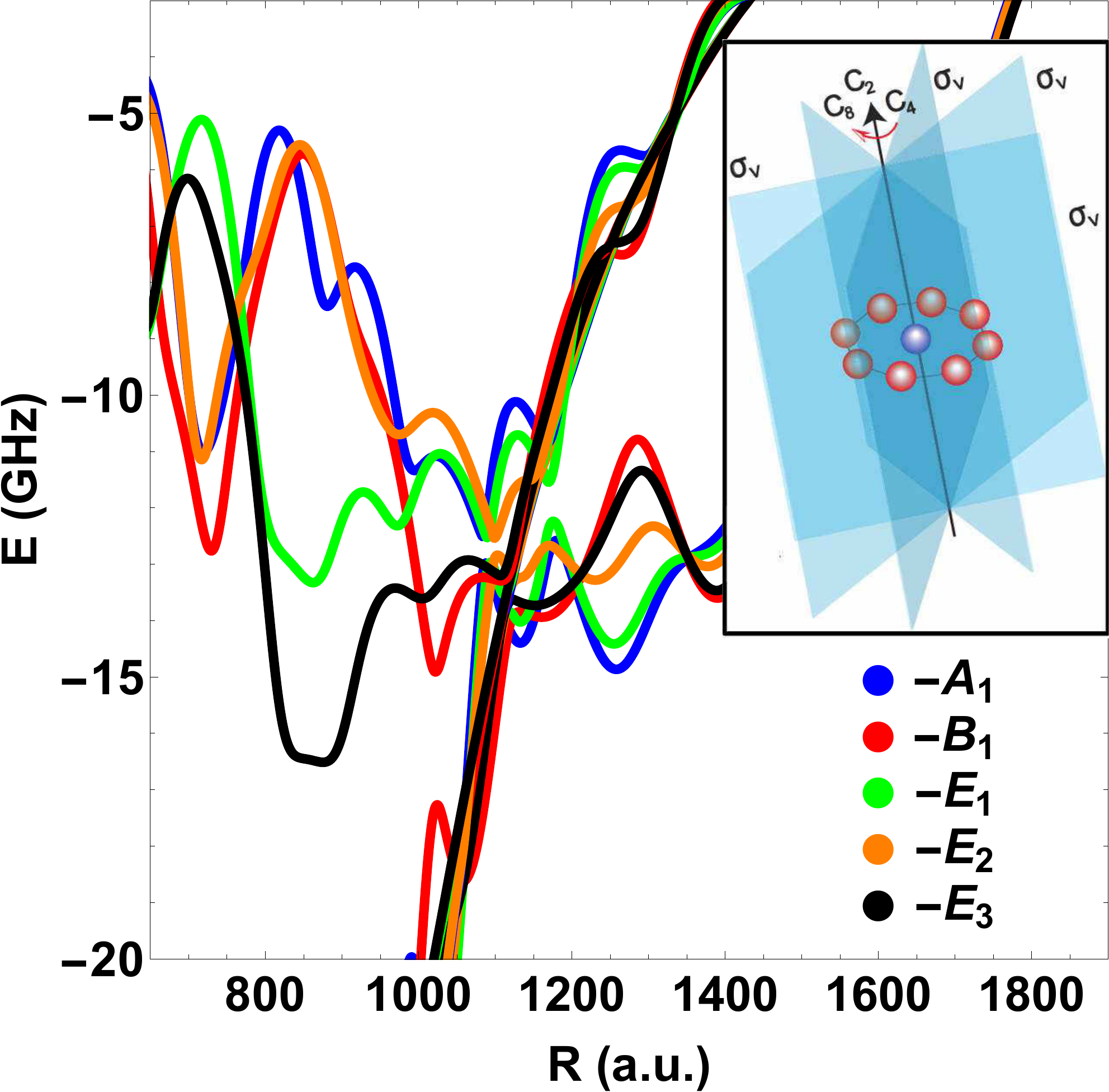}}
\subfigure[]{\label{hoodoo2}\includegraphics[scale = 0.4]{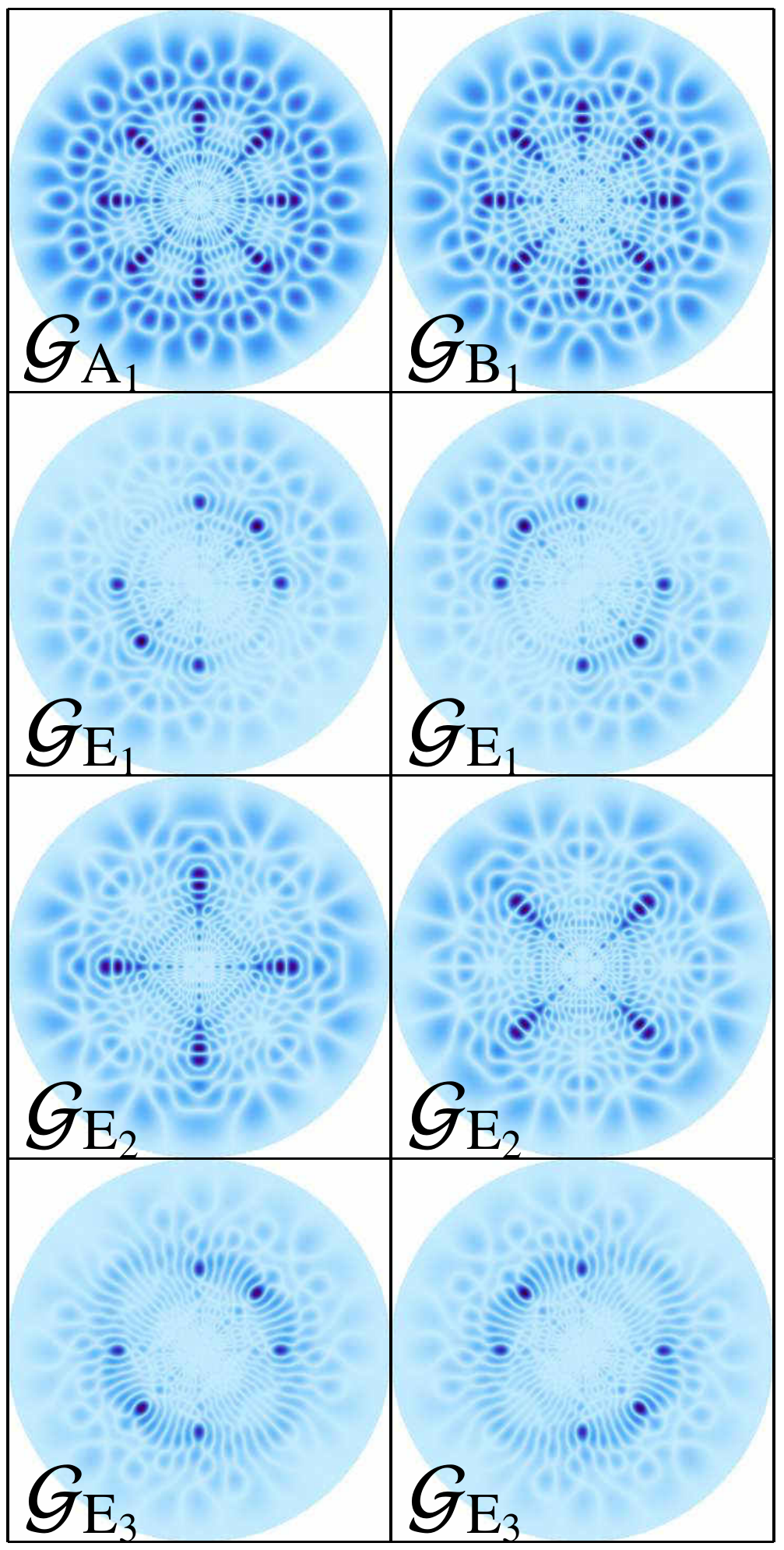}}\subfigure[]{\label{hoodoo3}\includegraphics[scale = 0.4]{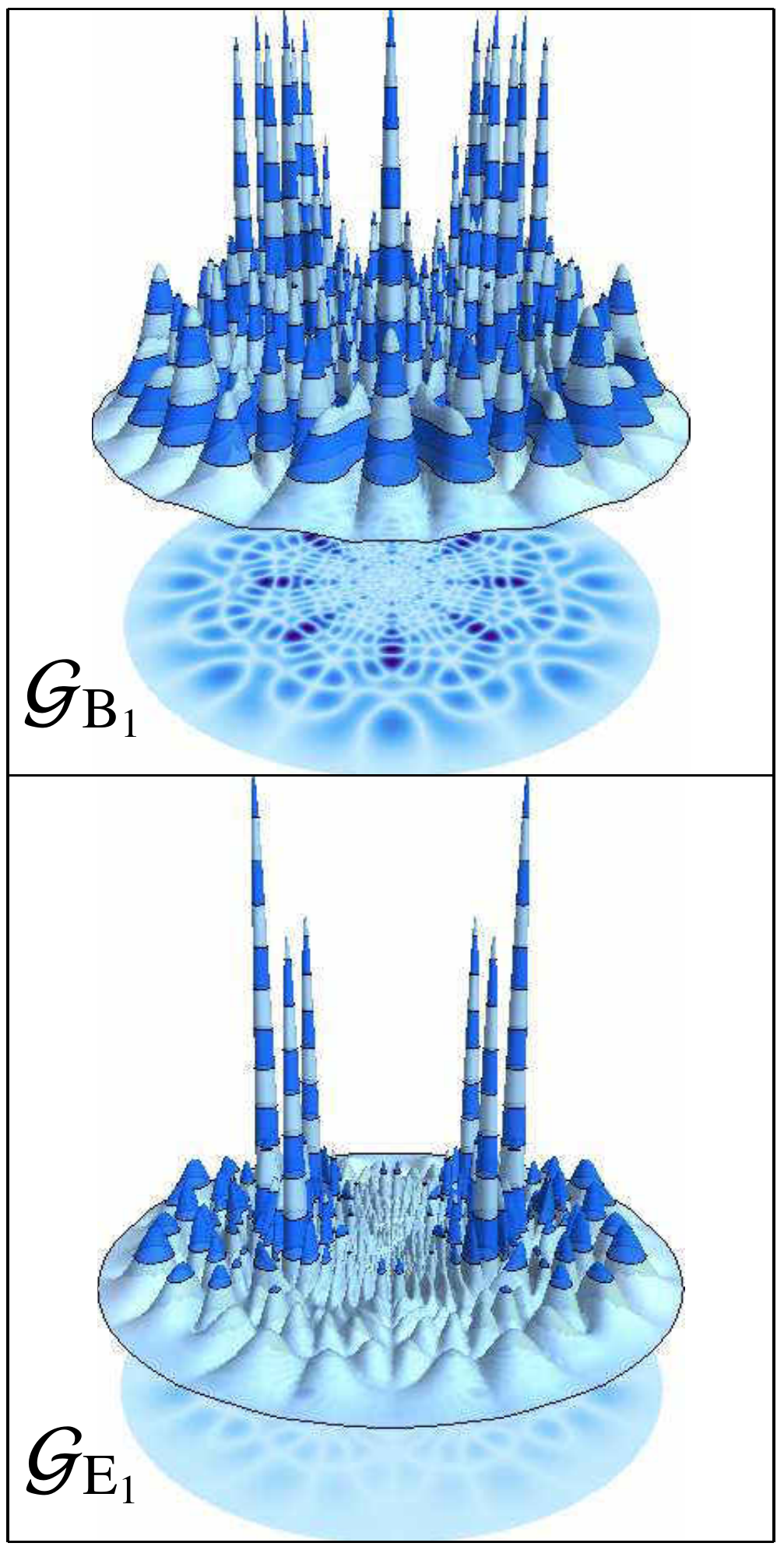}}
\end{center}
\begin{center}
\caption{(a) PECs for an octagonal molecular geometry, color-coded by their irrep. The symmetry operations for $C_{8v}$ symmetry are shown in the inset.  (b) ``Hoodoo'' symmetry adapted trilobite orbitals for evaluated at $R = 840$ $a_0$. The probability amplitude $\sqrt{r^2|\psi(x,y,0)|^2}$ is plotted in the $xy$ plane. (c) The  electron probability corresponding to the one-dimensional irrep $B_1$ (top) and one of the doubly-degenerate $E_1$ irreps (bottom) are plotted.  This figure is modified from Ref. \cite{JPBdens}.  }
\end{center}
\label{mainhoodoofigure1}
\end{figure*}

The methodology developed in the present section applies  to an arbitrary arrangement of perturbers, and shows that the spectroscopic signatures of polyatomic formation in non-isotropic Rydberg states will not be so clear as in $nS$ Rydberg states. Some additional understanding of the dependence of these potential curves on the geometry is gained by characterizing the symmetry group of the molecule. The molecular symmetry group is a 
subgroup of the complete nuclear permutation inversion group of 
the molecule~\cite{Bunker, Bunker2}, which commutes with the molecular 
Hamiltonian in free space. Therefore, the eigenstates of such a 
Hamiltonian can be classified in terms of the irreducible 
representations (irreps) of the given molecular symmetry group, called {\it symmetry-adapted orbitals} (SAOs). Given a molecular symmetry 
group, it is possible to calculate the SAOs associated 
with each irrep of the group using the projection operator method \cite{Bunker}. The projection operator also gives the coefficients $\mathcal{A}_p^{(\alpha,j)}$ for the SAO $\mathcal{G}^{(\alpha,j)}(\vec r)$ corresponding to the $\alpha^\text{th}$ orbital and $j^\text{th}$ irrep:
\be
\label{sao}
\mathcal{G}^{(\alpha,j)}(\vec r)=\sum_{p = 1}^N\+{\Upsilon_{pr,n}^{\pmb\alpha\pmb1}}\mathcal{A}_p^{(\alpha,j)}.
 \ee
The prescription for calculating the projection operator depends on the orbital in question, and becomes quite involved for the angular butterfly states since they mix together under symmetry operations. Ref. \cite{JPBdens} describes this calculation in full detail; here we report only the particularly elegant expression for the PEC for the trilobite molecule of the $j^\text{th}$ irrep:
\be
\label{saoswave}
E^{(j)} = 2\pi a_s(k)\sum_{p,q=1}^N\mathcal{A}_p^{j}\+{\Upsilon_{pq}^{11}}\mathcal{A}_q^{j}.
\ee
The PECs of an octagonal arrangement of atoms, characterized by their irreps, and symmetry adapted trilobite orbitals are displayed in Fig. \ref{mainhoodoofigure1} . Panel a) shows the PECs, color-coded by their symmetry irrep. This shows how the coupling strengths between trilobite and butterfly potentials depend on the irrep and how PECs of different irreps have real crossings. They symmetry irreps also determine the degeneracy remaining in the system: the $E_x$ irreps are all doubly degenerate, and hence only five rather than the anticipated eight PECs are visible. Panels b and c show representative symmetry adapted orbitals as density plots at the internuclear distance $R = 840$ $a_0$.   These explicitly exhibit the allowed symmetries and the interference between trilobite orbitals in their beautiful nodal patterns.

 \begin{figure}[t]
 \begin{center}
\includegraphics[width= \columnwidth]{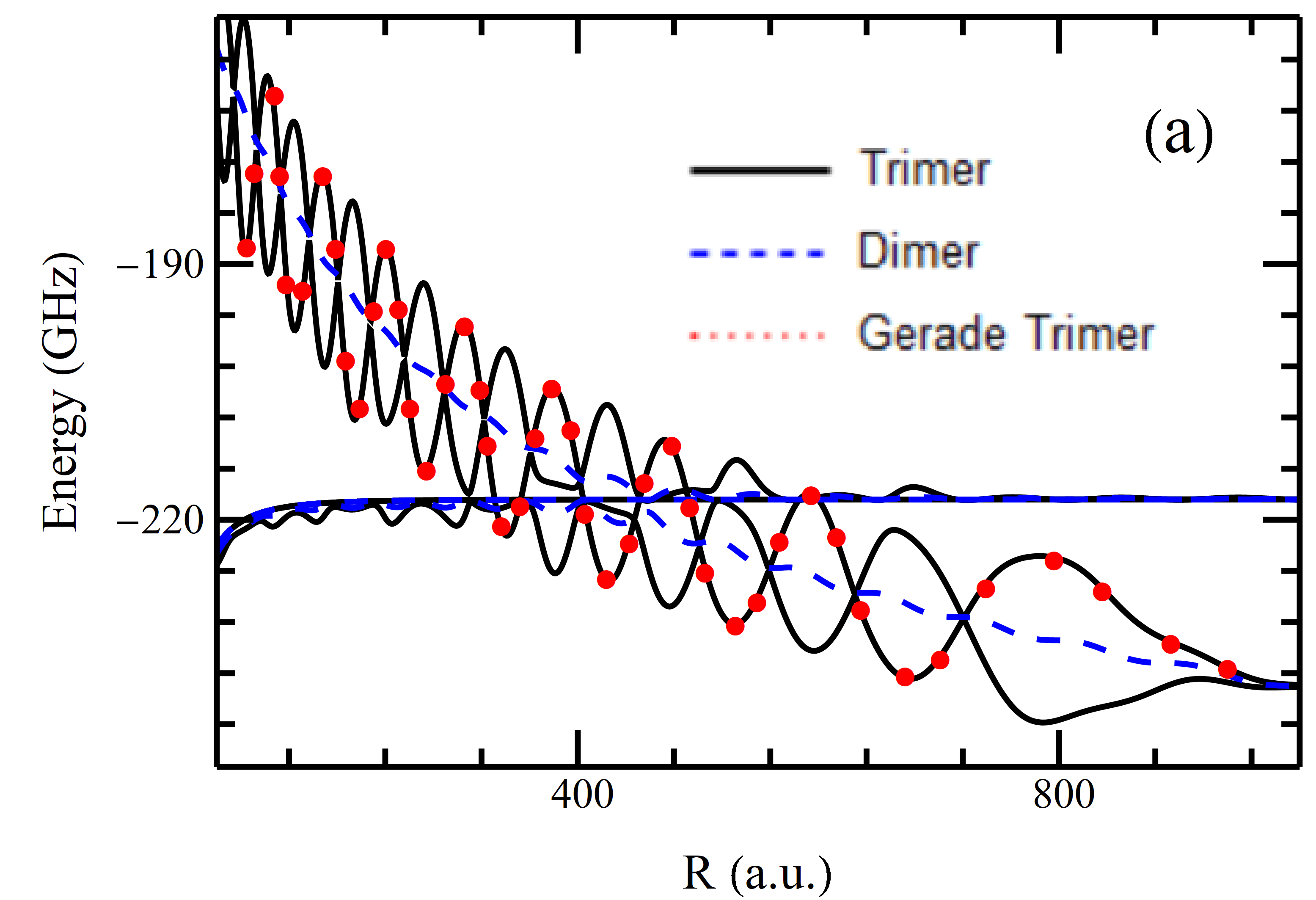}\\
\includegraphics[width= \columnwidth]{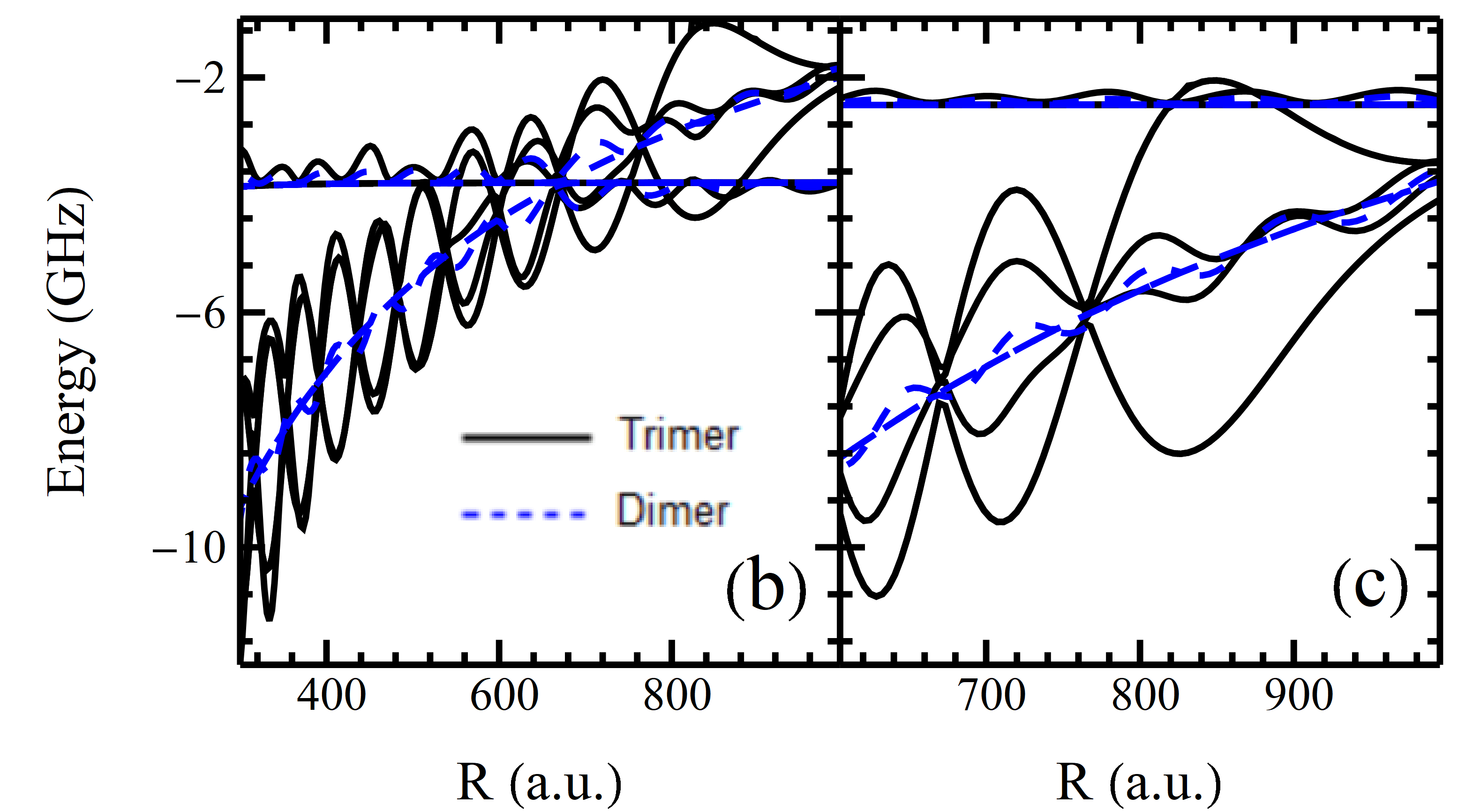}
\caption{Top: PECs showing the Borromean collinear trimer state in Na$_3$.  Bottom: $^1P$ and borromean trimer curves in Na$_3$ (left) and K$_3$ (right). The energy scale is relative to the $n=30$ Rydberg energy.}
\label{fig:trimers} 
\end{center}
\end{figure} 

The large oscillations in the breathing mode potentials in polyatomic molecules appear strongest in these symmetric configurations, and more stable LRRMs can thus be engineered by exploiting these features. They can also lead to exotic Borromean trimer\footnote{Borromean rings are three interlaced rings which, although any two of them are not attached together, cannot as a trio be separated.} states, originally predicted by Ref. \cite{Rost2009} to occur in neon. Ne has a small positive scattering length, and thus has weakly repulsive trilobite PECs which cannot support bound dimer states. However, when a third atom is introduced to form a collinear trimer with the Rydberg atom in the middle, the large oscillatory gerade/ungerade splitting creates deeper potential wells which can support bound states \cite{Rost2009}. Sodium is a more experimentally viable species to realize this scenario. It has a larger singlet scattering length and a nearly integer ($\sim$0.86) $nP$ quantum defect which brings this state energetically close to the $(n-1)$ hydrogenic manifold. It therefore couples to the ungerade trilobite trimer as shown in in Fig. \ref{fig:trimers}a. The gerade trimer remains uncoupled due to its opposite parity.  This Borromean trimer could be excited via the $nP$ admixture, circumventing the need to excite a high-$l$ state in the original proposal. Fig. \ref{fig:trimers}b,c show that this same phenomena occurs in the singlet butterfly curves, which have potential wells nearly an order of magnitude deeper in the trimer configuration (black) than in the dimer (dashed blue). 

What can we learn in general as $N$ tends towards larger values? The fact that the trilobite polymer PECs average to diatomic PEC reveals that the trilobite interaction is profoundly non-additive. The sophisticated quantum many-body treatment of Refs. \cite{WhalenPoly,Whalen2} has performed excellently in predicting the observed line shapes, but as it relies on the additive scaling of the $nS$ vibrational states it is not clear how to interpret spectra from states with higher $l$. In particular, the dramatic localization of the trilobite state about the location of the perturber causes the overlap matrix elements to be particularly large along the breathing mode, lending some physical import to the breathing mode potentials studied here. The lack of large overlap elements causes the spectrum in a random gas, even for fairly large $N$, to resemble that of the dimer \cite{EilesHyd}. As $N$ increases further the trilobite states fill the Rydberg volume and begin to have significant overlap. One now expects the presence of so many perturbers to destabilize the trilobite, leading to only delocalized polyatomic wave functions heavily perturbed by the dense gas. However, a second process competes with this in a truly random disordered environment. As explained in Ref. \cite{RostLuukko}, as $N$ increases so does the probability that two or more perturbers are in close proximity until it is essentially guaranteed\footnote{This is analogous to the ``birthday paradox:'' there is a 50\% probability that two people in a room of twenty-three will share a birthday.}. When a cluster of just two nearby atoms forms, the Hamiltonian matrix contains a $2\times 2$ sub-block that, to first order, has identical elements. This sub-block approximately decouples from the rest of the matrix, and has one vanishing eigenvalue and one at twice the dimer energy. A cluster of $N$ atoms thus behaves like a perturber with $N$ times the scattering length, which attracts the wave function more strongly and resurrects the trilobite molecule  \cite{RostLuukko}.  
  \section{Spin and relativistic effects}
  \label{sec:spinintro}

In all calculations until now we ignored spin-dependent terms in the electronic Hamiltonian, and thus the triplet and singlet phase shifts were studied independently. However, the energy scale of the resulting PECs and vibrational spectrum is similar to the size of the fine and hyperfine structure. Thus, an accurate description of Rydberg molecules must include the relevant spin-dependent interactions. This section discusses each of these new interactions separately, focusing mostly on the fine structure of the $p$-wave scattering phase shifts, as these have either been ignored or treated only approximately in most calculations until recently \cite{EilesSpin}. The trilobite overlap approach used in the rest of the tutorial is not adopted here, as these spin-dependent interactions are simpler to describe in the Rydberg basis. 

  \subsection{Construction of the Hamiltonian}
  \label{sec:spinham}
\begin{figure}[t]
{\normalsize 
\begin{centering}
\begin{center}
\includegraphics[width=\columnwidth]{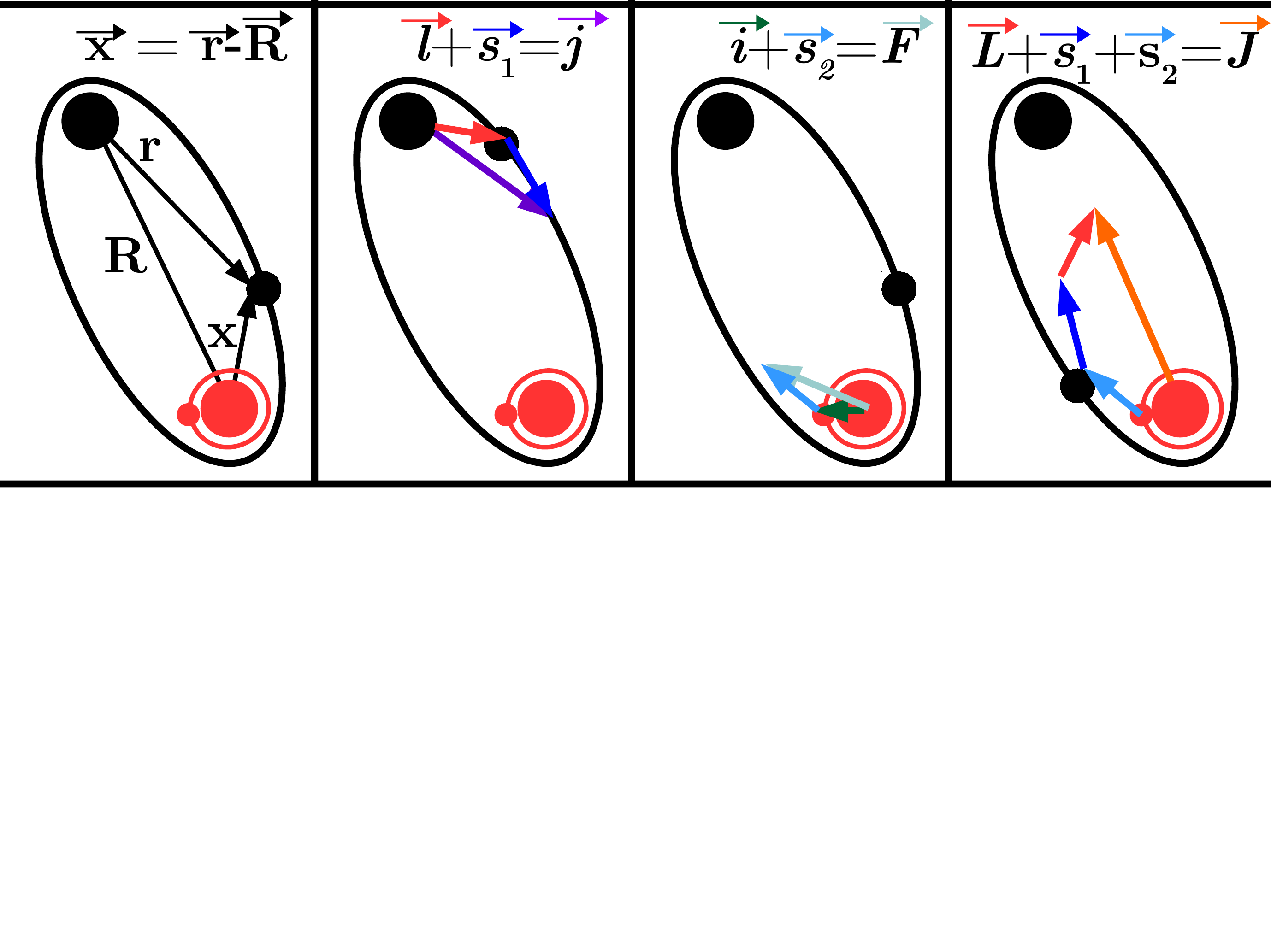}
\vspace{-110pt}
\end{center}
\end{centering}
}
\caption{The relevant coordinates and angular momenta. The internuclear axis lies parallel to the body-frame $z$ axis passing through the ionic core (black) and the perturber (red). The black (red) orbits represent the semiclassical orbits of the Rydberg (perturber's) electron. The Rydberg electron is located at $\vec r$ relative to the core and at $\vec X = \vec r - \vec R$ relative to the perturber.  The spin of the Rydberg electron, $\vec s_1$, couples to its orbital angular momentum relative to the core, $\vec l$, to give a total angular momentum $\vec j$. The spin of the perturber's outer electron, $\vec s_2$ (cyan) interacts with the perturber's nuclear spin, $\vec i$  to form $\vec F$. The electron-atom scattering potential depends on the total electronic spin, $\vec S = \vec s_1+ \vec s_2$, coupled to the orbital angular momentum $\vec L$ relative to the perturber to form total angular momentum $\vec J$.  }
\label{fig:cartoonPRA}
\end{figure}

The fine structure of the Rydberg atom, caused by the spin-orbit splitting, was discussed in Sec. \ref{sec:inter}. As depicted in the second panel of Fig. \ref{fig:cartoonPRA}, it can be added to our calculations by simply extending the Rydberg basis $\ket{nlm}$ to the spin-dependent basis $\ket{n(ls_1)jm_j}$. This almost doubles the number of low-$l$ potentials shown previously, but has very little effect on the trilobite and butterfly potentials due to the negligible fine structure of the high$-l$ states. 

The hyperfine structure of the perturber\footnote{The hyperfine splitting depends on the electron's wave function amplitude at the nucleus \cite{AllHF}. Since the Rydberg wave function has very little overlap with the ionic core the hyperfine splitting of Rydberg states decays rapidly with $n$ and can be safely ignored. \cite{AllHF}. Likewise, since the trilobite wave function has such a large overlap with the perturber, one might expect that this contributes an additional energy shift. We neglect this as well as the trilobite amplitude at the perturber is several orders of magnitude smaller than the amplitude of the perturber's valence electron, but this effect could merit further quantitative exploration. Finally, we ignore the dependence on the hyperfine state in the electron phase shifts. This is relevant near threshold where the electron's energy depends on the hyperfine state, and could therefore modify slightly the PECs near the classical turning point.}  couples the nuclear spin $i$ to the perturber's electronic spin $s_2$ (See Fig. \ref{fig:cartoonPRA}). This was first included in the theoretical treatment of Ref. \cite{AndersonPRA}, which showed that it mixes singlet and triplet symmetries. These states were measured in subsequent spectroscopy of $nD$ and $nP$ Rb states \cite{AndersonPRL,SingTripMix,MacLennan}. Measurements in Cs have also revealed this hyperfine-induced mixing \cite{Sass}.  One curious effect of  the hyperfine structure is its interplay with other energy splittings, particularly the fine structure. As the hyperfine splitting is $n$-independent while the fine structure splitting changes with $n$, degeneracies between molecular states can be engineered by changing principal quantum number. This was utilized in Ref. \cite{Niederprum} to induce spin-flips in the perturber, and in Ref. \cite{PfauRaithel} to excite trilobite molecules in Rb due to a ``spin-bridge'' when the hyperfine splitting becomes comparable to the splitting between the trilobite state and the $(n-3)S$ state. 

The hyperfine Hamiltonian is $ H_{HF}=A\vec i\cdot \vec s_2$, where the values of $A$ for several isotopes are given in Table \ref{tab:datatableHF}. Since this Hamiltonian commutes with the Rydberg Hamiltonian $H_0$ we choose an uncoupled basis  $\ket{n(ls_1)jm_j}\times\ket{s_2m_2;im_i}$, and use $\alpha = \{n,l,s_1,j,m_j\}$ to describe the Rydberg quantum numbers. The matrix elements of $ H_{HF}$ are
    \begin{align}
 & \bra{\alpha im_i,s_2m_2}A\vec I\cdot\vec S_2 \ket{\alpha' i'm_i',s_2'm_2'}\\&= \frac{A}{2}\delta_{\alpha\alpha'}\sum_{FM_F}C_{s_2m_2,im_i}^{FM_F} C_{s_2m_2',im_i'}^{FM_F}\nonumber\\&\times\left[(F(F+1)-i(i+1)-s_2(s_2+1)\right].\nonumber
  \end{align}

\noindent These terms, along with the Fermi pseudopotential, give the relativistic Hamiltonian 
\be
\label{eq:Hamiltonian}
 H(\vec r;\vec R) =  H_{Ryd}(\vec r) +  V^J_\text{fermi}(\vec R,\vec r) +  H_{HF}.
\ee
$ V_\text{fermi}^J(\vec R,\vec r)$ is the $J$-dependent Fermi pseudopotential which we derive now. The spin orbit coupling in the scattering interaction couples the orbital angular momentum $L$ to the total spin $S$ and results in split $^3P_J$ scattering states. Ref. \cite{EilesSpin} presents two complementary derivations of the spin-dependent pseudopotential. The first involves a recoupling of the tensorial operators in the Fermi pseudopotential using Wigner-Racah angular momenta algebra. The second, expanded here, reformulates the pseudopotential so that it is diagonal in the representation $\ket{(LS)J\Omega}$, just like the scattering Hamiltonian. Since the operator $\vec L^2$ does not commute with the Rydberg Hamiltonian, which has an operator $\vec l^2$ defining the angular momentum about the Rydberg core rather than about the perturber, we must develop a frame-transformation like that encountered in Sec. \ref{sec:inter}. While that one transformed between $LS$ and $jj$ coupling schemes, here we derive one to transform $l$ to $L$.

\begin{table}
\begin{center}
\begin{tabular}{| c | c | c |}
 \hline \hline
 Atom & $i$ & A (MHz)\\
    $^{6}$Li &$1$ & $152.137$\\
         $^7$Li & ${3}/{2}$ & $ 401.752 $\\
     $^{23}$Na&  ${3}/{2}$ & 885.813\\
      $^{39}$K&  ${3}/{2}$& 230.86\\
      $^{41}$K& ${3}/{2}$ & 127.01\\
      $^{85}$Rb & $5/{2}$ & 1011.9  \\
   $^{87}$Rb & ${3}/{2}$ & 3417\\
   $^{133}$Cs &${7}/{2}$ & 2298\\
   \hline
 \end{tabular}
 \end{center}
   \caption{Hyperfine constants $A$ and nuclear spin $i$ for the common isotopes of these alkali metals. \protect\cite{Beckmann1974,RbHF,AllHF,CsHF}.}
 
 \label{tab:datatableHF}
\end{table}

Our first goal is to write the Fermi pseudopotential so that it is diagonal in the proper representation, i.e. it has the form $\sum_J\ket{(LS)J\Omega}A_{LSJ}\bra{(LS)J\Omega}$, where $A_{LSJ}$ is a scattering length function depending on $L$, $S$, and now $J$. Thus far we have used a pseudopotential that leaves $S$ and $L$ uncoupled,
\begin{align}
\label{der1}
V_\text{fermi} &= \sum_{L'',S}\mathcal{A}(L''S,k) \cev\nabla {}^{L'' }\delta^3(\vec X)\cdot \vec\nabla^{L''}\\&\times\nonumber\sum_{M_{S}}\ket{SM_S}\bra{SM_S},
\end{align}
 where $\vec X = \vec r - \vec R$ and $\mathcal{A}(LS,k) = (2L+1)2\pi a_{LS}(k)$. We apply the projection operator $\sum_{LM_L}\ket{LM_L}\bra{LM_L}$ to both sides of Eq. \ref{der1}:
\begin{align}
V_\text{fermi}&=\sum_{L'',S,{M_{S}}}\sum_{\substack{LM_L\\L'M_L'}}\mathcal{A}(L''S,k)\ket{SM_S}\bra{SM_S}\\&\times \ket{LM_L}\bra{LM_L}\cev\nabla {}^{L'' }\delta^3(\vec X)\cdot \vec\nabla^{L''}\ket{L'M_L'}\bra{L'M_L'}\nonumber
\end{align}
We next perform the inner integration over angular coordinates $\hat X$ 
\begin{align}
\label{angint11}
&\bra{LM_L}\cev\nabla {}^{L'' }\delta(\vec X)\cdot \vec\nabla^{L''}\ket{L'M_L'}\\&= \int Y^*_{LM_L}(\hat X)\cev\nabla{}^{L''}\cdot \delta^3(\vec X)\vec\nabla^{L''}Y_{L'M_L'}(\hat X)d\hat X.\nonumber
\end{align}
Since this integration is over the solid angle $\hat X$, the final operator still retains a derivative with respect to $X$. With the benefit of foresight, we know that the basis states will eventually be written as a power series in $X$ (see Eq. \ref{firstorder}). The action of the derivative operator on this power series is
\begin{align}
\label{der2}
\partial_X^{L''}\left.\sum_La_LX^L\right|_{X=0} &= \left.\sum_L\frac{a_LL!X^{L-L''}}{(L-L'')!}\right|_{X = 0}\\
&= a_LL!\delta_{LL''}.\nonumber. 
\end{align}
Similarly, the conjugate of this acting on the bra will give a Kronecker delta $\delta_{L'L}$. This product of these two Kronecker deltas gives $\delta_{L'L''}$, i.e. the operator is diagonal in $L$. Thus, we integrate Eq. \ref{angint11} separately for $L=0$,
\begin{align}
 \int Y^*_{00}(\hat X) \delta^3(\vec X)Y_{00}(\hat X)d\hat X&=\frac{\delta(X)}{X^2}\left|Y_{00}(0,0)\right|^2,
\end{align}
and for $L=1$,
\begin{align}
 &\int Y^*_{1M_L}(\hat X)\cev\nabla\cdot \delta^3(\vec X)\vec\nabla Y_{1M_L'}(\hat X)d\hat X\\&=\frac{\delta(X)}{X^2}\Bigg(\partial_X'\partial_X Y_{1M_L}(0,0)Y_{1M_L'}(0,0) \nonumber\\&+\frac{1}{X^2}\frac{(2L+1)(L+1)L}{8\pi}\delta_{M_L,M_L'}\delta_{|M_L|,1} \Bigg).\nonumber
\end{align}
Here $\partial_X'\partial_X$ is the radially-dependent term of the dot product of the two gradient operators, where $\partial_X'$ acts to the left. For $L=1$, inspection of Eq. \ref{der2} shows that the effect of each of these derivatives is equivalent to multiplying by $X^{-1}$. We therefore replace $\partial_X'\partial_X\to X^{-2}$ factor to obtain a compact form for the  pseudopotential:
\begin{align}
V_\text{fermi}&=2\pi\sum_{LM_L,M_L'}\sum_{SM_S}\ket{LM_L,SM_S} \frac{(2L+1)^2}{4\pi}\\&\mathcal{A}(LS,k)\frac{\delta(X)}{X^{2(L+1)}}\delta_{M_L,M_L'}\bra{LM_L',SM_S}.
\end{align}

\begin{figure}[t]
\begin{center}
{\normalsize 
\begin{centering}
\includegraphics[scale =0.09]{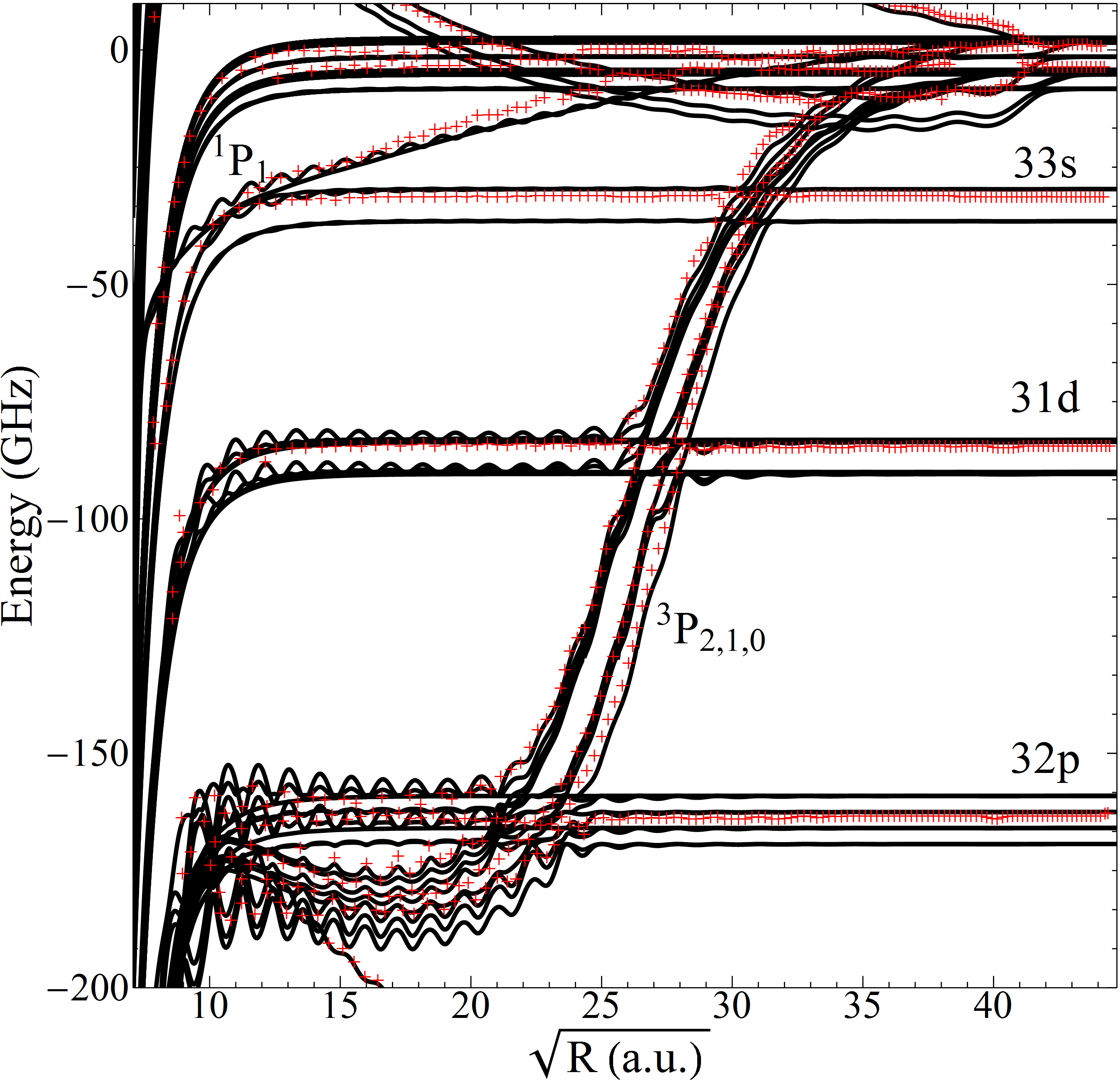}
\end{centering}
}
\end{center}
\caption{PECs of Rb$_2$, $\Omega = 1/2$. The results of Ref. \cite{KhuskivadzePRA} (red crosses) are overlaid in red. This figure is taken from Ref. \cite{EilesSpin}. }
\label{fig:RbSpinFig}
\end{figure}

\noindent We now couple $L$ and $S$, let the scattering length factor depend on $J$, and sum over $M_L'$, giving
\begin{align*}
V_\text{fermi}&= 2\pi\sum_{LM_L}\sum_{SM_S}\sum_{J\Omega,J'\Omega'}\frac{(2L+1)^2\delta(X)}{4\pi X^{2(L+1)}}\mathcal{A}(LSJ,k)\\\,\,\,\,&\times\ket{(LS)J\Omega} C_{LM_L,SM_S}^{J\Omega} C_{LM_L',SM_S}^{J'\Omega'}\bra{(LS)J'\Omega'}
\end{align*}
The sum over $M_L$ and $M_S$ gives
\be
\sum_{M_L,M_S}C_{LM_L,SM_S}^{J\Omega} C_{LM_L',SM_S}^{J'\Omega'}=\delta_{JJ'}\delta_{\Omega\Omega'},
\ee
which imposes the triangularity condition between $L$, $S$, and $J$. Finally, in terms of the collective quantum number for the scattering interaction, $\ket\beta=\ket{(LS)J\Omega}$, we have
\begin{align}
\label{finalprojectionform}
 V_\text{fermi}&=\sum_{\beta}\ket{\beta}\frac{(2L+1)^2}{2}\mathcal{A}(LSJ,k)\frac{\delta(X)}{X^{2(L+1)}} \bra{\beta}.
\end{align}
 The quantum numbers $\beta$ are incompatible with $\alpha$, which characterize the Rydberg eigenstates. We expand the Rydberg wave function of Eq. (\ref{eq:jdepefuncs}) to first order about the position of the perturber:
 \begin{align}
 \label{firstorder}
 &\psi_{nlm}(\vec r) \approx\left[\phi_{nlm}(\vec R) + \vec\nabla\left(\phi_{nlm}(\vec R)\right)\cdot\vec X\right].
 \end{align}
We define the following $Q$-functions,
   \begin{align}
 \label{eqn:Qdef}
  Q_{LM_L}^{nl}(R) &=\delta_{m,M_L} \left[\vec \nabla^L\left(\phi_{nlm}(R)\right)\right]_{M_L}^L.\\
\label{eqn:Qfuncs}
Q_{00}^{nl}(R)&= \frac{u_{nl}(R)}{R}\sqrt{\frac{2l+1}{4\pi}},\\
Q_{10}^{nl}(R) &= \sqrt{\frac{2l+1}{4\pi}}\partial_R\left(\frac{u_{nl}(R)}{R}\right),\\
Q_{1\pm 1}^{nl}(R)&=\frac{u_{nl}(R)}{R^2}\sqrt{\frac{(2l+1)(l+1)l}{8\pi}},l>0.
\end{align}  

\begin{figure*}[t]
\begin{center}
{\normalsize 
\begin{centering}
\vspace{-10pt}
\includegraphics[width=0.9\textwidth]{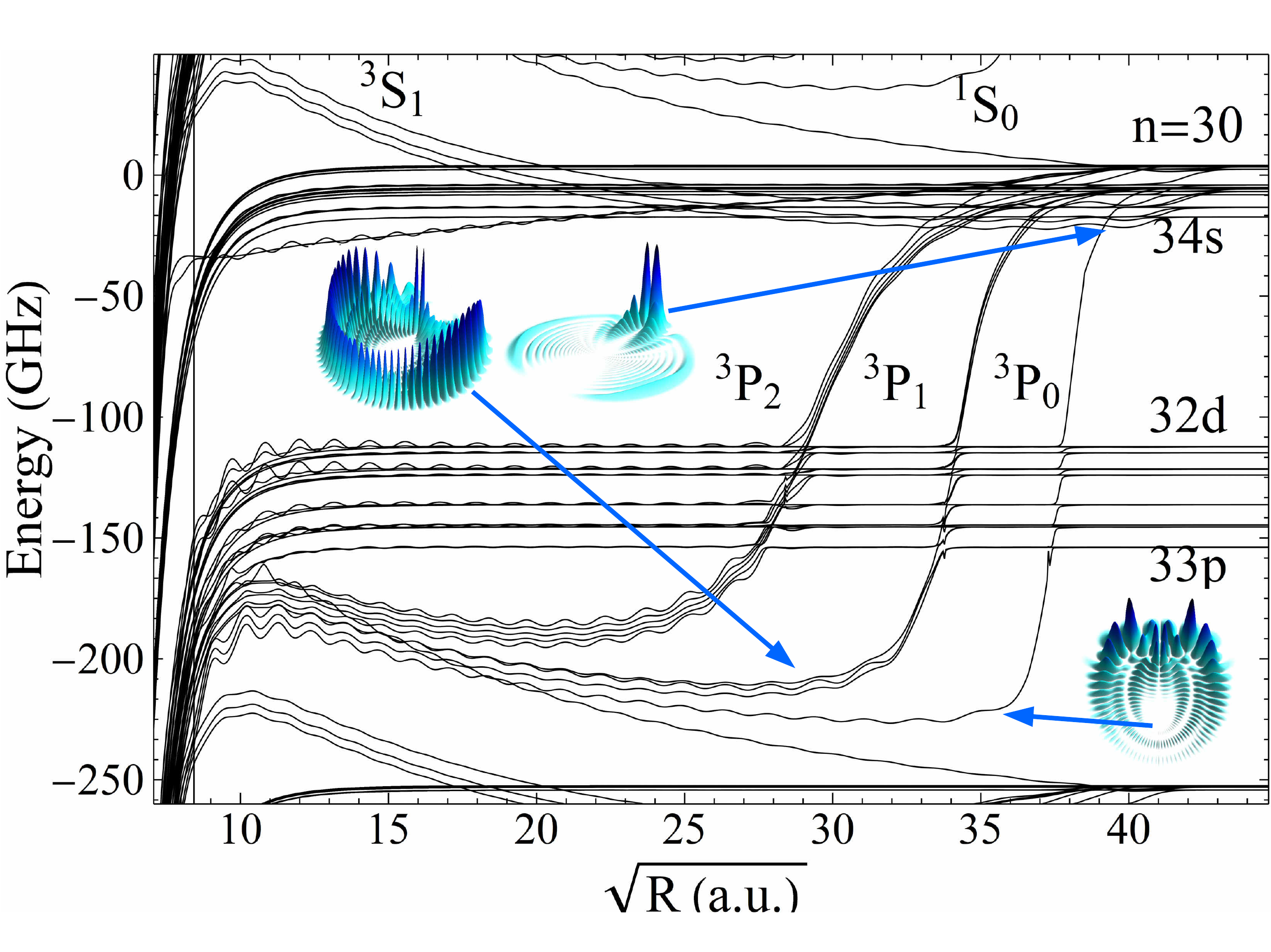}
\vspace{-20pt}
\end{centering}
}
\end{center}
\caption{PECs of Cs$_2$ for the projection $\Omega = 1/2$ are plotted in black. Several characteristic electronic wave functions are shown.  This figure is modified from Ref. \cite{EilesSpin}.   }
\label{fig:CsSpinFig}
\end{figure*}
  \noindent After using the spherical tensor representation of $\vec\nabla\phi_{nlm}(\vec R)$ given by the $Q$ functions and expressing $\vec X$ in terms of spherical harmonics $Y_{LM}(\hat X)$ centered at the perturber, it becomes clear that this expansion mediates the transformation from spherical harmonics relative to the Rydberg atom, $Y_{lm}(\hat r)$, to $S$ and $P$ partial waves relative to the perturber, $Y_{LM}(\hat X)$:
\begin{align}
  \label{taylorexpform}
  \psi_{nlm}(\vec r)&\approx \sum_{L=0}^1X^Lf_L Q_{LM_L}^{nlj}(R)Y_{LM_L}(\hat X),
  \end{align}
  where $f_L = \sqrt{\frac{4\pi}{(2L+1)}}$. 
We can now form matrix elements of Eq. \ref{finalprojectionform} in this representation, requiring only a trivial integration over the radial coordinate $X$. These are compactly expressed by first constructing the matrix representation of Eq. (\ref{finalprojectionform}) in the $\ket{\beta}$ basis
  \begin{align}
  \label{eqn:diagpotential}
U_{\beta,\beta'} &= \delta_{\beta,\beta'}\frac{(2L+1)^2} {2} a(SLJ,k).
  \end{align}
 The transformation of this diagonal matrix into one in the $\ket{\alpha s_2m_2}$ basis is mediated by a frame-transformation matrix $\mathcal{A}$, which transforms between the $\ket{\alpha s_2m_2}$ and $\ket{\beta}$ representations, analogous to what is done in multiple scattering theory \cite{DillDehmerJCP}. $\mathcal{A}$ is readily deduced from the prior steps of the derivation:
  \begin{align}
 \mathcal{A}_{\alpha s_2m_2,\beta}  &  =\sum_{M_L=-L}^{M_L=L}C_{LM_L,SM_S} ^{JM_J}f_{L} Q_{LM_L}^{nlj}(R) . \nonumber
  \end{align}
The final scattering matrix in the Rydberg basis consists of a block matrix
 \begin{align}
 \label{eqn:scattpotential}
 &V_{ii'}=\sum_{jj'}{\mathcal A_{ij}}{U_{jj}}{\mathcal A_{ji'}^\dagger}
 \end{align}
 for every $n$ and $l$. The mixing of $M_L,M_L'$ implied by Eq. \ref{eqn:scattpotential} is critical for an accurate physical description of this splitting, since the total spin vector $\vec{S}$ and total
orbital $\vec{L}$ precess during each
$P$-wave collision. This was recognized and incorporated in the Green's function calculation of Ref. \cite{KhuskivadzePRA}, but all subsequent work has neglected this detail. We expect that the much simpler description developed here will correct this oversight. This mixing of $M_L$ projections invalidates the use of $\Sigma$ and $\Pi$ symmetry labels to categorize the $^3P_J$ PECs. Incidentally, the Clebsch-Gordan coefficients vanish for $M_L = 0$ for the $^3P_1$ state, so that it remains a $\Pi$ state. Inclusion of the hyperfine interaction eliminates even this symmetry.

  \subsection{Spin-dependent potential energy curves and dipole moments}
To confirm the validity of this approach and to allow for a direct comparison with the spin-independent potentials of Fig. \ref{fig:Rbmanifold}, Rb$_2$ PECs are shown in Fig. \ref{fig:RbSpinFig}. The calculations of Ref. \cite{KhuskivadzePRA}, which ignored hyperfine and fine structure and thus cannot be compared directly, are overlayed. The main features of Ref. \cite{KhuskivadzePRA} are reproduced excellently, validating the accuracy of our $^3P_J$ pseudopotential.   The hyperfine structure adds significant complexity, increasing the multiplicity of the trilobite and butterfly states and splitting the low-$l$ states by several GHz.

 Fig. \ref{fig:CsSpinFig} shows PECs for Cs$_2$, which reveal the impact of this relativistic splitting in this molecule.  The positions of the $^3P_J$ shape resonances and their energy dependences strongly modify the butterfly potential wells. The $^3P_0$ resonance in cesium occurs at such a low electronic energy that the associated PECs cross the low-l states at very large internuclear distances, affecting the vibrational states to a greater degree than in Rb. The much larger $^3P_J$ splittings in Cs greatly spread the butterfly wells, limiting the density of avoided crossings. Some of these butterfly states are plotted in the insets. In the bottom right an unusual butterfly, situated in the deepest $^3P_0$ well at a large $\sim 1250a_0$, is shown. Its size and overall shape resemble a trilobite due to the similar bond length, but the specifics of its nodal character, particularly the node at the perturber, reveal its butterfly nature. Although in Cs$_2$ this state is challenging to excite since it contains very little low-$l$ character, Ref. \cite{EilesHetero} showed that in the \textit{heteronuclear} molecule NaCs the outer wells of this potential intersect the $nP$ Rydberg state and thus admix $P$ character into the butterfly. The top left inset shows a butterfly with mixed $\Pi$ and $\Sigma$ symmetry, and hence it has peaks of electron density both near the perturber and on the opposite side of the Rydberg core. The middle inset shows the mixed trilobite -$(n+4)S$ state, whose $S$ character is accessible experimentally\cite{TallantCS,BoothTrilobite}.  The improved description of the nearly-degenerate high-$l$ manifold with the very close $(n+4)s$ state given here lends a more complete theoretical description of this state that should encourage further exploration of the trilobite state in Cs.

\begin{figure}[t]
\begin{center}
{\normalsize 
\begin{centering}
\includegraphics[width=\columnwidth]{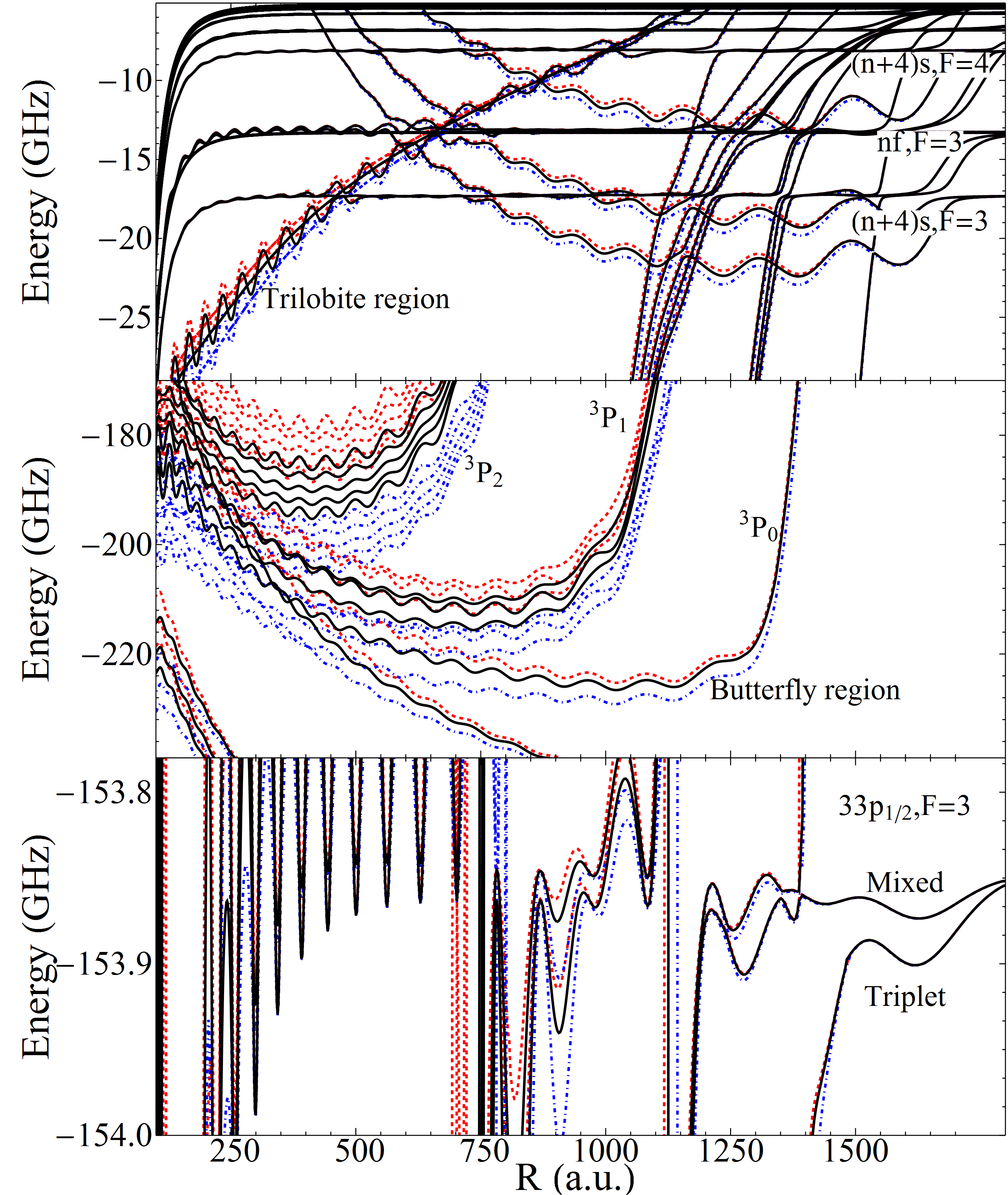}
\end{centering}
}
\end{center}
\caption{ PECs of Cs$_2$, $\Omega = 1/2$. The results using the $\{29,30,31\}$ basis (dot-dashed, blue), the $\{28,29,30,31\}$ basis (solid, black) and the $\{27,28,29,30,31\}$ basis (dashed,red) are plotted. Each panel displays a different regime, showing that at long-range the calculation is quite well converged with either basis, but the short-range butterfly curves in particular vary severely with the basis size.  This figure is taken from Ref. \cite{EilesSpin}.  }
\label{fig:convergencecomparison}
\end{figure}

Fig. \ref{fig:convergencecomparison} serves two purposes. First, it demonstrates the dependence of the PECs on basis size. Three different basis sets ($\{n_H-q,...,n_H,n_H+1\}$, with $q = 3,2,1$) are used. At large $R$ the inclusion of additional manifolds {\it below} the level of interest does not contribute to the non-convergent increase in well depth seen by \cite{Fey}, but at short range these additional manifolds have a strong effect on the potential wells, particularly the butterfly wells, repulsing them upwards. Setting $q=2$ agrees well with Ref. \cite{KhuskivadzePRA} and with the BK model. Nevertheless, the large variation in butterfly PECs with basis size reveals the convergence difficulties caused by the shape resonance. 

Second, Fig. \ref{fig:convergencecomparison} highlights some special features of this spin coupling. In the top panel, showing the trilobite potentials, we see that the hyperfine coupling leads to three trilobite potentials \cite{AndersonPRA}, which intersect and mix with the $(n+4)S$ state. In the middle panel we see that many butterfly potentials are associated with each scattering phase shift, depending on the other spin quantum numbers. It has recently been argued that the multiplicity of such potential curves leads to singlet, doublet, and triplet vibrational lines observed in Rb \cite{SpinFey}. Finally, the bottom panel shows how the hyperfine splitting mixes the singlet and triplet potential curves of Fig. \ref{fig:Rbmanifold}, resulting in one deep pure triplet state and a shallow mixed state.

\begin{figure}[b]
{\normalsize 
\begin{center}
\includegraphics[width=0.45\textwidth]{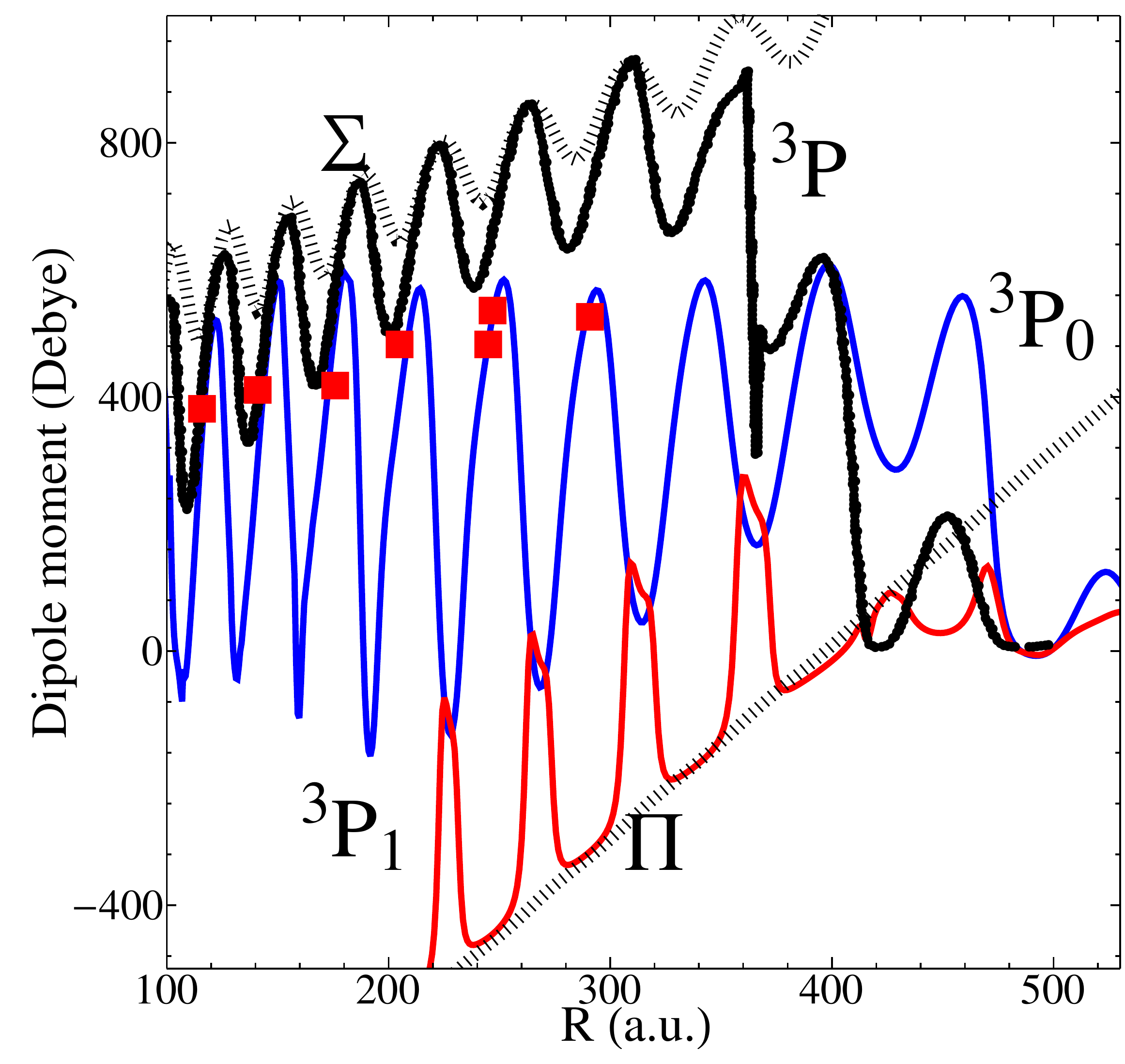}
\end{center}
}
\vspace{-20pt}
\caption{ Analytic permanent electric dipole moments (black, dashed), permanent electric dipole moments ignoring the $^3P_J$ splitting (black, solid, labeled $^3P$), and permanent electric dipole moments from the full spin model for electronic states dominated by $^3P_0$ scattering (blue,solid), and $^3P_1$ scattering (red,solid), are plotted. The red squares are placed at the observed bond lengths and permanent electric dipole moments \cite{Butterfly}.  The $^3P_0$ and $^3P_1$ permanent electric dipole moments correspond to states of mixed $M_L$, although the mixing is quite weak for $^3P_1$ scattering and the analytic and exact results agree more closely.  The $^3P_2$ case is not shown, for simplicity.   This figure is taken from Ref. \cite{EilesSpin}. }
\label{fig:dipoles}
\end{figure}

We conclude by showing an experimental signature of this fine structure in the butterfly molecule dipole moments, shown as a function of bond length in Fig. \ref{fig:dipoles}. The dashed black lines are the predictions of Eq.  \ref{eq:dipolehydrogen}, which assume no coupling between butterfly states and zero quantum defects. The solid black line neglects the $^3P_J$ splitting. The red points are measured  \cite{Butterfly}. The blue and red curves are calculated from the $^3P_0$ and $^3P_1$ electronic eigenstates. The $^3P_0$ dipole moments are weaker than the $J$-independent ones since the spin orbit splitting mixes $M_L$. The $M_L = 0$ butterfly molecules focus the electronic wave function near the perturber, while $|M_L| = 1$ states maximize the wave function closer to the Rydberg core; this is reflected in their dipole moments (positive for $M_L=0$, negative for $|M_L| = 1$), and can be seen in the exemplary wave functions sketched in  Fig. \ref{fig:fieldbutterflies}.   Quantitative agreement is seen between the experimental values and the $J$-dependent calculation, which are both systematically smaller (by $\sim$25\%) than the $J$-independent calculation. This is evidence that even though the relatively small e-Rb $^3P_J$ scattering splittings do not dramatically shift the PECs, these splittings do have significant impact on observables such as the dipole moments.

\section{Interactions with external fields}
\label{sec:fieldstudies}
The past three sections described the properties of LRRMs in increasing detail.  Throughout a major theme was how these theoretical descriptions were paired with experimental developments, resulting in a fruitful cooperation between these twin pillars of physics which led to the observation and characterization of these molecules.  We now turn to a more practical matter which was one of the initial motivations for studying these molecules: their large permanent dipole moments, which make them remarkably sensitive to external control. We focus first on weak fields which primarily address the rotational structure of the molecules; this discussion follows Refs. \cite{Butterfly,EilesPendular} and demonstrates the pendular nature of the rotational spectrum of these molecules\footnote{Dipolar molecules librate around the field axis analogously to a pendulum, and have an evenly spaced level structure like the quantum harmonic oscillator \cite{Friedrich}.}. Following this we study how such pendular LRRMs interact \cite{EilesPendular}.  Finally, we develop and present a semi-perturbative treatment of external fields which are strong enough to shape the vibrational and electronic structure of the LRRMs, and apply this to some exemplary trimer configurations. These final calculations are presented here for the first time.  Note that in the following we neglect the spin terms used in the previous section. 
\subsection{Pendular spectrum in a weak electric field}

We first consider an external electric field $F<1$V/cm which is too weak to modify the electronic structure of the Rydberg molecule. The field-free PECs discussed previously are therefore still applicable, as are the vibrational states which only couple to the external field via the electronic PECs. The field-free rotational spectrum is set by the rotational constant $B_\nu = \frac{1}{m\langle R^2\rangle}$ in the rigid rotor approximation, where $m$ is the molecule's reduced mass and $\langle R^2\rangle$ is the vibrationally-averaged squared bond length. Due to the large size of LRRMs this constant is usually only some hundreds of kHz, smaller than the energy shift of the dipole-field coupling $-\vec{d}\cdot \vec{F}$, where $\vec{d}$ is the dipole moment. We therefore want to determine the effect of this field on the rotational spectrum. In the absence of external fields, polar molecules rotate freely with random orientations. Setting the quantization axis parallel to the electric field, the rotational Hamiltonian in an electric field is 
\be
\label{eq:rotationalham}
H_\text{rot}=B_\nu\hat{N}^2-dF\cos\theta.
\ee
The rotational angular momentum operator is $\hat{N}$. We define a dimensionless parameter $\omega=\frac{dF}{N_\nu}$. The large dipole moments and small rotational constants of LRRMs conspire together to make this parameter very large,  $\omega\sim 10^2-10^3$, about four orders of magnitude larger than in typical heteronuclear molecules at the same field strength ~\cite{KRb}. Trilobite or butterfly molecules are therefore ideal candidates to realize \text{pendular molecules}, since in this high $\omega$ limit the eigenstates of Eq. \ref{eq:rotationalham} resemble harmonic oscillator states, which librate like a pendulum about the electric field axis ~\cite{Rost,Friedrich}. These pendular states are obtained by diagonalizing $H_\text{rot}$ in the basis of spherical harmonics $Y_{NM_{N}}(\theta,\phi)$, $N=0,1,...$. $H_\text{rot}$ is diagonal in $M_N$ since the quantization axis is  parallel to the electric field.  The eigenstates $|\tilde NM_N\rangle$ are thus characterized by $M_N$ and their librational quantum number, $\tilde N$.

\begin{figure}[b]
{\normalsize 
\begin{center}
\includegraphics[width=\columnwidth]{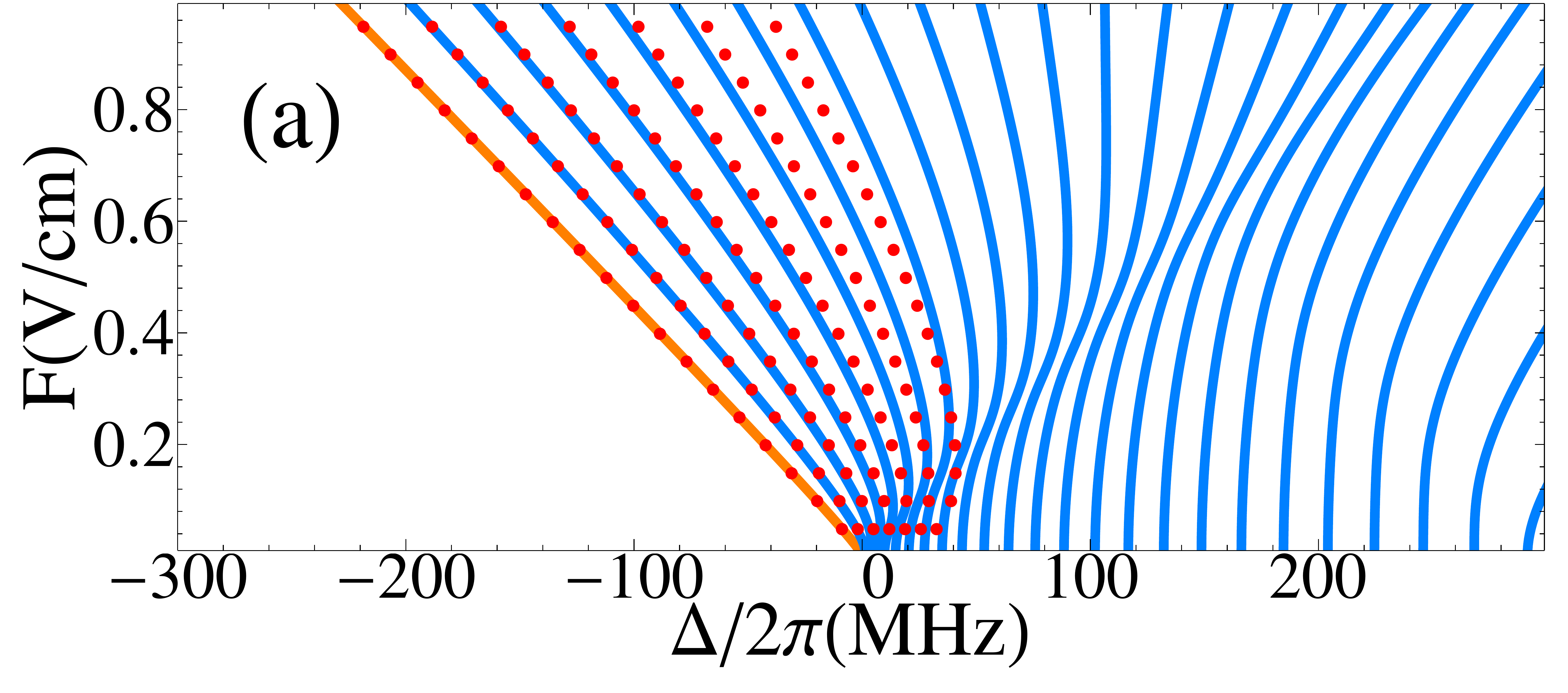}
\includegraphics[width=\columnwidth]{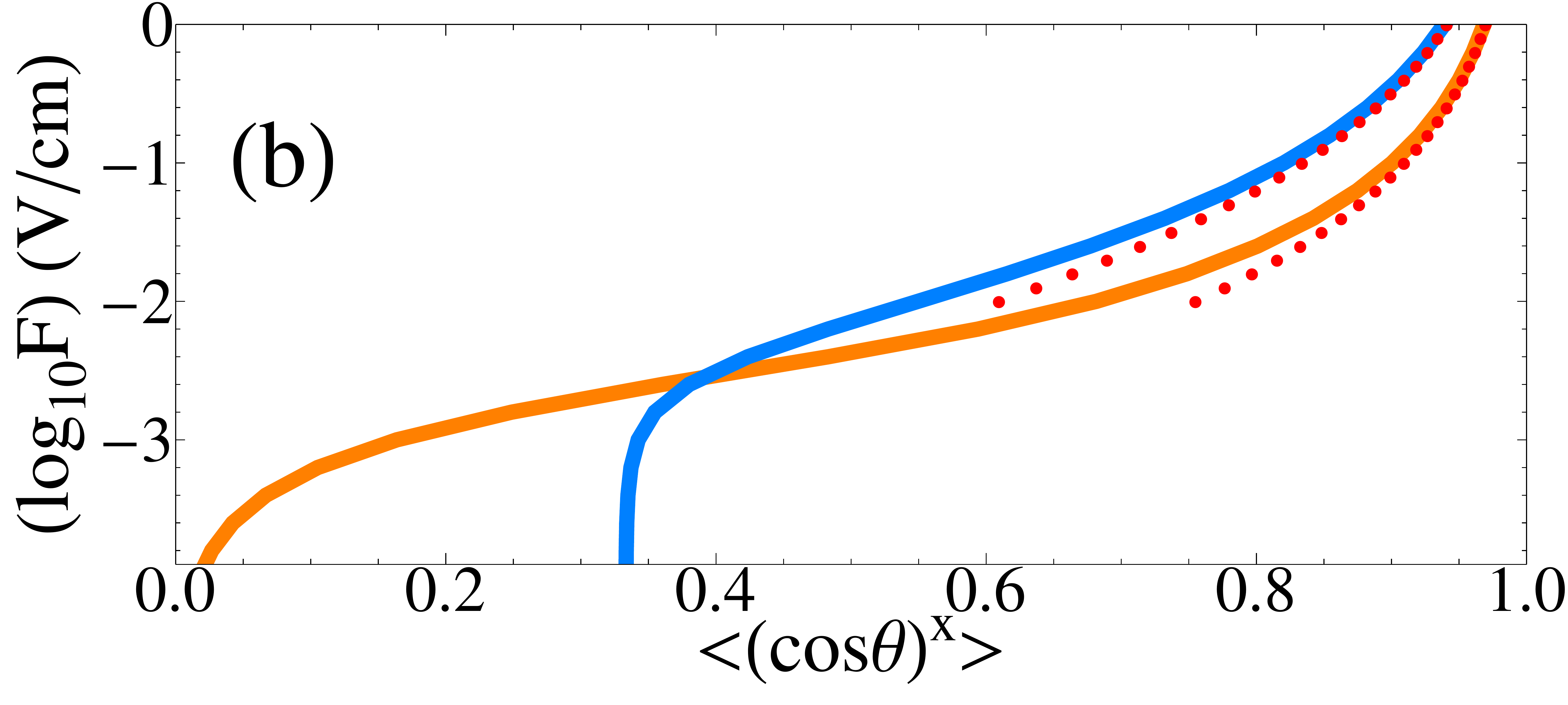}
\end{center}
}
\caption{(a) Stark spectrum of the $n=24$ pendular states, showing the energy shift $\Delta$ as a function of the applied electric field for $M_{N}=0$. The red points correspond to the two-dimensional harmonic oscillator approximation. (b) Orientation $(x=1)$ (orange) and alignment $(x=2)$ (blue) of the lowest pendular state [orange line in (a)]. This figure is taken from Ref. \cite{EilesPendular}. 
 }
\label{fig:pendular}
\end{figure}

 Fig. \ref{fig:pendular}a shows the resulting Stark spectrum for $n=24$ butterfly molecules. At zero field, the typical $N(N+1)$ spacing of rotational states can be seen; this adiabatically switches into the equally-spaced energy levels of a harmonic oscillator. The ground state,  $\Psi_{00}(\theta,\phi)$ highlighted in orange, is the most aligned.  Fig. \ref{fig:pendular}b shows the orientation $\langle \cos{\theta} \rangle$ and alignment $\langle \cos^2{\theta} \rangle$ of this state. The almost perfect alignment at $F=1$V/cm is unmatched in previous efforts with traditional molecules. 

\begin{figure}[t]
{\normalsize 
\begin{center}
\includegraphics[scale =0.12]{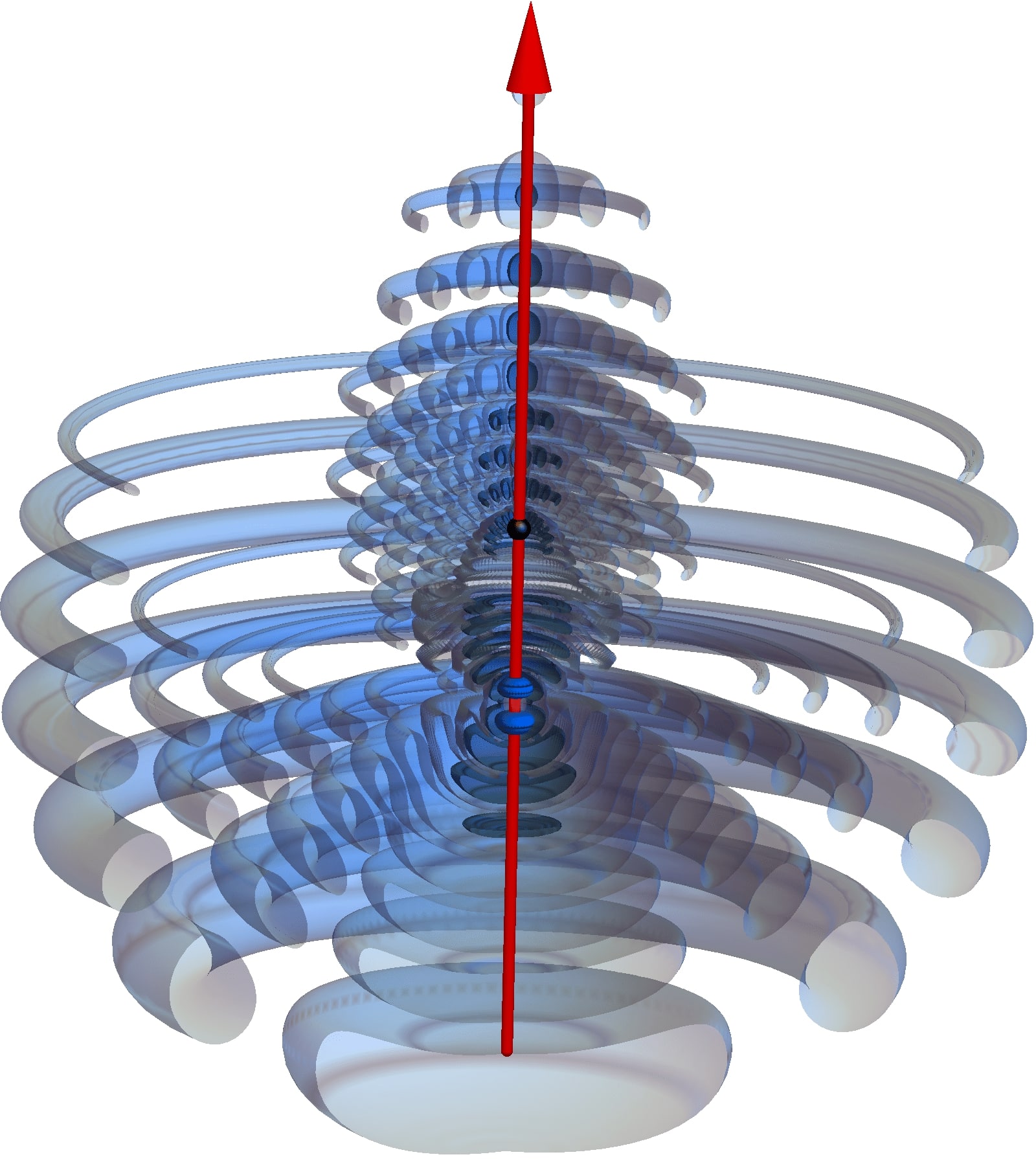}
{\vspace{-0pt}\scalebox{1}[-1]{\includegraphics[scale =0.12]{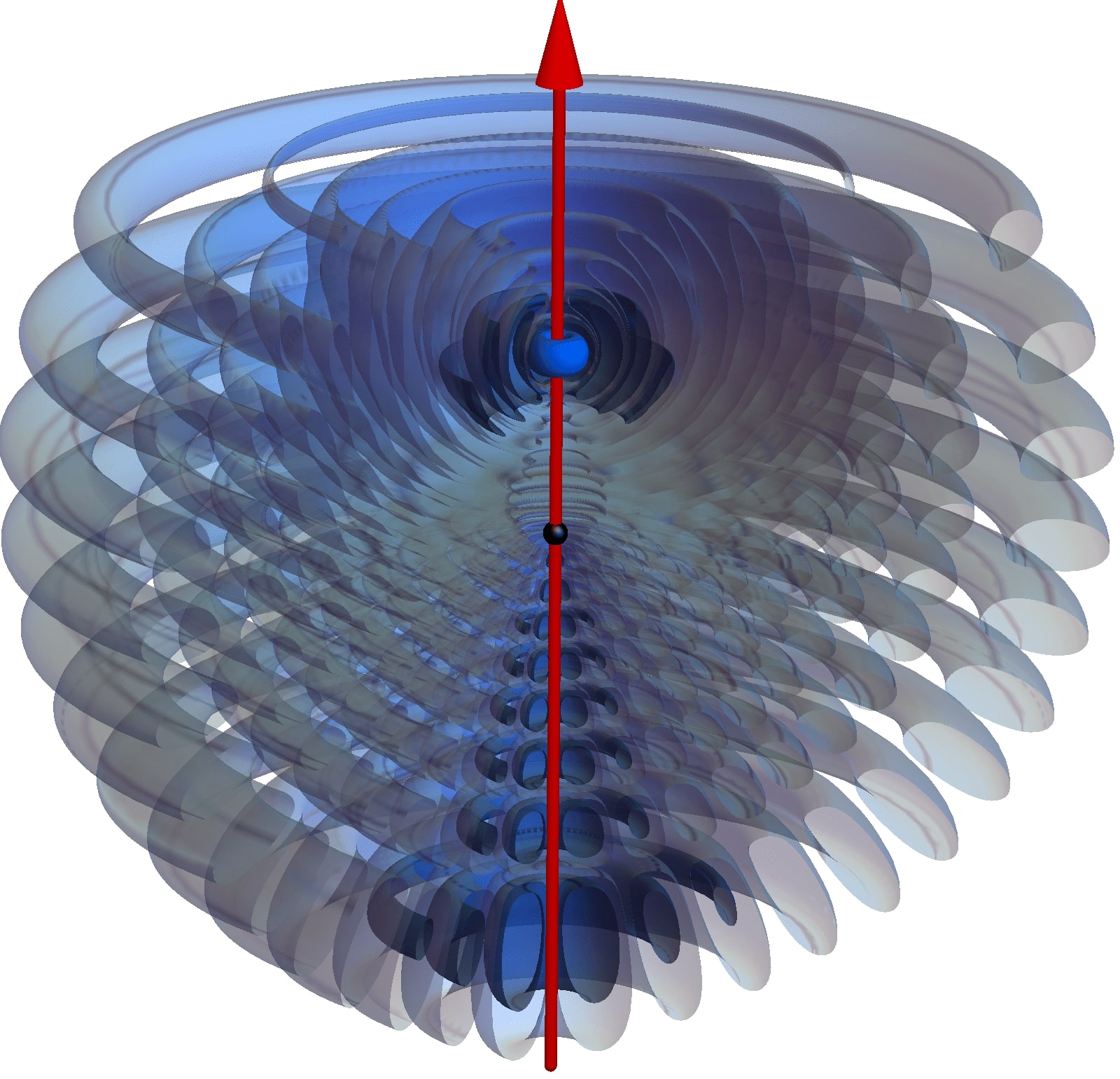}}}
\end{center}
}
\caption{Top) A $\Sigma$ butterfly molecule shown using isosurfaces as in Fig. \ref{fig:trilobitebasis}. The small black sphere represents the Rydberg core; the perturber is placed in between the small blue lobes below it. The red arrow denotes the direction of the dipole moment. Bottom) A $\Pi$ butterfly molecule, plotted in the same way. The electron's wave function is now localized opposite the perturber, causing the dipole moment to point in the opposite direction as above.   }
\label{fig:fieldbutterflies}
\end{figure}

Greater  insight into the character of these pendular states is given in the limit $\omega\to\infty$. Using the explicit form for $\hat N^2$ in spherical coordinates, the Schr\"{o}dinger equation defined by $H_\text{rot}$ is
\begin{align}
&0=\\&\left(\frac{\partial^2}{\partial\theta^2} +\cot\theta\frac{\partial}{\partial\theta} + \frac{1}{\sin^2\theta}\frac{\partial^2}{\partial\phi^2}+\omega\cos\theta + W\right)\Psi(\theta,\phi),\nonumber
\end{align}
where $W = E/B_\nu$. This equation maps onto the 2D harmonic oscillator by setting $\xi = 2\alpha\tan(\theta/2)$, where $\alpha = \sqrt{\omega/2}$. A separable solution in $\xi$ and $\phi$ is then obtained, where $\Psi(\xi,\phi) = U(\xi)\frac{1}{\sqrt{2\pi}}e^{i|M_N|\phi}$. $U(\xi)$ is then given by:
\begin{align}
0&=  \left(1 + \frac{\xi^2}{4\alpha}\right)^2\left[\frac{d^2}{d\xi^2} + \frac{1}{\xi}\frac{d}{d\xi} - \frac{M_N^2}{\xi^2}\right] U(\xi)\\&\nonumber+\frac{WU(\xi)}{\alpha}+ \frac{\omega}{\alpha}\frac{4 - \xi^2/\alpha}{4 + \xi^2/\alpha}U(\xi).
\end{align}
Since $\alpha\gg 1$ in the pendular regime we discard all terms of order $1/\alpha$ to obtain the Schr\"{o}dinger equation of a harmonic oscillator
\begin{equation}
\left[\left(\frac{d^2}{d\xi^2} + \frac{1}{\xi}\frac{d}{d\xi} - \frac{M_N^2}{\xi^2}\right) + \beta - \xi^2\right]U(\xi) = 0,
\end{equation}
where $\alpha\cdot\beta = W + \omega$. The energies of the pendular states are
\begin{equation}
E = B_\nu(\sqrt{2\omega}(2\tilde N + |M_N| + 1)-\omega),
\end{equation}
 and the pendular states are
\begin{equation}
\Psi_{\tilde N,M_N}(\xi)=(-1)^{M_N}\mathcal{N}e^{-\frac{\xi^2}{2}}\xi^{M_N}L_{\tilde N}^{M_N}(\xi^2)e^{i|M_N|\phi},
\end{equation}
where $L_{N}^M(x)$ is a Laguerre polynomial and $\mathcal{N}=\sqrt{\frac{\tilde N!}{\pi\Gamma(\tilde N+M_N+1)}}$. The red points in Fig. \ref{fig:pendular} were computed using this approximation. This Stark map has been experimentally mapped out in butterfly molecules in Ref. \cite{Butterfly}, providing clear evidence of the polar nature of these molecules. The theory discussed in this section allows for the extraction of dipole moments as well as bond lengths from such measurements, which was how the dipole moments presented in Fig. \ref{fig:dipoles} were measured \cite{Butterfly}. In Fig. \ref{fig:fieldbutterflies} the $r$ and $\theta$-type butterfly molecules pendular molecules are shown using the same plotting procedure as in Fig. \ref{fig:trilobitebasis}. This figure illustrates better than Fig. \ref{fig:trilobitebasis} the reason for the ``butterfly'' moniker since the shorter internuclear distance exaggerates the ``wings''. Another interesting facet of these molecules is that, due to the extended electron cloud beyond the perturber, the dipole moments can actually exceed $R$, the classical limit \cite{Butterfly}.

\subsection{Intermolecular interactions}
The large multipole moments of LRRMs also imply that they can interact at long range. This is very similar to the interactions between isolated Rydberg atoms giving rise to the famous Rydberg blockade, which prevents two Rydberg atoms from both being excited if their internuclear separation is less than the blockade radius. This is the distance at which their mutual interaction exceeds the linewidth of the laser, and hence shifts the state with two excitations out of resonance \cite{DipoleBlockade}. At larger internuclear separations the Rydberg-Rydberg interaction can form attractive potential wells at long range which support the Rydberg macrodimer states \cite{Boisseau,DeiglmayrRydRyd}.

\begin{figure}[t]
{\normalsize 
\begin{center}
\includegraphics[width=\columnwidth]{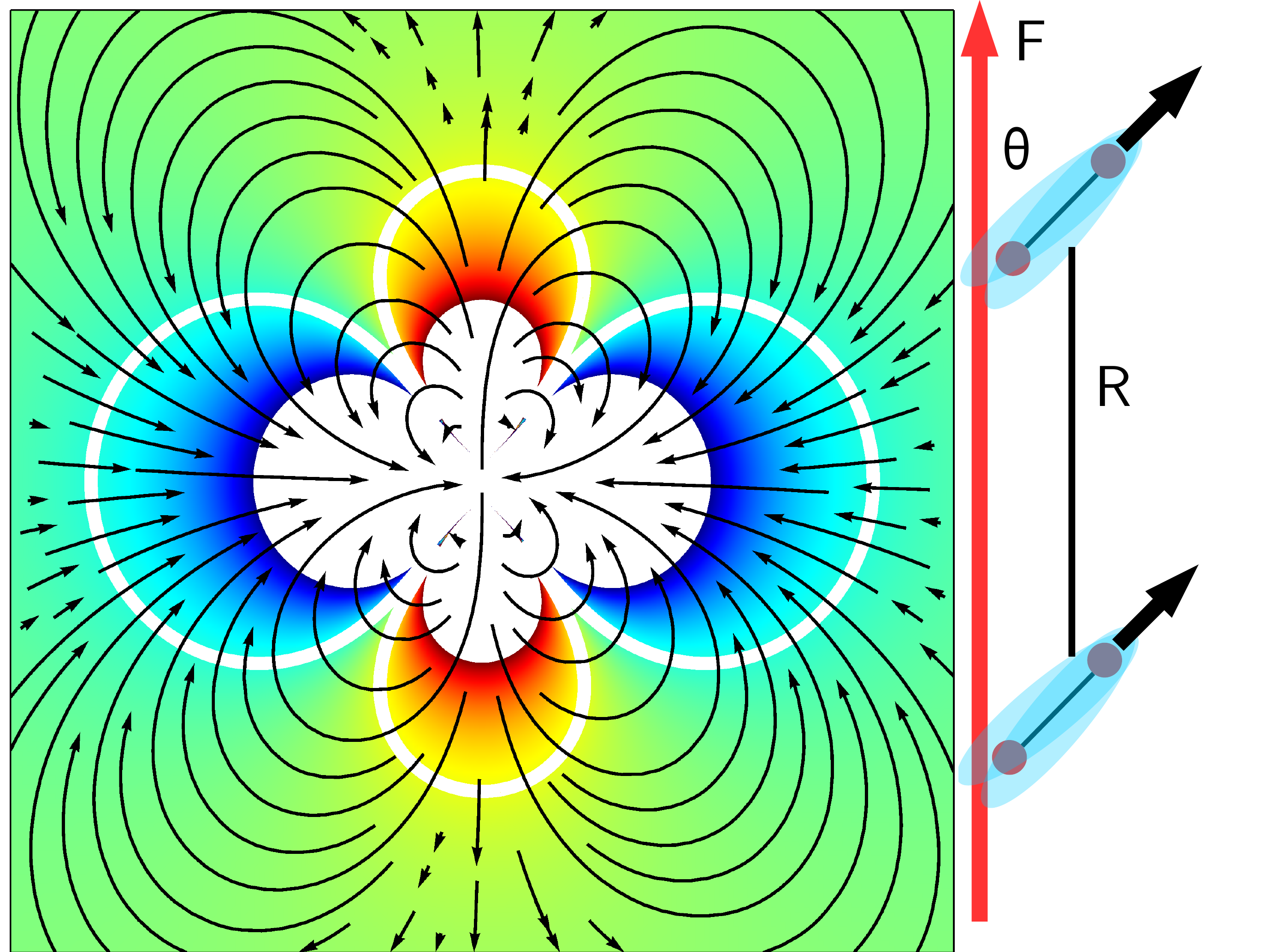}
\end{center}
}
\caption{ The interaction potential $V(x,y)$ between two aligned pendular butterfly states with $n=24$. Blue(red) regions are attractive(repulsive). The inner white region (outer white line) demark contours at $|V(x,y)|= 0.5(0.1)$MHz. Lines of force are overlayed in black arrows. The length of a side of the figure is $2\times 10^5$ a.u. On the right the basic geometry is sketched, showing the electric field direction, the two butterfly molecules, and defining the intermolecular distance $R$ and interaction angle $\theta$. This figure is modified from Ref. \cite{EilesPendular}.   }
\label{fig:potentials}
\end{figure}

These same effects also apply in LRRMs. Whereas in atoms the interactions are typically isotropic van der Waals potentials \footnote{It is possible in special circumstances to realize anisotropic dipole-dipole interactions with Rydberg atoms, such as in the presence of an external field \cite{Marcassa} or near a F\"{o}rster resonance \cite{SaffmanBlockade}.}, in LRRMs these permanent dipole moments are present automatically due to the molecular structure. Their internal molecular structure also provides  vibrational and rotational degrees of freedom that can be accessed to tune the interactions. It is simplest and most experimentally interesting to explore these interactions between the pendular molecules discussed above, since the direction of the applied electric field provides an easily accessible experimental knob. 

The potential between two pendular molecules, aligned by an electric field at an angle $\theta$ relative to the axis connecting their center of masses, is given by the two-center multipolar expansion of the Coulomb force \cite{MargenauInt,Stone,vanderAvoird1980}, 
\begin{align}
\label{eqn:interaction}
\hat V&=\sum_{L_A,L_B}q(L_A)q(L_B)\frac{f_{L_A,L_B}^{n}}{R^{L+1}}\sum_{m_A,m_B,m}\begin{pmatrix}L_A & L_B &L\\m_A & m_B & m\end{pmatrix}\nonumber\\ &\times D_{m_A0}^{L_A}(\theta_A,\phi_A)^*D_{m_B0}^{L_B}(\theta_B,\phi_B)^*D_{m0}^{L}(\theta,0)^*.
\end{align}
Here $q(L)$ is the molecular multipole moment of order $L$ averaged over the vibrational molecular wave function and rescaled by the principle quantum number $n$, $q(L)=Q_0^{L}/n^{2L}$, and $L = L_A+L_B$. Several Wigner D-matrices are used: $D_{m_X0}^{L_X}(\theta_X,\phi_X)^*$ rotates the multipole operator between the lab and molecule frame, and $D_{m0}^L(\theta,0)$ rotates between the lab (field) axis and the intermolecular axis. We also have defined
\begin{align}
f_{L_A,L_B}^{n}&= (-1)^{L_A}n^{2L}\left[\frac{(2L+1)!}{(2L_A)!(2L_B)!}\right]^{1/2}.
 \end{align} 
 Eq. \ref{eqn:interaction}  is valid provided the electron clouds do not overlap \cite{LeRoy}. This same potential  diagonalized within a basis of Rydberg states to compute the Born-Oppenheimer PECs for Rydberg macrodimers \cite{Boisseau,Farooqi, CsReview,DeiglmayrRydRyd,RaithelRydRyd}. In our case this basis becomes prohibitively large due to the additional molecular configurations in the basis. However, we expect the dominant term to be that of the dipole-dipole interaction which is obtained immediately in first-order perturbation theory. We thus proceed perturbatively, using the pendular ground state calculated previously for the molecular state (see the diagram in Fig. \ref{fig:potentials} for the geometry).  We include terms up to $L=4$ in first order perturbation theory and up to $L= 2$ in second order perturbation theory. This gives the potential as a power series in $1/R$ accurate to order $R^{-6}$. The van der Waals term is proportional to $n^{11}/R^6$, familiar from Rydberg scaling laws, and contains induction $(C_{6i})$ and dispersion $(C_{6d})$ contributions \cite{vanderAvoird1980}. In total, the intermolecular potential is
\begin{align}
\label{eqn:interactionresults}
&V(R,\theta)= -\frac{2C_3d^2n^4}{R^3}P_2(x)-\frac{8n^8}{R^5}P_4(x)\left(C_{5a}do-C_{5b}q^2\right) \nonumber\\&-2\frac{4d^4n^{11}}{R^6}\left(C_{6i}^a[P_2(x)]^2+C_{6i}^b\frac{(xy)^2}{4}\right)\\
&-\frac{4d^4n^{11}}{R^6}\left(C_{6d}^a[P_2(x)]^2+\frac{C_{6d}^c}{4}y^4+C_{6d}^b\frac{(xy)^2}{4}\right),\nonumber
\end{align}
where $x=\cos\theta,y=\sin\theta$, and where all coefficients $C$ are positive  \cite{EilesPendular}.  Using the  large $\omega$ harmonic oscillator approximation applicable for the pendular states the coefficients $C_3$, $C_{5a}$, and $C_{5b}$ are
\begin{align}
C_3&=  1 + \frac{3}{2\omega}  - \sqrt{\frac{2}{\omega}}\\C_{5a} &= 1 + \frac{14}{\omega}-\frac{7}{\sqrt{2\omega}}\\C_{5b} &=\frac{3}{8\omega}(21 - 6\sqrt{2\omega} + 2\omega).
\end{align}  
The coefficients of the second-order terms are complicated and can be found in Ref. \cite{EilesPendular}.  This potential surface is displayed for $n=24$ in Fig. \ref{fig:potentials}. It has the characteristic anisotropic shape of the dipole-dipole interaction, and even for the relatively low $n=24$ has 100kHz size shifts at distances exceeding $10\mu$m. At this scale the presence of higher-order potentials is only seen for $\theta$ near the ``magic angle'' where $P_2(x_\text{magic})=0$. In Ref. \cite{EilesPendular} this anisotropy was used to propose an anisotropic blockade effect between LRRMs that could be used to create a crystalline formation of LRRMs with tunable separations.

\subsection{Influence of fields on electronic potential energy curves}
\label{sec:strongfields}
We now consider electric and magnetic fields which can modify the electronic structure of the Rydberg atom, and hence change the PECs. These fields break the cylindrical symmetry present in dimers and add a second preferred direction to the system; the resulting potential energy surfaces depend on both $R$ and $\theta$, the internuclear distance and its angle relative to the field axis, respectively. This turns a rotational degree of freedom into a vibrational one and makes it possible for the fields to align LRRMs. This has been observed in Rb $nD$ states in a magnetic field Ref. \cite{PfauKurz} Ref. \cite{KurzSchmelcherPRA} investigated these potential energy surfaces and alignment in an electric field, while Refs. \cite{Lesanovsky,KurzSchmelcherJPB,Hummel,Hummel2} concentrate on magnetic fields effects. These studies showed that the strong coupling to external fields can prove destructive to LRRMs, as the Zeeman splitting can eliminate the trilobite states \cite{Lesanovsky}. By applying both an electric and a magnetic field new control possibilities open up \cite{KurzSchmelcherJPB}, making it possible to control the hybridization of Rydberg orbitals \cite{GajKrupp}. Ref. \cite{Rosario2016} presents an exploratory investigation into the effect of an electric field on trimer molecules.

In the following we approach this problem by extending the trilobite overlap formalism to include the effect of external fields. Since this approach hinged on the degeneracy of high-$l$ Rydberg states, external fields are clearly problematic since they destroy this atomic degeneracy. The Zeeman Hamiltonian splits equally all of the Rydberg $m_l$ levels, while the Stark Hamiltonian adds a linear shift to each high-$l$ state.  Nevertheless, we find that it this method remains surprisingly accurate so long as the field influences remain largely perturbative, i.e. we can use the eigenstates of the Fermi pseudopotential as the zeroth-order states to calculate the energy shifts of the external fields.

We first write down the field operators, focussing on the simple scenario of a single applied field\footnote{The formalism we develop is general to any form of the field operators, but this offers an illustrative case.}. We set the quantization axis parallel to the field, and obtain
\be
H_B = \frac{B}{2}\hat L_z + \frac{B^2}{8}r^2\sin^2\theta
\ee
for the magnetic field and 
\be
H_F = Fr\cos\theta.
\ee
for the electric field. In the generic case with $N$ perturbers the Hamiltonian is $H = H_N({\vec r};\{\vec R_i\})  + H_B + H_F$; the matrix elements of $H_N$ are computed as in Sec. \ref{sec:polyintro} and the new matrix elements are written below. The matrix elements of the linear Zeeman term are straightforward to calculate:
\begin{align}
\bra{\+{\Upsilon_{p r,n}^{\pmb\alpha\pmb 1}}}\frac{B}{2}\hat L_z\ket{\+{\Upsilon_{qr,n'}^{\pmb\beta \pmb1}}}&=\frac{B\delta_{nn'}}{2}\sum_{lm}m\phi_{nlm}^\alpha(\vec R_p)\phi_{nlm}^\beta(\vec R_q)\\
\bra{\+{\Upsilon_{p r,n}^{\pmb\alpha \pmb1}}}\frac{B}{2}\hat L_z\ket{\phi_{n'l'm'}^1}&=\frac{B\delta_{nn'}}{2}\phi^\alpha_{nlm}(\vec R_p)m\\
\bra{\phi_{nlm}^1}\frac{B}{2}\hat L_z\ket{\phi_{n'l'm'}^1}&=\frac{B}{2}\delta_{nn'}\delta_{ll'}\delta_{mm'}m.
\end{align}
The spatial dependence of the Stark operator and the quadratic $B$-field operator complicates the computation of their matrix elements. Since these operators separate in spherical coordinates we can write them in general as $H_\text{field} = a(r)b(\theta)$. Their matrix elements, $a_{nl,n'l'}$ and $b_{lm,l'm'}$ are found in standard references \cite{radmatel,Varsh}, and so
\begin{align}
&\bra{\+{\Upsilon_{pr,n}^{\pmb\alpha 1}}} H_\text{field}\ket{\+{\Upsilon_{qr,n'}^{\pmb\beta 1}} }\\&=\sum_{lm}\sum_{l'm'}\left[\phi_{n'l'm'}^\beta(\vec R_q)\right]^* a_{nl,n'l'}b_{lm,l'm'}\phi_{nlm}^\alpha(\vec R_p)\nonumber\\
&\bra{\+{\Upsilon_{p r,n}^{\pmb\alpha \pmb1}}}H_\text{field}\ket{\phi_{n'l'm'}^1}\\&=\sum_{lm}\left[\phi_{n'l'm'}^\beta(\vec R_q)\right]^* a_{nl,n'l'}b_{lm,l'm'}\nonumber\\
&\bra{\phi_{nlm}^1}H_\text{field}\ket{\phi_{n'l'm'}^1}= a_{nl,n'l'}b_{lm,l'm'}.
\end{align}
The matrices $b_{lm,l'm'}$ are often sparse due to angular momentum selection rules so that these summations can be efficiently evaluated. To improve the accuracy of this approximation and reduce the inaccuracies created by the symmetry-breaking of the external fields, it is advantageous to increase $l_\text{min}$ (we set $l_\text{min}=5$) to help compensate for the degeneracy-breaking of the fields.

 \begin{figure}[t]
 \begin{center}
\includegraphics[width=1.2\columnwidth]{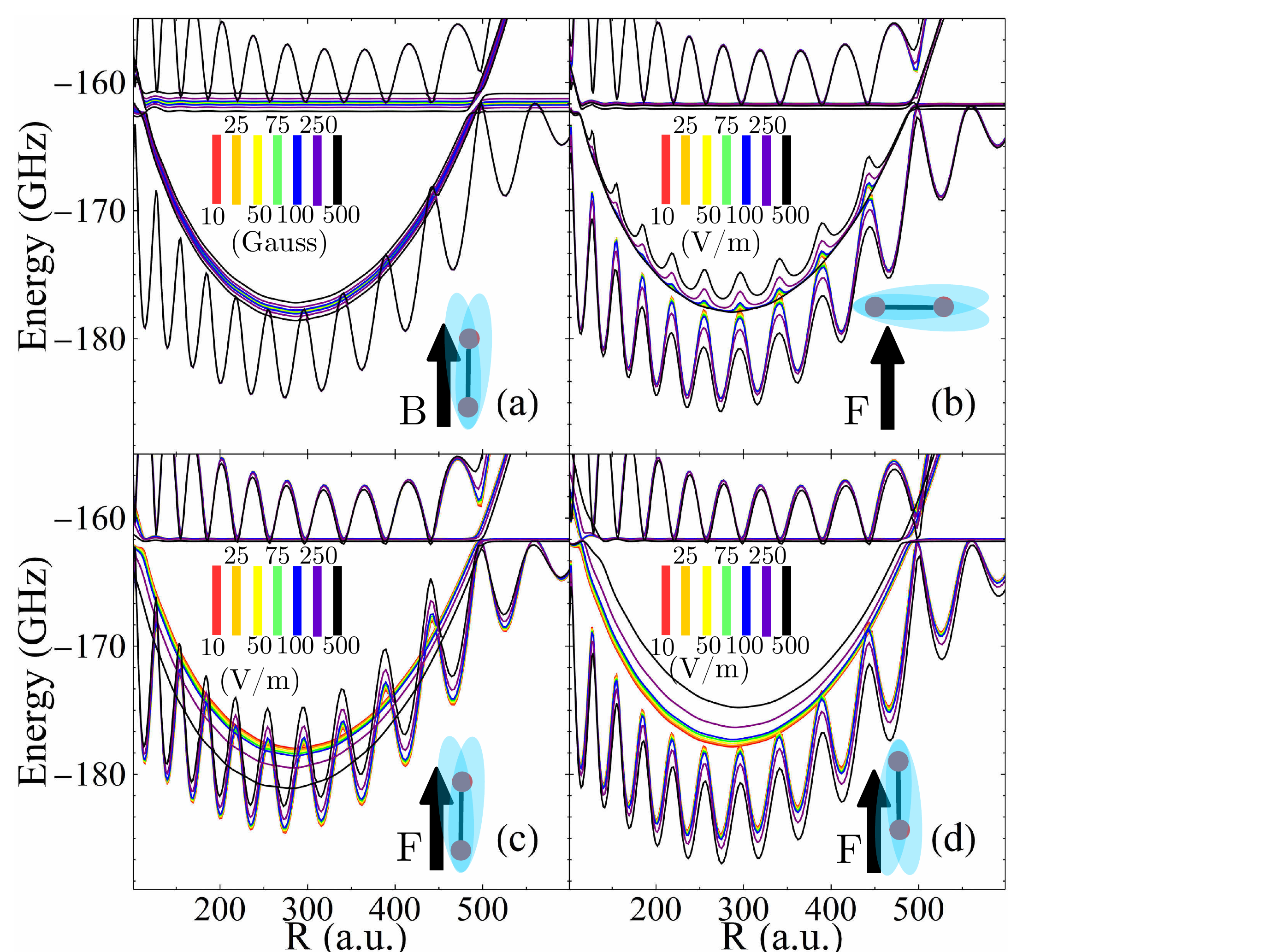}
\hspace{-50pt}
\caption{Radial cuts in the $n=30$ Rb$_2$ butterfly potential energy surfaces in the presence of a) a magnetic field parallel to the internuclear axis, b) an electric field perpendicular to the internuclear axis, c) an electric field parallel to the internuclear axis, d) an electric field anti-parallel to the internuclear axis. Several different fields strengths are plotted.  }
 \label{fig:butterflyfields}
\end{center}
 \end{figure}

 \begin{figure*}[h]
 \begin{center}
\includegraphics[width=\textwidth]{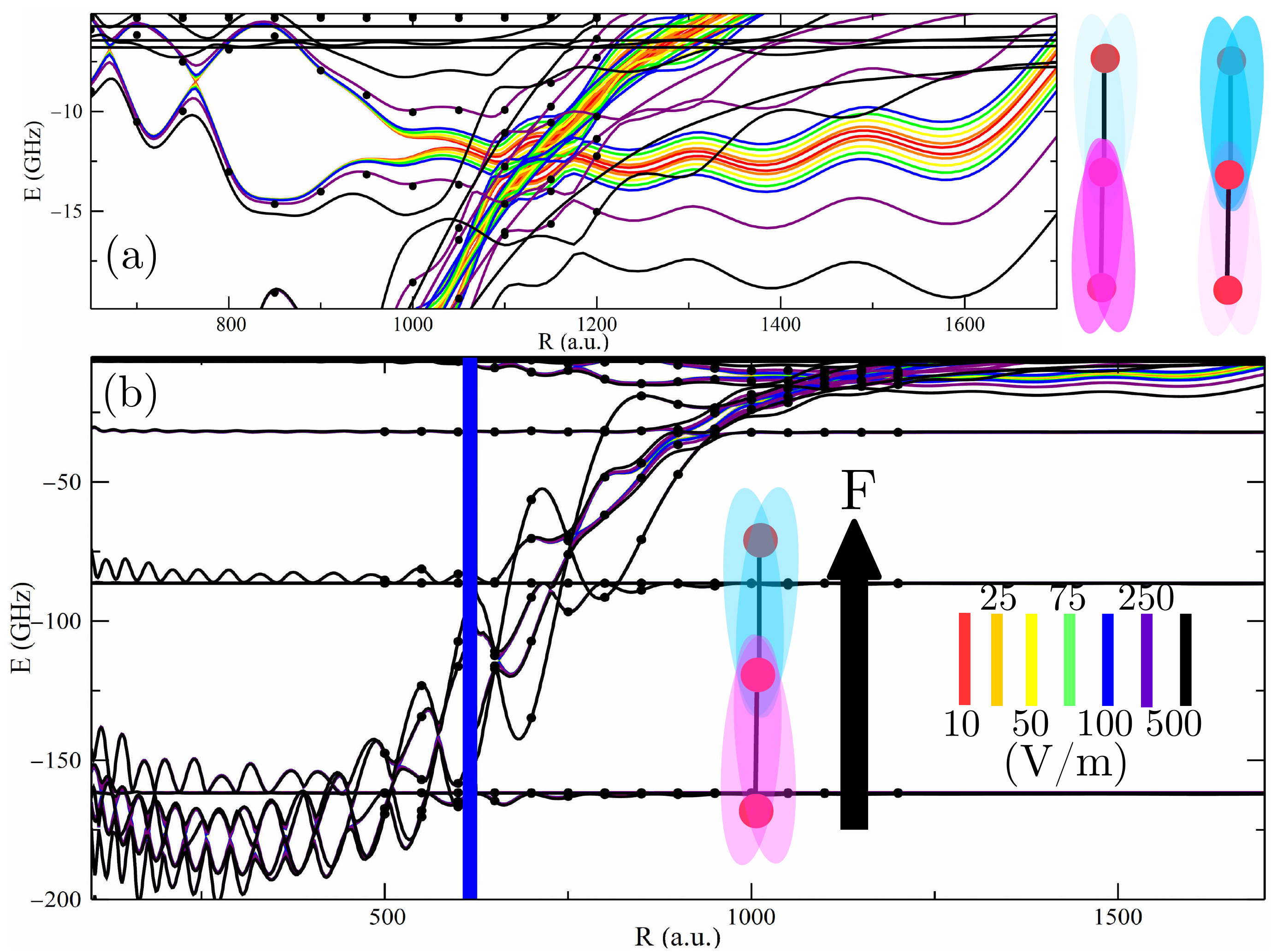}
\caption{Rb$_3$ collinear trimer breathing mode PECs in the presence of an electric field aligned with the internuclear axis. The curved lines were calculated with the approximate trilobite basis approach, while the dots for the 250V/cm calculation were computed using exact diagonalization. The top panel shows the trilobite region. The bottom panel shows the full range of PECs. The vertical blue line marks the shape resonance position.\editm{This figure and B field figure changed to remove arrows}}
 \label{fig:efieldtrilobites}
\end{center}
 \end{figure*}

 \begin{figure*}[h]
 \begin{center}
\includegraphics[width=\textwidth]{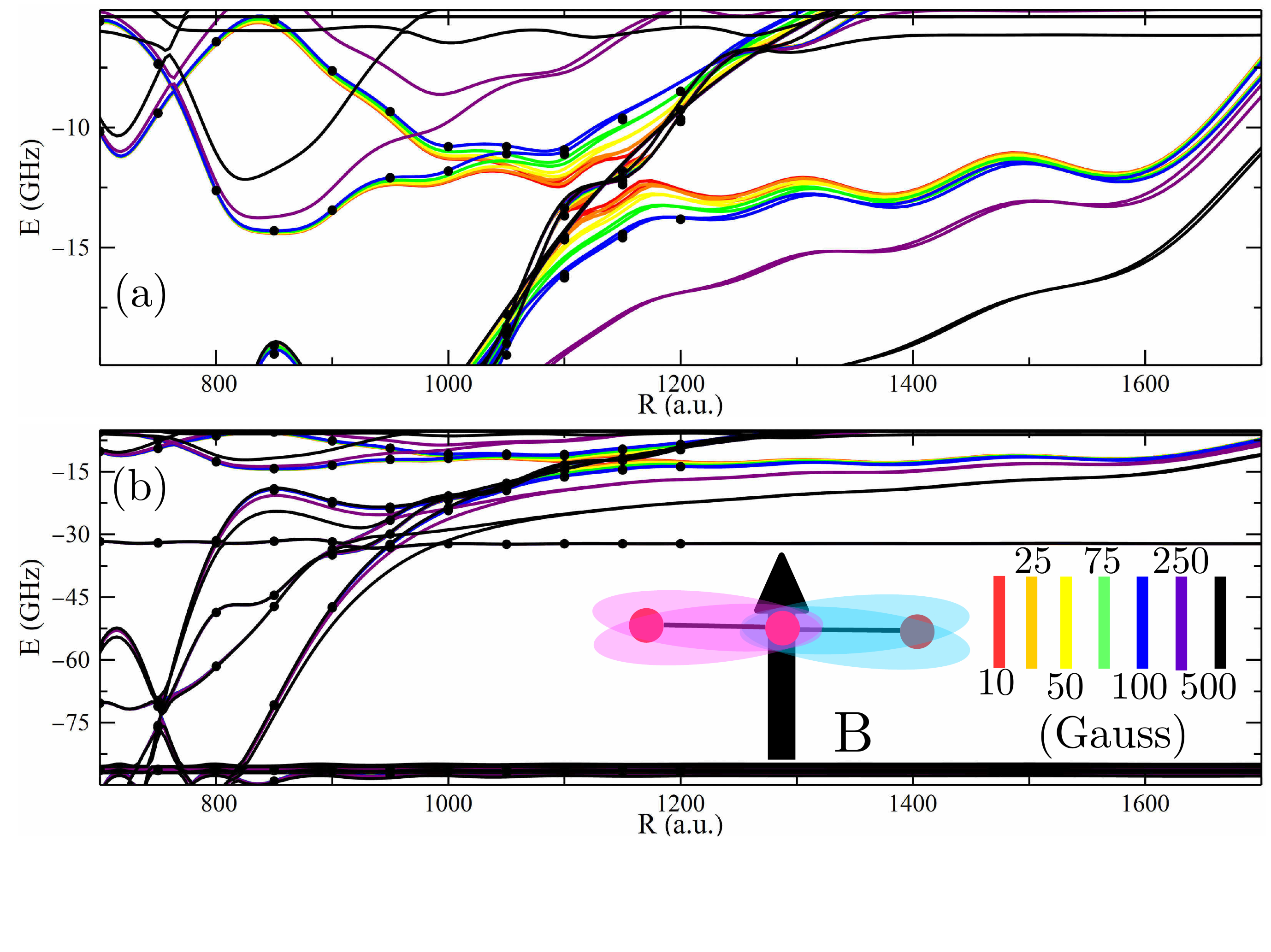}
\vspace{-50pt}
\caption{Rb$_3$ collinear trimer breathing mode PECs in the presence of an magnetic field aligned perpendicular to the internuclear axis. The curved lines were calculated with the approximate trilobite basis approach, while the dots for the 100G calculation were computed using exact diagonalization. The top panel shows the trilobite region. The bottom panel shows an enlarged view, but the butterfly region is not shown here as it is barely affected as in the electric field case. }
 \label{fig:bfieldtrilobites}
\end{center}
 \end{figure*} 
 
 In Fig. \ref{fig:butterflyfields} we present, for the first time, radial cuts through the potential energy surfaces for butterfly molecules in the presence of external fields of varying strength. We chose three different geometries. In Fig. \ref{fig:butterflyfields}a we show that the magnetic field has an almost completely negligible effect on the butterfly potentials in this geometry. This is because the diamagnetic term is very small at these field strengths, and so when the magnetic field and internuclear axes are parallel only non-$\Sigma$ states are shifted by the Zeeman effect. Thus the $P$ Rydberg levels and $\Pi$ butterfly curves evince linear shifts in $B$, but the $\Sigma$ butterfly wells are unchanged.  The shift in the molecular lines is in accordance with the expected scale of the Zeeman splitting $B/2$, which on the order of 1GHz for the highest $B$ field considered here. 
 
The other three panels show the behavior of the PECs when exposed to an electric field. We see in Fig. \ref{fig:butterflyfields}b that the perpendicular configuration, through coupling between $\Sigma$ and $\Pi$ molecular states, does reveal some field dependence through pronounced avoided crossings despite having to first order a vanishing Stark shift in this geometry. These avoided crossings lead to potential wells in the $\Pi$ PECs due to the oscillations in the $\Sigma$ PECs. In Fig. \ref{fig:butterflyfields} c and d the linear Stark shift $d\cdot F$ is quite obvious, as is the change in dipole direction between $\Sigma$ and $\Pi$ states, which shift in opposite directions. Based on the energy shifts witnessed in these figures it is clear that the electric field can provide angular alignment through its effect on the electronic states. 
 
Our results for the dimer states are in good agreement with previous studies \cite{KurzSchmelcherPRA,KurzSchmelcherJPB}, and so we turn to field effects on triatomic molecules. These help elucidate some of the qualitative effects of  different external fields, and have not been studied in detail before \cite{Rosario2016}. We show breathing mode PECs for a collinear Rydberg trimer in the presence of a parallel electric field (Fig. \ref{fig:efieldtrilobites}); this is an interesting study because the trimer, unlike the dimer, has no dipole moment and thus should respond very differently to the external field. We show the same potentials in Fig. \ref{fig:bfieldtrilobites}, but now in the presence of a perpendicular magnetic field.  In each case we also compare with the exact diagonalization for a single field strength, $250$V/cm for the electric field and $100$G for the magnetic field. Even for these high field strengths the agreement is excellent, proving the robustness of the trilobite overlap method. Due to the much lower numerical demands of the trilobite overlap method all PECs could be calculated in less time than was required to compute the few points using the atomic basis, and this allowed us to study butterfly states which could not be converged in earlier, more heavily truncated, calculations \cite{KurzSchmelcherJPB,Rosario2016}. 

The bottom panel of Fig. \ref{fig:efieldtrilobites} gives the global picture of the PECs. Since it has no dipole moment the trimer is unaffected by the electric field over nearly the whole range of $R$. Curiously, the trilobite states respond strongly to the field, as shown in the enlarged top panel view. Here we see yet again the importance of degeneracy in determining energy shifts. For $R<1100$, the PECs are essentially unchanged from the trimer curves seen in Fig. \ref{fig:poly1}. These PECs correspond to non-degenerate gerade and ungerade states. However, as the coupling between this states drops off at large $R$ these states collapse onto the dimer potential curve as shown in Sec. \ref{sec:polyintro}. The electric field now takes advantage of this degeneracy to mix the gerade/ungerade trimer states into upstream/downstream trilobites pointing against/along the electric field. These states now respond to the electric field essentially as the dimer does \cite{KurzSchmelcherPRA}. The size of this splitting is on par with the expected energy shift of a dipole, $dF$. 

The collinear trimer also responds very little to the external magnetic field except near the trilobite wells, and so in Fig. \ref{fig:bfieldtrilobites} we focus on this region. Here, as was also observed in dimers \cite{KurzSchmelcherJPB}, the magnetic field magnifies the coupling between trilobite and butterfly curves, which leads to very large avoided crossings at around $R = 1100$ which strongly destabilize the long-range trilobite wells. This remains dominant here, and in fact one can see that in this orientation the long-range trimer states remain basically degenerate, and thus must not be coupled by the magnetic field. 

Our analysis of these molecules is intended to illustrate how the trilobite overlap method, although approximate in the presence of fields, results in very accurate calculations and has significant technical advantages over the standard atomic basis approach. Furthermore, we discussed some of the structural implications of these fields, but certainly there is far more to study.  As the fields provide angular confinement this section has important implications for the controllability and structure of these molecules in laboratory environments and in future applications.

 \section{Conclusions}
\label{sec:conclusions}

The study of long-range Rydberg molecules has matured into a vibrant  field since its origin almost two decades ago. This has occurred due to theoretical efforts to widen the scope of this topic and especially to the tremendous experimental accomplishments of the last decade. This tutorial has focused on  fundamental aspects of these molecules.  Section \ref{sec:inter} detailed their ``ingredients:'' the highly excited bound states of the Rydberg atom and the very low-energy scattering states of the perturber-electron complex. These concepts are united in the Fermi pseudopotential, which then determines the structure of the Born-Oppenheimer PECs. The fundamental characteristics of these potentials were described in Section \ref{sec:primer}. Experimental effort has demanded a deeper theoretical understanding of these molecules, particularly in the three areas described in Sections \ref{sec:polyintro}-\ref{sec:fieldstudies}. We have therefore built on the foundation of Sec. \ref{sec:primer} to include multiple perturbers, the perturber's fine and hyperfine structure, and external fields. 

What is next for trilobites? Given the unpredictable paths that research takes through nature's mysteries, it is dangerous to speculate too much on what the future of this field might resemble. However, there are several paths based on the major themes of this tutorial that seem likely to even the most cautious prognosticator. 
\begin{itemize}
\item Although the qualitative success of the theory described here is well documented, it still cannot calculate vibrational bound states with an accuracy matching that of modern experimental spectra. This is especially evident in regimes complicated by the $p$-wave shape resonance. The origin of this difficulty is still unclear. One major source of error is that these calculations are only as accurate as the scattering phase shifts used as input. Small errors in the position of the shape resonance and the value of the zero-energy scattering lengths correlate directly with shifts in the PECs. Several groups have attempted to fit their experimental spectra to zero-energy scattering lengths or even to fully energy-dependent phase shift calculations \cite{Sass,DeSalvo2015,MacLennan}, but the diverging values that they have obtained suggest that other uncertainties cloud this approach. Many of these certainly stem from the Fermi pseudopotential, given its problematic $p$-wave divergence and the subtle pathologies of the three-dimensional delta function potential. These issues must be understood before these fits can be expected to yield consistent results.

\item This desire for higher accuracy should motivate the development and use of more accurate techniques, for example the Green's Function methods described in Sec. \ref{sec:primer}. At present, the three generalizations -- polymers, spin, and fields -- discussed here are still dependent on the Fermi model, so whatever alternative techniques are developed must also be able to incorporate these effects.
\item There is still a need for a theoretically rigorous calculation of rovibrational line strengths. This has been explored previously \cite{Markson,OttGroup}, but these studies still lack conclusive assigments of vibrational states, particularly excited ones. The understanding of the mean field lineshapes in dense environments is an impressive success  \cite{Schlag, Whalen2}, but this theory must be generalized to non-$S$ state LRRMs. 
\item Due to the many degrees of freedom of the polyatomic molecules presented here, detailed information about the spectra of trimers or tetramers in trilobite-like states is still absent even from a theoretical perspective. Since these polymers will have many metastable configurations their spectra will be very complicated and probably only observable as broadened dimer lines, as was observed in Cs $nD$ trimers \cite{FeyNew}. Since nearly all experimental exploration of polyatomic molecules has thus far been restricted only to isotropic Rydberg $S$ states, there is very little known about polyatomic signatures in anisotropic Rydberg states. 
\item  Studies in structured environments -- such as a Rydberg atom in an optical lattice or near engineered mesoscopic structures like a tube or a torus -- could take advantage of the enhanced sensitivity to the environment's symmetry seen in polyatomic systems.  Non-adiabatic effects and conical intersections stemming from symmetry breaking, as in the Jahn-Teller effect, could be studied. LRRMs can span multiple lattice sites and hence could be used to probe lattice disorder. 
\item The spin-dependent approach of Sec. \ref{sec:spinintro} has not yet been generalized so that it can be unified with the trilobite overlap method. This should be fairly straightforward to accomplish, and it will have a significant impact on the practicality of including spin into a study of polyatomic molecules. As the present moment the enlarged Hilbert space including the hyperfine structure of each perturber rapidly increases the basis size needed for diagonalization \cite{FeyNew,Hummel}, and the massive reduction in basis size within the trilobite overlap approach would dramatically reduce the computational effort. 
\item Building off of the interactions between aligned butterfly molecules described in Sec. \ref{sec:fieldstudies}, one could explore the consequences of the inherently mixed nature of a system of both Rydberg atoms and LRRMs immersed in a sea of bosonic (or fermionic) atoms. This would be a new regime for studying impurity physics with strong interactions in the ultracold. Likewise, it could even be possible to form more exotic systems consisting of Rydberg atoms bound to LRRMs, an extreme version of the Rydberg-polar perturber molecules predicted in Refs. \cite{Rittenhouse2010,Rittenhouse2011,Mayle2012,Rosario2015,polarperturberpoly} . Mixtures of Rydberg atoms with dipolar molecules like those studied in Refs. \cite{Kuznetsova1,Kuznetsova2,Kuznetsova3} could be realized at exaggerated scales in such a system. Finally, the intermolecular interactions could even lead to the formation of tetramers, bound states between two LRRMs. This would be a fascinating unification of LRRMs and macrodimers.
\item The decay mechanism of these molecules is still not fully understood, and could be influenced by a variety of non-adiabatic processes. 
\end{itemize}

We conclude with a rapid summary of recent developments that have arisen as researchers have begun to study LRRMs not for their intrinsic interest but for their utility as microscopic tools. This reveals the versatility of these molecules in many new regimes. For example, LRRMs formed in an optical lattice can non-destructively probe the Mott transition  \cite{NiederpruemMott}. They could be used to probe density distributions in ultracold mixtures \cite{EilesHetero}. Recently it was proposed that a Rydberg-molecule dressed state could act as a tunable optical Feshbach resonance, and this has already been successfully implemented \cite{RosarioFR,OttFR}. LRRMs can serve as microscopic ``colliders'' to study atom-ion \cite{NewPfau} and electron-atom  \cite{CsReview,MacLennan} scattering. Rydberg atoms can interact with many perturbers in a BEC and can either induce condensate losses or imprint a phase on the condensate\cite{Wuester1,Karpiuk2017,PfauBEC,TiwariWuester}. In a mixed system with two different species of atoms they can probe local density distributions \cite{EilesHetero}\editm{, or they can study the effects of fermionic or bosonic statistics by probing the pair correlation function\cite{WhalenFermions}}. They are the launching point for studies of ultracold ion-atom hybrid systems  \cite{IonColdMeinert,IonRydBlockade}. Formation of LRRMs has been linked to oscillations in photon retrieval measurements in Rydberg electromagnetically-induced transparency experiments \cite{RydbergEIT}.  Finally, some exotic predictions have been made supporting even more unusual types of LRRMs. One such type is a ``ghost'' molecule formed when ultrafast electric and magnetic field pulses are used to shape a Rydberg wave function so that it is identical to the trilobite wave function, despite the absence of a perturber atom. It then looks as if the Rydberg atom is bound to a point in empty space \cite{EilesGhost}. A second type occurs in a totally different context. Theoretical evidence supports the existence and stability of ``Rydberg nuclear molecules'' composed of Rydberg-like states of halo nuclei \cite{NucRydMols}. This last point illustrates the remarkable convergence of diverse physical ideas, and allows us to conclude this tutorial on a a symmetric note. Just as we began with Fermi's pioneering work, we conclude unexpectedly with a connection to nuclear physics, another field where Fermi -- who was instrumental in creating the first nuclear reactor -- also played an oversized role.

\ack {I am greatly indebted to Chris Greene for his invaluable support and guidance through the zoo of Rydberg physics, and for his keen physical insight and intuition into all of these intriguing topics.  I thank F. Robicheaux for encouraging me to take on the project of turning my dissertation into this tutorial and for many valuable ideas and advice. I studied several aspects of Rydberg physics in close collaboration with Jesus Perez-Rios, whose enthusiastic help was greatly appreciated. I also thank Christian Fey and Frederic Hummel for many helpful comments and enlightening discussions about polyatomic molecules.  At the MPI-PKS I have benefited from many discussions with A. Eisfeld, J.M. Rost, and P. Giannakeas. This work is supported in part by the National Science Foundation under Grants PHY-1306905 and PHY-1404419, and by the Max Planck Society.}


\bibliographystyle{iopart-num}
\providecommand{\newblock}{}

\end{document}